\documentclass[12pt]{article}
\usepackage{latexsym}
\usepackage{amsmath,amsfonts}
\usepackage{times}
\allowdisplaybreaks[4]

\catcode`\@=11

\newcount\hour
\newcount\minute
\newtoks\amorpm \hour=\time\divide\hour by 60\minute
=\time{\multiply\hour by 60 \global\advance\minute by-\hour}
\edef\standardtime{{\ifnum\hour<12 \global\amorpm={am}%
        \else\global\amorpm={pm}\advance\hour by-12 \fi
        \ifnum\hour=0 \hour=12 \fi
        \number\hour:\ifnum\minute<10
        0\fi\number\minute\the\amorpm}}
\edef\militarytime{\number\hour:\ifnum\minute<10
0\fi\number\minute}

\def\draftlabel#1{{\@bsphack\if@filesw {\let\thepage\relax
   \xdef\@gtempa{\write\@auxout{\string
      \newlabel{#1}{{\@currentlabel}{\thepage}}}}}\@gtempa
   \if@nobreak \ifvmode\nobreak\fi\fi\fi\@esphack}
        \gdef\@eqnlabel{#1}}
\def\@eqnlabel{}
\def\@vacuum{}
\def\marginnote#1{}
\def\draftmarginnote#1{\marginpar{\raggedright\scriptsize\tt#1}}
\def\draft{
        \pagestyle{plain}
        \overfullrule=2pt
        \oddsidemargin -.5truein
        \def\@oddhead{\sl \phantom{\today\quad\militarytime} \hfil
        \smash{\Large\sl DRAFT} \hfil \today\quad\militarytime}
        \let\@evenhead\@oddhead
        \let\label=\draftlabel
        \let\marginnote=\draftmarginnote
        \def\ps@empty{\let\@mkboth\@gobbletwo
        \def\@oddfoot{\hfil \smash{\Large\sl DRAFT} \hfil}
        \let\@evenfoot\@oddhead}
        \def\@eqnnum{(\theequation)\rlap{\kern\marginparsep\tt\@eqnlabel}%
        \global\let\@eqnlabel\@vacuum}  }

\newcommand{\rf}[1]{(\ref{#1})}
\renewcommand{\theequation}{\thesection.\arabic{equation}}
\renewcommand{\thefootnote}{\fnsymbol{footnote}}
\newcommand{\newsection}{    
\setcounter{equation}{0}\section}

\def\appendix#1{\addtocounter{section}{1}\setcounter{equation}{0}
\renewcommand{\thesection}{\Alph{section}}
\section*{Appendix \thesection\protect\indent \parbox[t]{11.15cm}{#1}}
\addcontentsline{toc}{section}{Appendix \thesection\ \ \ #1}}

\def\nline{\,\nabla\kern -0.7em\raise0.2ex\hbox{/}\,\,}
\def\yline{\,y\kern -0.47em /}
\def\aline{\,a\kern -0.49em /}
\def\parline{\,\partial\kern -0.55em /\,\,}

\newcommand{\No}{\mathbb{N}}
\newcommand{\Po}{\mathbb{P}}

\newcommand{\Zo}{\mathbb{Z}}

\def\be{\begin{equation}}
\def\ee{\end{equation}}
\def\beq{\begin{eqnarray}}
\def\eeq{\end{eqnarray}}

\def\Rsm{{\scriptscriptstyle R}}
\def\Lsm{{\scriptscriptstyle L}}

\def\smone{{\scriptscriptstyle (1)}}
\def\smtwo{{\scriptscriptstyle (2)}}

\def\smpt{{\scriptscriptstyle [2]}}
\def\smp3{{\scriptscriptstyle [3]}}

\def\smpn{{\scriptscriptstyle [n]}}

\def\Jbf{{\bf J}}
\def\Ebf{{\bf E}}

\def\Pbf{{\bf P}}
\def\Sbf{{\bf S}}
\def\Xbf{{\bf X}}
\def\Ubf{{\bf U}}

\def\ibf{{\bf i}}
\def\iibf{{\bf ii}}
\def\iiibf{{\bf iii}}

\def\oplussm{{\scriptscriptstyle \oplus}}
\def\ominussm{{\scriptscriptstyle \ominus}}

\def\MM{{\cal M}}

\def\PP{{\cal P}}
\def\VV{{\cal V}}

\def\half{\frac{1}{2}}

\def\Cb{{\bar{C}}}

\def\Vb{{\bar{V}}}
\def\Xb{{\bar{X}}}
\def\Yb{{\bar{Y}}}
\def\vb{{\bar{v}}}

\def\irm{{\rm i}}

\def\dyn{{\rm dyn}}

\def\fix{{\rm fix}}

\def\diff{{\rm diff}}
\def\GP{{\rm GP}}

\def\betach{\check{\beta}}

\def\noinbf#1{\noindent {\bf #1}}

\begin{document}


\begin{flushright}
FIAN-TD-2021-13  \ \ \ \\
arXiv: yymm.nnnnn
\end{flushright}

\vspace{1cm}

\begin{center}

{\Large \bf Superfield approach to interacting N=2  massive and massless

\medskip
supermultiplets in 3d flat space}

\vspace{2.5cm}

R.R. Metsaev\footnote{ E-mail: metsaev@lpi.ru }

\vspace{1cm}

{\it Department of Theoretical Physics, P.N. Lebedev Physical
Institute, \\ Leninsky prospect 53,  Moscow 119991, Russia }

\vspace{3cm}

{\bf Abstract}

\end{center}

Massive arbitrary spin supermultiplets and massless (scalar and spin one-half) supermultiplets of the N=2 Poincar\'e superalgebra in three-dimensional flat space are considered. Both the integer spin and half-integer spin supermultiplets are studied.  For such massive and massless supermultiplets, a formulation in terms of light-cone gauge unconstrained superfields defined in a momentum superspace is developed. For the supermultiplets under consideration a superspace first derivative representation  for all cubic interaction vertices is obtained. A superspace representation for dynamical generators of the N=2 Poincar\'e superalgebra is also found.

\vspace{3cm}


\newpage
\renewcommand{\thefootnote}{\arabic{footnote}}
\setcounter{footnote}{0}

\section{ \large Introduction}

In view of simplicity and aesthetic features of field theories in three dimensions these theories have attracted a considerable interest during long period of time. Lagrangian formulation of free arbitrary spin massive bosonic and fermionic irreducibe fields propagating in the flat $R^{2,1}$ space was obtained in Ref.\cite{Tyutin:1997yn}.%
\footnote{ Discussion of various interesting aspects of massless and massive higher-spin dynamics in three dimensions may be found in the (incomplete) list of Refs.\cite{Prokushkin:1998bq}-\cite{Kuzenko:2021hyd}.
}
In Ref.\cite{Metsaev:2020gmb}, by using the light-cone gauge formalism, we begun a systematic study of interacting massive arbitrary spin and massless (scalar and spin one-half) fields in three dimensions.%
\footnote{In $3d$, all massless unitary irreps of the Poincar\'e algebra can be described by the spin-0 and spin-$\half$ fields. For this reason, for the case of massless fields, we deal only with spin-0 and spin-$\half$ fields.}
We studied both the fermionic and bosonic fields. We introduced a classification for cubic interactions and, by using a first-derivative representation for cubic interactions,  obtained the explicit expressions for all cubic interaction vertices.
In this paper, we study interacting arbitrary spin massive supermultiplets and massless (scalar and spin one-half) supermultiplets of the N=2 Poincar\'e superalgebra in the $R^{2,1}$. Both the integer spin and half-integer spin supermultiplets are studied.
For such supermultiplets, our aim in this paper, is to provide a first-derivative superspace representation for all cubic interactions. To this end we use a light-cone momentum superspace and unconstrained light-cone gauge superfields defined in such superspace. We note that, it is the light-cone gauge unconstrained superfields we introduce in this paper that allow us to construct a simple superspace representation for all cubic interaction vertices of the supermultiplets under consideration.
For the reader convenience, we note that, in the past, the light-cone momentum superspace has been used successfully  in various interesting studies of supergravity  and superstring  theories. We mention the use of the momentum superspace in 10d extended supergravity in Ref.\cite{Green:1982tk} and superstring field theories in Refs.\cite{Green:1983hw}. Application of the light-cone momentum superspace for the study of light-cone gauge 11d supergravity may be found in Ref.\cite{Metsaev:2004wv}.

In conclusion of the Introduction, let us briefly mention our main two long term motivations for our study of supersymmetric theories of massive and massless fields in the $R^{2,1}$ space. First,  we believe that our results in this paper may be helpful in the search of yet unknown models of supersymmetric higher-spin massive fields in the $R^{2,1}$ space and their cousins in higher dimensions. Second, one expects that a spectrum of states of the superstring in the $AdS_3$ space is realized by arbitrary spin massive fields and low spin massless fields. We think then that our superspace representation for cubic vertices of light-cone gauge fields in the $R^{2,1}$ space might be a good starting point to understand cubic interactions of light-cone gauge fields in the $AdS_3$ space and hence might find applications in various studies of superstring in the $AdS_3$ space.%
\footnote{ In the framework of light-cone gauge approach, superstring in the $AdS_3$ space was studied in Ref.\cite{Metsaev:2000mv}. See also Ref.\cite{Klose:2011rm} for studying 3-point correlator functions of AdS superstring. Interesting discussion of the $N=2$ superstring model in the $R^{2,1}$ space may be found in Ref.\cite{Mezincescu:2011nh}.
}
Further discussion of our long term motivations for our research in this paper can be found in Conclusions.

This paper is organized in the following way.

In Sec.\ref{sec-02}, we start with the presentation of a light-cone frame of the N=2 Poincar\'e superalgebra. After that,  by using light-cone gauge components fields,  we review integer spin and half-integer spin massive and massless supermultiplets of the N=2 Poincar\'e superalgebra.

Sec.\ref{sec-03} is devoted to a superfield description of the N=2 Poincar\'e superalgebra supermultiplets. We introduce a momentum superspace and define the corresponding light-cone gauge unconstrained superfields.
In terms of our superfields,  we describe then a realization of the N=2 Poincar\'e superalgebra on a space of the massive and massless supermultiplets under consideration.

In Sec.\ref{sec-04} we discuss $n$-point interaction vertices. We present constraints on  $n$-point interaction vertices which are obtained by using kinematical symmetries of the N=2 Poincar\'e superalgebra.

Sec.\ref{sec-05} is devoted to cubic vertices. We consider constraints on cubic vertices which are obtained by using  dynamical and kinematical symmetries of the N=2 Poincar\'e superalgebra. After that, we formulate a light-cone gauge dynamical principle and discuss the complete system of equations which are required to determine the cubic interaction vertices uniquely.

In Sec.\ref{sec-06}, we present a superspace form for all our cubic vertices which describe interactions of arbitrary spin massive and massless supermultiplets of the N=2 Poincar\'e superalgebra.
We introduce a notion of critical and non-critical masses of fields entering the cubic vertices.

In Sec.\ref{sec-07}, we present solution for so called $V$-vertices entering superspace form of cubic vertices for the case of the non-critical masses, while, in Sec.\ref{sec-08}, we present solution for $V$-vertices entering superspace form of cubic vertices for the case of the critical masses.

Sec.\ref{sec-09} is devoted to the study of a superspace representation of cubic vertices for the case when all three N=2 supermultiplets entering the cubic vertices are massless.

In Sec.\ref{sec-10}, we present our conclusions.

Our notation and conventions are presented in Appendix A. Properties of our unconstrained superfields are discussed in Appendix B. In appendix C, we outline the derivation of the superspace representation of our cubic vertices.

\newsection{ \large Light-cone gauge formulation of free massive and massless $N=2$ supermultiplets }\label{sec-02}

 We use a method suggested in Ref.\cite{Dirac:1949cp}. According to this method, the finding of a new light-cone gauge dynamical system  amounts to the finding of solution for (anti)commutators of a symmetry algebra.
For N=2 supersymmetric theories of massive and massless fields in the flat space $R^{2,1}$, basic symmetries are associated with the N=2 Poincar\'e superalgebra. In this section, we start therefore with the presentation of (anti)commutators of the N=2 Poincar\'e superalgebra in light-cone frame. After that, we review N=2 supermultiplets in light-cone frame and realization of Poincar\'e algebra on space of massive and massless component fields. Realization of the N=2 Poincar\'e superalgebra on space of supermultiplets will be discussed in the next section by using unconstrained light-cone gauge superfields.

\noindent {\bf Light-cone frame for N=2 Poincar\'e superalgebra in $3d$}. For the flat space $R^{2,1}$, the N=2 Poincar\'e superalgebra consists of the three translation generators $P^\mu$, the three generators of the $so(2,1)$ Lorentz algebra $J^{\mu\nu}$, the $U(1)$ R-symmetry generator denoted as $U$, and four
supercharges. For the Poincar\'e algebra generators, we assume the following commutators:
\be \label{26082021-man02-01}
[P^\mu,\,J^{\nu\rho}]=\eta^{\mu\nu} P^\rho - \eta^{\mu\rho} P^\nu\,,
\qquad {} [J^{\mu\nu},\,J^{\rho\sigma}] = \eta^{\nu\rho} J^{\mu\sigma} + 3\hbox{ terms}\,,
\ee
where $\eta^{\mu\nu}$ stands for the mostly positive Minkowski metric. Explicit form of the remaining (anti)commutators of the N=2 Poincar\'e superalgebra will soon be given below.

In place of the Lorentz frame coordinates $x^\mu$, $\mu=0,1,2$, we use the light-cone basis coordinates $x^\pm$, $x^1$, where the coordinates $x^\pm$ are defined as
\be \label{26082021-man02-02}
x^\pm \equiv \frac{1}{\sqrt{2}}(x^2  \pm x^0)\,,
\ee
and the coordinate $x^+$ is taken to be an evolution parameter. In the light-cone frame, the $so(2,1)$ Lorentz algebra vector $X^\mu$ is decomposed as $X^+,X^-,X$. A scalar product of the $so(2,1)$ Lorentz algebra vectors $X^\mu$ and $Y^\mu$ takes the form
\be \label{26082021-man02-03}
\eta_{\mu\nu}X^\mu Y^\nu = X^+Y^- + X^-Y^+ + X^1 Y^1\,.
\ee
Relation \rf{26082021-man02-03}  tells us that, in the light-cone frame, the metric $\eta_{\mu\nu}$ has the following non-vanishing elements: $\eta_{+-} = \eta_{-+}=1$, $\eta_{11} = 1$. This implies the rules $X^+=X_-$, $X^-=X_+$, $X^1=X_1$.

In the light-cone frame, commutators of the Poincar\'e algebra can simply be obtained from \rf{26082021-man02-01}  by using the flat metric $\eta^{\mu\nu}$ which has the following non-vanishing elements $\eta^{+-}=\eta^{-+}=1$, $\eta^{11}=1$. Using notation $Q^{\pm \Rsm}$, $Q^{\pm \Lsm}$ for the four supercharges, we then note that, in the light-cone frame, the remaining (anti)commutators of the N=2 Poincar\'e superalgebra takes the form,%
\footnote{ We use the N=2 Poincar\'e superalgebra without central charges. Dimensional reduction of the N=1 Poincar\'e superalgebra from $4d$ to $3d$ leads to the appearance of central charges. Therefore we do not expect that our vertices in this paper can be obtained via dimensional reduction from light-cone gauge vertices of N=1 massless supermultiplets obtained in Ref.\cite{Metsaev:2019dqt}. Study of light-cone gauge vertices for N-extended massless supermultiplets in $4d$ may be found in Refs.\cite{Bengtsson:1983pg,Metsaev:2019aig}. For discussion of Lorentz covariant vertices for various massless supermultiplets in $4d$, see, e.g., Refs.\cite{Gates:2019cnl}-\cite{Khabarov:2020bgr} (and references therein)}
\beq
\label{26082021-man02-04} && [J^{+-},Q^{\pm \Rsm} ] = \pm \half Q^{\pm \Rsm}\,, \hspace{1.5cm}
[J^{+-},Q^{\pm \Lsm} ] = \pm \half Q^{\pm \Lsm}\,,
\\
\label{26082021-man02-05} && [Q^{-\Rsm},J^{+1}] = - \frac{1}{\sqrt{2}} Q^{+\Rsm}\,,\hspace{1.3cm} [Q^{-\Lsm},J^{+1}] = - \frac{1}{\sqrt{2}} Q^{+\Lsm}\,,
\\
\label{26082021-man02-06} && [Q^{+\Rsm},J^{-1}] = \frac{1}{\sqrt{2}} Q^{-\Rsm}\,, \hspace{1.6cm}  [Q^{+\Lsm},J^{-1}] = \frac{1}{\sqrt{2}} Q^{-\Lsm}\,,
\\[10pt]
\label{26082021-man02-07} && [U,Q^{\pm \Rsm}]= Q^{\pm \Rsm}\,, \hspace{2.7cm} [U,Q^{\pm \Lsm}]= - Q^{\pm \Lsm}\,,
\\[10pt]
\label{26082021-man02-08} && \{Q^{+\Rsm},Q^{+\Lsm}\} =  P^+\,,  \hspace{2.3cm} \{Q^{-\Rsm},Q^{-\Lsm}\} = - P^-\,,
\\
\label{26082021-man02-09} && \{Q^{+\Rsm},Q^{-\Lsm}\} =  \frac{1}{\sqrt{2}} P^1\,,\hspace{1.8cm} \{Q^{+\Lsm},Q^{-\Rsm}\} =  \frac{1}{\sqrt{2}} P^1\,,
\eeq
where, we recall, that the $U$ appearing in \rf{26082021-man02-07} is a generator of $U(1)$ symmetry ($R$-symmetry). For the generators, we assume the following hermitian conjugation rules:
\beq
&& \hspace{-1.4cm}
\label{26082021-man02-10} P^{\pm \dagger} = P^\pm, \qquad \ \
P^{1\dagger} = P^1, \qquad
J^{+-\dagger} = - J^{+-}\,, \quad
J^{\pm 1\dagger} = -J^{\pm 1}\,, \quad U^\dagger =  U\,,\quad
\nonumber\\
&& \hspace{-1.4cm}
Q^{+\Rsm\dagger} = Q^{+\Lsm}\,, \hspace{0.6cm}  Q^{-\Rsm\dagger} = Q^{-\Lsm}\,. \qquad
\eeq
Generators of the N=2 Poincar\'e superalgebra can be separated into the following two groups:
{\small
\beq
\label{26082021-man02-11}  && \hspace{-1.6cm}
P^+,\quad
P^1,\hspace{0.6cm}
J^{+1},\quad
J^{+-},\quad
U, \quad
Q^{+\Rsm}, \quad
Q^{+\Lsm}, \hspace{1.5cm}
\hbox{ kinematical generators};
\\
\label{26082021-man02-12}   && \hspace{-1.6cm}
P^-, \quad
J^{-1}, \quad
Q^{-\Rsm}, \quad
Q^{-\Lsm}, \quad \hspace{4.5cm}
\hbox{ dynamical generators}.
\eeq
}
In a field realization, with the exception of $J^{+-}$, the kinematical generators \rf{26082021-man02-11} are quadratic in fields%
\footnote{ For interacting fields, the $J^{+-}$ takes the form $J^{+-} = G_0 + \irm x^+ P^-$, where a generator $G_0$ is quadratic in fields, while the dynamical generator  $P^-$ (Hamiltonian) involves quadratic and higher order terms in fields.},
while, the dynamical generators \rf{26082021-man02-12} involve quadratic and higher order terms in fields.

To build a field theoretical realization of the N=2 Poincar\'e superalgebra generators on massive and massless fields, we use a light-cone gauge formulation of the fields. Let us first explain our notation we use for component fields entering the N=2 supermultiplets.
Component fields we use in this paper are denoted as $\phi_{m,\lambda,u}(x)$ and $\psi_{m,\lambda,u}(x)$, where $x$ stands for the space time-coordinates $x^+$, $x^-$, $x^1$,
while the labels $m$, $\lambda$, and $u$ denote the respective mass parameter, spin, and $U(1)$ charge of the fields.
We recall that, for fields in three dimensions, the mass parameter $m$ is allowed to be negative or positive. Sometimes, in the literature, fields with $m>0$ and $m<0$ are refereed to as the respective self-dual and anti-self-dual massive fields. Obviously, the self-dual and anti-self-dual massive fields are associated with different irreps of the Poincar\'e algebra $iso(2,1)$. For bosonic fields, the spin parameter $\lambda$ takes integer values, $\lambda =s$, $s\in \No_0$, while for fermionic fields, the $\lambda$ takes  half-integer values, $\lambda=s+\half$, $s\in \No_0$. The $U(1)$ charge takes values $u\in \Zo$.
We now proceed with the description of massive and massless supermultiplets of the N=2 Poincar\'e superalgebra.

\noindent {\bf $N=2$ massive spin-$s$ supermultiplet}. $N=2$ massive spin-$s$ supermultiplet  which we denote as $\{m,s,u\}_{N=2}$ is described by the following set of massive fields:
\beq
\label{26082021-man02-15} && \{m,s,u\}_{N=2}^{\vphantom{10pt}} = \phi_{m,s,u-1}^\oplussm\,,\quad \phi_{m,s+\half,u}^\oplussm\,,\quad \phi_{m,s-\half,u}^\ominussm\,,\quad \phi_{m,s,u+1}^\ominussm\,,
\nonumber\\
&& \hspace{2.9cm} m\ne 0\,, \qquad s\in \No_0\,,\qquad u\in \Zo\,,
\eeq
where $\phi_{m,s,u-1}^\oplussm$, $\phi_{m,s,u+1}^\ominussm$ are bosonic fields, while $\phi_{m,s+\half,u}^\oplussm$, $\phi_{m,s-\half,u}^\ominussm$ are fermionic fields.%
\footnote{For $s=0$, the field $\phi_{m,s-\half,u}^\ominussm$ in \rf{26082021-man02-15} should read as $\phi_{-m,\half,u}$.}
For the fields in \rf{26082021-man02-15}, we assume the following hermitian conjugation rules
\be \label{26082021-man02-17}
\phi_{m,s,u}^{\oplussm\dagger} = \phi_{m,s,-u}^{\ominussm}\,, \qquad
\phi_{m,s+\half,u}^{\oplussm\dagger} = \phi_{m,s+\half,-u}^{\oplussm}\,, \qquad
\phi_{m,s-\half,u}^{\ominussm\dagger} = \phi_{m,s-\half,-u}^{\ominussm}\,.
\ee
From \rf{26082021-man02-17}, we see that, for $u=0$, the fermionic fields are real-valued and the supermultiplet \rf{26082021-man02-15} is self-adjoint. For $u\ne 0$, all fields in \rf{26082021-man02-15} are complex-valued.

\noindent {\bf $N=2$ massive spin-$(s+\half)$ supermultiplet}.  $N=2$ massive spin-$(s+\half)$ supermultiplet denoted as $\{m,s+\half,u\}_{N=2}$ is described by the following set of massive fields:
\beq
\label{26082021-man02-18} && \{m,s+\half,u\}_{N=2}^{\vphantom{10pt}} = \psi_{m,s+\half,u-1}^\oplussm\,,\quad \psi_{m,s+1,u}^\oplussm\,,\quad \psi_{m,s,u}^\ominussm\,,\quad \psi_{m,s+\half,u+1}^\ominussm\,,
\nonumber\\
&& \hspace{3.7cm} m\ne 0\,, \qquad s\in \No_0\,,\qquad u\in \Zo\,,
\eeq
where $\psi_{m,s+1,u}^\oplussm$, $\psi_{m,s,u}^\ominussm$ are bosonic fields, while
$\psi_{m,s+\half,u-1}^\oplussm$, $\psi_{m,s+\half,u+1}^\ominussm$ are fermionic fields.
For the fields in \rf{26082021-man02-18}, we use the following hermitian conjugation rules:
\be \label{26082021-man02-20}
\psi_{m,s+\half,u}^{\oplussm\dagger} = \psi_{m,s+\half,-u}^{\ominussm}\,, \qquad
\psi_{m,s+1,u}^{\oplussm\dagger} = \psi_{m,s+1,-u}^{\oplussm}\,, \qquad
\psi_{m,s,u}^{\ominussm\dagger} = \psi_{m,s,-u}^{\ominussm}\,.
\ee
Relations \rf{26082021-man02-20} tell us that, for $u=0$, the bosonic fields are real-valued and the supermultiplet \rf{26082021-man02-18} is self-adjoint. . For $u\ne 0$, all fields in \rf{26082021-man02-18} are complex-valued.

\noindent {\bf $N=2$ massless spin-$0$ supermultiplet}. Fields of $N=2$ massless spin-0 supermultiplets are introduced by analogy with the ones for the massive supermultiplets in \rf{26082021-man02-15},
\be  \label{26082021-man02-21}
\{0,0,u\}_{N=2} = \phi_{0,0,u-1}^\oplussm\,,\quad \phi_{0,\half,u}^\oplussm\,,\quad \phi_{0,\half,u}^\ominussm\,,\quad \phi_{0,0,u+1}^\ominussm\,,\hspace{1cm} \qquad u\in \Zo\,,
\ee
where $\phi_{0,0,u-1}^\oplussm$, $\phi_{0,0,u+1}^\ominussm$ are bosonic fields, while $\phi_{0,\half,u}^\oplussm$, $\phi_{0,\half,u}^\ominussm$ are fermionic fields.
For the fields in \rf{26082021-man02-21}, we assume the following hermitian conjugation rules
\be \label{26082021-man02-23}
\phi_{0,0,u}^{\oplussm\dagger} = \phi_{0,0,-u}^{\ominussm}\,, \qquad
\phi_{0,\half,u}^{\oplussm\dagger} = \phi_{0,\half,-u}^{\oplussm}\,, \qquad
\phi_{0,\half,u}^{\ominussm\dagger} = \phi_{0,\half,-u}^{\ominussm}\,.
\ee
Fields in \rf{26082021-man02-21} are decomposed into two invariant subspaces. These invariant spaces will be described in Sec.\ref{sec-03}.

\noindent {\bf $N=2$ massless spin-$\half$ supermultiplet}. Fields of $N=2$ spin-$\half$ massless  supermultiplets are introduced by analogy with the ones for the massive supermultiplets in \rf{26082021-man02-18},
\be \label{26082021-man02-24}
\{0,\half,u\}_{N=2} = \psi_{0,\half,u-1}^\oplussm\,,\quad \psi_{0,0,u}^\oplussm\,,\quad \psi_{0,0,u}^\ominussm\,,\quad \psi_{0,\half,u+1}^\ominussm\,, \hspace{1cm} u\in \Zo\,,
\ee
where $\psi_{0,0,u}^\oplussm$, $\psi_{0,0,u}^\ominussm$ are bosonic fields, while
$\psi_{0,\half,u-1}^\oplussm$, $\psi_{0,\half,u+1}^\ominussm$ are fermionic fields.
For the fields in \rf{26082021-man02-24}, we use the following hermitian conjugation rules
\be \label{26082021-man02-26}
\psi_{0,\half,u}^{\oplussm\dagger} = \psi_{0,\half,-u}^{\ominussm}\,, \qquad
\psi_{0,0,u}^{\oplussm\dagger} = \psi_{0,0,-u}^{\oplussm}\,, \qquad
\psi_{0,0,u}^{\ominussm\dagger} = \psi_{0,0,-u}^{\ominussm}\,.
\ee
Fields in \rf{26082021-man02-24} are decomposed into two invariant subspaces. These invariant spaces and some equivalencies between the massless spin-0 and the spin-$\half$ supermultiplets  are described in Sec.\ref{sec-03}.

In what follows, in place of the fields defined in $x$-space, we prefer to use fields defined in a momentum space. Namely, using the shortcut notation $\chi_{m,\lambda,u}(x)$ for all fields above discussed,
\be \label{26082021-man02-27}
\chi_{m,\lambda,u} = \phi_{m,\lambda,u}^{\oplussm,\ominussm}\,, \ \psi_{m,\lambda,u}^{\oplussm,\ominussm}\,,
\ee
we introduce fields $\chi_{m,\lambda,u}(x^+,p,\beta)$ by using the Fourier transform with respect to the coordinates $x^-$, $x^1$,
\be \label{26082021-man02-28}
\chi_{m,\lambda,u}(x) = \int \frac{ d^2p }{ 2\pi } e^{\irm(\beta x^- +  p x^1)} \chi_{m,\lambda,u}(x^+,p,\beta)\,,\hspace{1.5cm} d^2p \equiv dp\,d\beta\,.
\ee
In what follows, in place of the field $\chi_{m,\lambda,u}(x^+,p,\beta)$,  we use the shortcut $\chi_{m,\lambda,u}(p)$. Note also that, throughout this paper, the field $\chi^\dagger_{m,\lambda,u}(p)$ is defined as $\chi_{m,\lambda,u}^\dagger(p)\equiv (\chi_{m,\lambda,u}(p))^\dagger$.

\noindent {\bf Realization of the Poincar\'e algebra in terms of the fields}. We now recall a field theoretical realization of the Poincar\'e algebra and $U(1)$ symmetry on the space of the component fields above discussed. As is known, the realizations of the Poincar\'e algebra \rf{26082021-man02-01} and $U(1)$ symmetry in terms of differential operators acting on the component fields $\chi_{m,\lambda,u}(p)$ \rf{26082021-man02-28} is given as follows.
\beq
&& \hbox{ \it Realizations on space of fields $\chi_{m,\lambda,u}(p)$:}
\nonumber\\[-3pt]
\label{26082021-man02-29} && P^1 = p\,, \hspace{0.8cm} P^+=\beta\,,\qquad
P^- = p^-\,, \qquad p^- \equiv - \frac{p^2 + m^2}{2\beta}\,,\qquad
\\
\label{26082021-man02-30} && J^{+1} = \irm x^+ P^1 + \partial_p \beta \,, \hspace{1cm} J^{+-} =  \irm x^+ P^- + \partial_\beta \beta - \half e_\lambda\,,
\\
\label{26082021-man02-31} && J^{-1} =- \partial_\beta p + \partial_p p^- +   \frac{\irm m\lambda}{\beta} + \frac{p}{2\beta} e_\lambda \,,
\\
\label{26082021-man02-32} && U = u\,,
\eeq
where partial derivatives $\partial_\beta$, $\partial_p$, and symbol $e_\lambda$ are defined as
\beq
\label{26082021-man02-33} && \partial_\beta\equiv \partial/\partial \beta\,, \quad \partial_p\equiv \partial/\partial p\,,
\\
\label{26082021-man02-34} && e_\lambda =0 \hspace{0.5cm} \hbox{ for integer } \lambda\,,\hspace{1.4cm}
e_\lambda = 1 \hspace{0.5cm} \hbox{ for half-integer } \ \lambda\,.
\eeq

Making use of the relations \rf{26082021-man02-29}-\rf{26082021-man02-34} we can build a field representation for the symmetry algebras. Namely, to quadratic order in the component fields, a field-theoretical realization for generators of the Poincar\'e algebra \rf{26082021-man02-01} and the $u(1)$ algebra is given by
\be \label{26082021-man02-35}
G_\smpt =  2 \int d^2p\,\, \beta^{e_{\lambda +\half}}  \chi_{m,\lambda,u}^\dagger(p) G_\diff \chi_{m,\lambda,u}(p)\,,
\ee
where $G_\diff$ are given in \rf{26082021-man02-29}-\rf{26082021-man02-34}, while
$G_\smpt$ stands for the field-theoretical realization for generators of the Poincar\'e algebra and the $u(1)$ algebra.  To avoid unnecessary complicated expressions in terms of the component fields, the field-theoretical realization of the supercharges $Q^{\pm \Rsm,\Lsm}$ will be given in Sec.\ref{sec-03}, by using superfield formulation of the N=2 supermultiplets.

The fields $\chi_{m,\lambda,u}$ satisfy the Poisson-Dirac equal-time (anti)commutation relations
\be \label{26082021-man02-36}
[\chi_{m,\lambda,u}^{\vphantom{\dagger}}(p),\chi_{m,\lambda',u'}^\dagger(p')]_\pm = \frac{1}{2} \beta^{-e_{\lambda+\half}}\,\, \delta^2(p-p') \delta_{\lambda\lambda'}\delta_{uu'} \,,
\ee
where the notation $[a,b]_\pm$ is used for a graded commutator, $[a,b]_\pm = (-)^{\varepsilon_a \varepsilon_b +1} [b,a]_\pm$.
Making use of the relations \rf{26082021-man02-35}{,\rf{26082021-man02-36}, it is easy to check the standard equal-time commutators between  the components fields and the Grassmann even generators
\be \label{26082021-man02-37}
[ \chi_{m,\lambda,u},G_\smpt\,] =  G_\diff \chi_{m,\lambda,u} \,.
\ee
Free light-cone gauge action for the component field $\chi_{m,\lambda,u}(p)$ takes the well known form
\be \label{26082021-man02-38}
S_\smpt = \int dx^+ d^2p\, \, \beta^{-e_\lambda} \chi_{m,\lambda,u}^\dagger(p) \big( 2\irm \beta \partial^- - p^2 - m^2\big)\chi_{m,\lambda,u}(p)\,,
\ee
where $\partial^-\equiv\partial/\partial x^+$, while $e_\lambda$ is given in \rf{26082021-man02-34}.

\newsection{ \large   Superfield formulation of free massive and massless $N=2$ supermultiplets}   \label{sec-03}

To develop a light-cone gauge superfield formulation of the N=2 supermultiplets under consideration we introduce two Grassmann-odd momenta denoted by $p_\theta$ and $p_\eta$. Our momentum superspace is parametrized then by the light-cone time $x^+$, the momenta $p$, $\beta$ and the two Grassmann momenta $p_\theta$, $p_\theta$,
\be \label{01092021-man02-01}
x^+\,, \ p\,, \ \beta\,, \ \ p_\theta\,, p_\eta\,.
\ee
We now introduce various unconstrained superfields on the superspace \rf{01092021-man02-01} in turn.

\noinbf{Superfield for N=2 massive spin-$s$ supermultiplet}. For N=2 massive spin-$s$ supermultiplet \rf{26082021-man02-15} we introduce Grassmann even superfield $\Phi_{m,s,u}$ defined as
\be
\label{01092021-man02-02} \Phi_{m,s,u}(p,p_\theta,p_\eta) = \beta \phi(p) + p_\theta \phi_\theta(p) +  p_\eta \phi_\eta(p) +  p_\theta p_\eta \phi_{\theta \eta}(p)\,,\qquad
\ee
where component fields entering the Grassmann momenta expansion of the superfield in \rf{01092021-man02-02} are expressed in terms of the component fields of the supermultiplets  \rf{26082021-man02-15} as
\beq
\label{01092021-man02-03} && \phi = \phi_{m,s,u-1}^\oplussm\,, \hspace{4.2cm} \phi_{\theta \eta}= \phi_{m,s,u+1}^\ominussm\,,
\\
\label{01092021-man02-05} && \phi_\theta = \frac{1}{\sqrt{2}}(\irm \phi_{m,s+\half,u}^\oplussm + \phi_{m,s-\half,u}^\ominussm)\,, \hspace{1cm}\phi_\eta = \frac{1}{\sqrt{2}}(\irm\phi_{m,s+\half,u}^\oplussm - \phi_{m,s-\half,u}^\ominussm)\,.
\eeq
An argument $p$ of superfields and fields stands for the momenta $p$ and $\beta$ entering the momentum superspace \rf{01092021-man02-01}.
In terms of the component fields in \rf{01092021-man02-01}, the hermitian conjugation rules in \rf{26082021-man02-17} take the form
\be \label{01092021-man02-07}
\phi_{\theta\eta}^\dagger(p) = \phi(-p)\big|_{u\rightarrow -u}\,, \hspace{1.5cm}
\phi_\theta^\dagger(p) = - \phi_\eta(-p)\big|_{u\rightarrow -u}\,.
\ee
We recall that, for all fields, we use the convention $\chi^\dagger(p)\equiv (\chi(p))^\dagger$.

\noinbf{Superfield for N=2 massive spin-$(s+\half)$ supermultiplet}. For N=2 massive spin-$(s+\half)$ supermultiplet \rf{26082021-man02-18} we introduce Grassmann odd superfield $\Psi_{m,s+\half,u}$ defined as
\be
\label{01092021-man02-10} \Psi_{m,s+\half,u}(p,p_\theta,p_\eta) =  \beta \psi(p) +  \beta p_\theta \psi_\theta(p) + \beta p_\eta \psi_\eta(p) +   p_\theta p_\eta \psi_{\theta \eta}(p)\,,
\ee
where component fields entering the Grassmann momenta expansion of the superfield in \rf{01092021-man02-10} are expressed in terms of the component fields of the supermultiplets defined in \rf{26082021-man02-18} as
\beq
\label{01092021-man02-11} && \psi = \psi_{m,s+\half,u-1}^\oplussm\,, \hspace{4.1cm} \psi_{\theta\eta} = \psi_{m,s+\half,u+1}^\ominussm\,,
\\
\label{01092021-man02-13} && \psi_\theta= \frac{1}{\sqrt{2}}(\irm \psi_{m,s+1,u}^\oplussm + \psi_{m,s,u}^\ominussm)\,, \hspace{1.9cm} \psi_\eta = \frac{1}{\sqrt{2}}(\irm\psi_{m,s+1,u}^\oplussm - \psi_{m,s,u}^\ominussm)\,.
\eeq
In terms of the component fields \rf{01092021-man02-10}, the hermitian conjugation rules  \rf{26082021-man02-20} take the form
\be \label{01092021-man02-15}
\psi_{\theta\eta}^\dagger(p) = \psi(-p)\big|_{u\rightarrow -u}\,, \hspace{1.5cm}
\psi_\theta^\dagger(p) = - \psi_\eta(-p)\big|_{u\rightarrow -u}\,.
\ee

\noinbf{Superfield for N=2 massless spin-$0$ supermultiplet}. For N=2 massless spin-$0$ supermultiplet \rf{26082021-man02-21}, we introduce Grassmann even superfield denoted as $\Phi_{0,0,u}$. This superfield takes the Grassmann momenta expansion identical to the one in \rf{01092021-man02-02}, where fields entering the Grassmann momenta expansion of the superfield $\Phi_{0,0,u}$ are expressed in terms of component fields of the supermultiplets defined in \rf{26082021-man02-21} as
\beq
\label{01092021-man02-16} && \phi = \phi_{0,0,u-1}^\oplussm\,, \hspace{3.4cm} \phi_{\theta \eta}= \phi_{0,0,u+1}^\ominussm\,,
\\
\label{01092021-man02-17} && \phi_\theta = \frac{1}{\sqrt{2}}(\irm \phi_{0,\half,u}^\oplussm + \phi_{0,\half,u}^\ominussm)\,, \hspace{1cm} \phi_\eta = \frac{1}{\sqrt{2}}(\irm\phi_{0,\half,u}^\oplussm - \phi_{0,\half,u}^\ominussm)\,.
\eeq
In terms of the component fields entering the superfield $\Phi_{0,0,u}$, hermitian conjugation rules in \rf{26082021-man02-23} take the same form as in \rf{01092021-man02-07}.
Note that the massless fields $\phi$, $\phi_\theta$ and  $\phi_\eta$, $\phi_{\theta\eta}$ in \rf{01092021-man02-16},\rf{01092021-man02-17} constitute two invariant subspaces under transformations of the $N=2$ Poincar\'e superalgebra.

\noinbf{Superfield for N=2 massless spin-$\half$ supermultiplet}.  For N=2 massless spin-$\half$ supermultiplet \rf{26082021-man02-24}, we introduce Grassmann odd superfield denoted as $\Psi_{0,\half,u}$. This superfield takes the Grassmann momenta expansion identical to the one in \rf{01092021-man02-10}, where fields entering the Grassmann momenta expansion of the superfield $\Psi_{0,\half,u}$ are expressed in terms of the component fields of the supermultiplets defined in \rf{26082021-man02-24} as
\beq
\label{01092021-man02-20} && \psi = \psi_{0,\half,u-1}^\oplussm\,, \hspace{3.3cm}  \psi_{\theta\eta} = \psi_{0,\half,u+1}^\ominussm\,,
\\
\label{01092021-man02-21} && \psi_\theta= \frac{1}{\sqrt{2}}(\irm \psi_{0,0,u}^\oplussm + \psi_{0,0,u}^\ominussm)\,, \hspace{1cm} \psi_\eta = \frac{1}{\sqrt{2}}(\irm\psi_{0,0,u}^\oplussm - \psi_{0,0,u}^\ominussm)\,.
\eeq
In terms of the component fields entering the superfield $\Psi_{0,\half,u}$, hermitian conjugation rules in \rf{26082021-man02-26} take the same form as in \rf{01092021-man02-15}.
The massless fields $\psi$, $\psi_\theta$ and  $\psi_\eta$, $\psi_{\theta\eta}$ in \rf{01092021-man02-20},\rf{01092021-man02-21} constitute two invariant subspaces under transformations of the $N=2$ Poincar\'e superalgebra. We note also some equivalencies between the N=2  massless spin-0 and spin-$\half$ supermultiplets. Namely, under action of the N=2 Poincar\'e  superalgebra generators, the fields $\phi,\irm \phi_\theta$ entering the superfield $\Phi_{0,0,u}$ transform in the same way as the fields $\irm\psi_\eta, \psi_{\theta\eta}$ entering the superfield $\Psi_{0,\half,u-1}$, while,
the fields $\irm\phi_\eta,\phi_{\theta\eta}$ entering the superfield $\Phi_{0,0,u}$ transform in the same way as the fields $\psi, \irm \psi_\theta$ entering the superfield $\Psi_{0,\half,u+1}$. This implies the following equivalence
\be
\sum_{u\in \Zo} \oplus\, \{0,0,u\}_{_{N=2}} \sim \sum_{u\in \Zo}\oplus\,\{0,\half,u\}_{_{N=2}} \hspace{1cm} \hbox{for N=2 massless supermultiplets}.
\ee

In order to treat the above discussed superfields $\Phi$ and $\Psi$ on an equal footing and to simplify our analysis we collect our superfields into new superfields denoted as $\Theta_{m,\lambda,u}$,
\beq
\label{01092021-man02-24} && \Theta_{m,\lambda,u}(p,p_\theta,p_\eta)\,, \qquad \lambda = \hbox{$0,\half, 1,\ldots , \infty$}\,, \qquad m\ne 0;
\\
\label{01092021-man02-25} && \Theta_{0,\lambda,u}(p,p_\theta,p_\eta)\,, \qquad \lambda = \hbox{$0,\half$}\,, \hspace{2.7cm} m=0\,,
\eeq
where, depending on the values of $\lambda$ and $m$, the new superfields  $\Theta$ are identified with the $\Phi$ and $\Psi$ in \rf{01092021-man02-02},\rf{01092021-man02-10} in the following way:
\beq
\label{01092021-man02-26} && \hspace{-0.5cm} \Theta_{m,s,u} \equiv \Phi_{m,s,u}\,, \hspace{1cm} \Theta_{m,s+\half,u} \equiv \Psi_{m,s+\half,u}\,, \hspace{1cm} m\ne 0\,, \hspace{1cm} s =0,1,2,\ldots, \infty;\qquad
\nonumber\\
\label{01092021-man02-27} &&  \hspace{-0.5cm} \Theta_{0,0,u} \equiv \Phi_{0,0,u}\,, \hspace{1.3cm} \Theta_{0,\half,u} \equiv \Psi_{0,\half,u}\,, \hspace{1.9cm} m =0 \,.
\eeq
We note that, for the integer $\lambda$, the new superfield $\Theta_{m,\lambda,u}$ is Grassmann even, while, for the half-integer $\lambda$, the new superfield $\Theta_{m,\lambda,u}$ is Grassmann odd. Therefore, making use of the notation $\GP(\Theta_\lambda)$ for the Grassmann parity of the new superfields $\Theta_{m,\lambda,u}$ and recalling the definition of $e_\lambda$ \rf{26082021-man02-34}, we note the obvious relation,
\be \label{01092021-man02-28}
\GP(\Theta_{m,\lambda,u}) = {e_\lambda}\,.
\ee

\noinbf{N=2 Poincar\'e superalgebra realization on superfield $\Theta_{m,\lambda,u}$}. We find the realization of the N=2 Poincar\'e superalgebra in terms of differential operators acting on the superfield $\Theta_{m,\lambda,u}(p,p_\theta)$:
\beq
\label{01092021-man02-29} && P^1 = p\,, \qquad P^+ = \beta\,, \qquad P^- = p^-\,, \qquad p^- = -\frac{p^2 + m^2}{2\beta}\,,
\\
 \label{01092021-man02-30} && J^{+1}  = \irm x^+ P^1 + \partial_p \beta\,, \hspace{2cm} J^{+-}  = \irm x^+ P^- + \partial_\beta \beta + M^{+-}\,,
\\
\label{01092021-man02-32} && J^{-1} = - \partial_\beta p + \partial_p p^- + \frac{\irm m}{\beta}\big( \lambda  + \half   p_\theta \partial_{p_\eta} + \half p_\eta \partial_{p_\theta}\big) - \frac{p}{\beta} M^{+-}\,,
\\
\label{01092021-man02-33} && U =  1 - u  - p_\theta \partial_{p_\theta} -  p_\eta \partial_{p_\eta} \,,
\\
\label{01092021-man02-34} && Q^{+\Rsm} = (-)^{e_\lambda}\beta \partial_{p_\theta}\,,
\\
\label{01092021-man02-35} && Q^{+\Lsm} = (-)^{e_\lambda} p_\theta\,,
\\
\label{01092021-man02-36} && Q^{-\Rsm} = \frac{(-)^{e_\lambda}}{\sqrt{2}} \big(p  \partial_{p_\theta} + \irm m \partial_{p_\eta}\big)\,,
\\
\label{01092021-man02-37} && Q^{-\Lsm} = \frac{(-)^{e_\lambda}}{\sqrt{2}\,\beta} \big( p p_\theta - \irm m p_\eta\big)\,,
\eeq
where operator $M^{+-}$ is defined as
\be \label{01092021-man02-38}
M^{+-} \equiv \half p_\theta \partial_{p_\theta}  + \half p_\eta \partial_{p_\eta} - 1 - \half e_\lambda\,,
\ee
while the definition of $\partial_\beta$, $\partial_p$ and $e_\lambda$ may be found in \rf{26082021-man02-33} and \rf{26082021-man02-34}. Quantities $\partial_{p_\theta}$, $\partial_{p_\eta}$ stand for the left derivatives of the respective Grassmann momenta $p_\theta$ and $p_\eta$.

To discuss a light-cone gauge action, we find it convenient to use, in addition to the superfields $\Phi_{m,\lambda,u}$ and $\Psi_{m,\lambda,u}$, superfields which we denote as $\Phi_{m,\lambda,u}^*$ and $\Psi_{m,\lambda,u}^*$ respectively. The superfields $\Phi_{m,\lambda,u}^*$, $\Psi_{m,\lambda,u}^*$ are built in terms of the hermitian conjugated component fields appearing in \rf{01092021-man02-02},\rf{01092021-man02-10}. Namely, for $m\ne 0$, the Grassmann even superfield $\Phi_{m,\lambda,u}^*$  and the Grassmann odd superfield  $\Psi_{m,\lambda,u}^*$ are defined by the relations
\beq
\label{01092021-man02-39}  && \hspace{-1.5cm}
\Phi_{m,s,u}^*(p,p_\theta,p_\eta) = \phi_{\theta\eta}^\dagger(p) - \frac{p_\theta}{\beta} \phi_\eta^\dagger(p) + \frac{p_\eta}{\beta} \phi_\theta^\dagger(p) + \frac{p_\theta p_\eta}{\beta} \phi^\dagger(p)\,,
\\
\label{01092021-man02-40} && \hspace{-1.5cm}  \Psi_{m,s+\half,u}^*(p,p_\theta,p_\eta) =   \psi_{\theta\eta}^\dagger(p) +   p_\theta \psi_\eta^\dagger(p) - p_\eta \psi_\theta^\dagger(p) +   \frac{p_\theta p_\eta}{\beta} \psi^\dagger(p)\,.
\eeq
For the massless case, we introduce $\Phi_{0,0,u}^*$ and $\Psi_{0,\half,u}^*$ which take the same Grassmann momenta expansion as in \rf{01092021-man02-39} and \rf{01092021-man02-40} respectively.
As before, in order to treat the superfields $\Phi^*$, $\Psi^*$ on an equal footing, we introduce superfields $\Theta_{m,\lambda,u}^*$,
\beq
\label{01092021-man02-41} && \Theta_{m,\lambda,u}^*(p,p_\theta,p_\eta)\,, \qquad  \lambda = \hbox{$ 0,\half,1,\ldots , \infty$}; \hspace{1cm} m\ne 0\;
\\
\label{01092021-man02-42} && \Theta_{0,\lambda,u}^*(p,p_\theta,p_\eta)\,, \qquad \lambda = \hbox{$0,\half$}\,, \hspace{2.8cm} m =  0\,,
\eeq
where, depending on the values of $m$ and $\lambda$, the superfield  $\Theta^*$ is identified with $\Phi^*$, $\Psi^*$  in the following way
\beq
\label{01092021-man02-43} && \Theta_{m,s,u}^* \equiv \Phi_{m,s,u}^*\,, \hspace{1.3cm} \Theta_{m,s+\half,u}^* \equiv \Psi_{m,s+\half,u}^*\,,
\\
\label{01092021-man02-44} && \Theta_{0,0,u}^* \equiv \Phi_{0,0,u}^*\,, \hspace{1.3cm} \Theta_{0,\half,u}^* \equiv \Psi_{0,\half,u}^*\,.
\eeq
Grassmann parity of the superfield $\Theta_{m,\lambda,u}^*$ is given by the relation
\be \label{01092021-man02-45}
\GP(\Theta_{m,\lambda,u}^*) = {e_\lambda}\,,
\ee
which tells us, for the integer $\lambda$, the $\Theta_{m,\lambda,u}^*$ is Grassmann even, while, for the half-integer $\lambda$, the $\Theta_{m,\lambda,u}^*$ is Grassmann odd.

Relations \rf{01092021-man02-29}-\rf{01092021-man02-37} allow  us to build a superfield representation for the N=2 Poincar\'e superalgebra. Namely, to quadratic order in the superfields, a field-theoretical superspace representation for
the generators of the N=2 Poincar\'e superalgebra is given by
\be \label{01092021-man02-46}
G_\smpt  =  \sum_{2\lambda \in \No_0,\, u\in \Zo} G_{\smpt,\,m,\lambda,u}  \qquad G_{\smpt,\, m, \lambda,u}  = \int d^2p\, dp_\eta dp_\theta \,\, \beta^{e_{\lambda+\half}} \Theta_{m,\lambda,u}^* G_{\diff}\Theta_{m,\lambda,u}\,,\qquad
\ee
where $G_\diff$ are given in \rf{01092021-man02-29}-\rf{01092021-man02-37}.
A realization of $G_\diff$ on space of the superfield $\Theta_{m,\lambda,u}^*$ and some other properties of the superfield $\Theta_{m,\lambda,u}^*$ are discussed in Appendix B.

The superfields $\Theta$, $\Theta^*$ satisfy the Poisson-Dirac equal-time commutation relations
\be \label{01092021-man02-47}
[\Theta_{m,\lambda,u}(p,p_\theta,p_\eta),\Theta_{m,\lambda',u'}^*(p',p_\theta',p_\eta')]_\pm = \half \beta^{- e_{\lambda+\half} } \delta^2(p-p')\delta(p_\theta-p_\theta')\delta(p_\eta-p_\eta') \delta_{\lambda\lambda'}\delta_{uu'}\,,
\ee
where, we recall, $[a,b]_\pm$ is a graded commutator, $[a,b]_\pm = (-)^{\varepsilon_a \varepsilon_b +1} [b,a]_\pm$. Making use of the relations \rf{01092021-man02-46},\rf{01092021-man02-47}, it is easy to verify the standard equal-time (anti)commutator between  the generators and the superfields
\be \label{01092021-man02-48}
[\Theta_{m,\lambda,u}G_\smpt]_{\pm} =  G_\diff \Theta_{m,\lambda,u} \,,
\ee
where expressions for the $G_\diff$ may be found in  \rf{01092021-man02-29}-\rf{01092021-man02-37}.

Finally we note that, in terms of the superfields above discussed, the light-cone gauge action takes the form
\be
S = \half\sum_{2\lambda\in \No_0,\, u\in \Zo} \int dx^+ d^2p\, dp_\eta dp_\theta \,\, \beta^{-e_\lambda} \Theta_{m,\lambda,u}^* \big( 2\irm \beta \partial^- - p^2 - m^2 \big)\Theta_{m,\lambda,u}  +\int dx^+ P_{\rm int}^-\,,
\ee
where $\partial^-\equiv\partial/\partial x^+$ and $P_{\rm int}^-$ stands for light-cone gauge Hamiltonian which describes interacting fields.

\newsection{ \large Kinematical symmetries of $n$-point dynamical generators of N =2 Poincar\'e superalgebra} \label{sec-04}

For interacting fields, the dynamical generators of the N=2 Poincar\'e superalgebra receive corrections having higher powers of the superfields. Namely, for the dynamical generators, one has the following expansion in the superfields
\be \label{27082021-man02-01}
G^\dyn
= \sum_{n=2}^\infty
G_\smpn^\dyn\,,
\ee
where $G_\smpn^\dyn$ is a functional having $n$ powers of the superfields $\Theta^*$. We now describe restrictions imposed on the dynamical generators $G_\smpn^\dyn$, $n\geq 3$, by the kinematical symmetries of the N=2 Poincar\'e superalgebra. We discuss the kinematical symmetry restrictions in turn.

\noindent {\bf Kinematical $P^1$, $P^+$, $Q^{+\Lsm}$ symmetries}. Making use of the (anti)commutators between the dynamical generators \rf{26082021-man02-12} and the kinematical generators $P^1$, $P^+$, $Q^{+\Lsm}$,  we find that the dynamical generators $G_\smpn^\dyn$ with $n\geq 3$ can be cast into the form:
\beq
\label{27082021-man02-02} && P_\smpn^- = \int\!\! d\Gamma_\smpn\,\,  \langle \Theta_\smpn^*  | p_\smpn^-\rangle\,,
\\
\label{27082021-man02-03} && Q_\smpn^{-\Rsm} = \int\!\! d\Gamma_\smpn\,\,  \langle \Theta_\smpn^* | q_\smpn^{-\Rsm}\rangle\,,
\\
\label{27082021-man02-04} && Q_\smpn^{-\Lsm} = \int\!\! d\Gamma_\smpn\,\,  \langle \Theta_\smpn^*  | q_\smpn^{-\Lsm} \rangle\,,
\\
\label{27082021-man02-05} && J_\smpn^{-1} = \int\!\! d\Gamma_\smpn\,\,  \langle\Theta_\smpn^* | j_\smpn^{-1}\rangle  +   \langle \Xbf_\smpn^1 \Theta_\smpn^* | p_\smpn^-\rangle
\nonumber\\
&& \hspace{3cM} + \,\, \frac{1}{\sqrt{2}} \langle \Xbf_{\smpn\,\theta } \Theta_\smpn^* |q_\smpn^{-\Lsm} \rangle - \frac{1}{\sqrt{2}\, n} \langle \PP_{\smpn\,\theta } \Theta_\smpn^* |q_\smpn^{-\Rsm} \rangle\,,\qquad
\eeq
where we are using the following notation
\beq
\label{27082021-man02-06} && d\Gamma_\smpn = d\Gamma_\smpn^p d\Gamma_\smpn^{p_\eta p_\theta} \,,
\\
\label{27082021-man02-07} && d\Gamma_\smpn^p =  \delta^{2}(\sum_{a=1}^n p_a)\prod_{a=1}^n \frac{d^2p_a}{2\pi }\,, \qquad d^2 p_a = dp_a d\beta_a\,,
\\
\label{27082021-man02-08} && d\Gamma_\smpn^{p_\eta p_\theta} \equiv dp_{\eta_1} dp_{\theta_1} \ldots dp_{\eta_n} dp_{\theta_n} \delta\big( \sum_{a=1}^n  p_{\theta_a} \big)\,,
\\
\label{27082021-man02-09} && \Xbf_\smpn^1 =  - \frac{1}{n}\sum_{a=1}^n \partial_{p_a}\,, \hspace{1cm} \Xbf_{\smpn\,\theta} = \frac{1}{n}\sum_{a=1}^n \partial_{p_{\theta a}}\,,\hspace{1cm} \PP_{\smpn \theta}=  \sum_{a=1}^n \frac{p_{\theta_a}}{\beta_a}\,.\qquad
\eeq
In \rf{27082021-man02-02}-\rf{27082021-man02-05}, the expressions $\langle \Theta_\smpn| p_\smpn^-\rangle$, $\langle \Theta_\smpn| q_\smpn^{-\Rsm,\Lsm}\rangle$, and $\langle \Theta_\smpn| j_\smpn^{-1}\rangle$ are shortcuts which are defined  as
\beq
\label{27082021-man02-10} && \langle \Theta_\smpn^*| p_\smpn^-\rangle \quad \equiv \sum_{\sigma_1\ldots\sigma_n} \Theta_{\sigma_1\ldots\sigma_n}^*  p_{\sigma_1\ldots\sigma_n}^-\,,
\\
\label{27082021-man02-11} && \langle \Theta_\smpn^*| q_\smpn^{-\Rsm,\Lsm} \rangle \equiv \sum_{\sigma_1\ldots\sigma_n} \Theta_{\sigma_1\ldots\sigma_n}^*  q_{\sigma_1\ldots\sigma_n}^{-\Rsm,\Lsm}\,,
\\
\label{27082021-man02-12} && \langle \Theta_\smpn^*| j_\smpn^{-1} \rangle \equiv \sum_{\sigma_1\ldots\sigma_n} \Theta_{\sigma_1\ldots\sigma_n}^*  j_{\sigma_1\ldots\sigma_n}^{-1}\,,
\\
\label{27082021-man02-13} && \hspace{2cm} \Theta_{\sigma_1\ldots\sigma_n}^* \equiv  \Theta_{\sigma_1}^*(p_1,p_{\theta_1},p_{\eta_1})  \ldots  \Theta_{\sigma_n}^*(p_n,p_{\theta_n},p_{\eta_n}) \,,
\eeq
where a label $\sigma_a$ is a shortcut for mass parameters, spins, and $U(1)$ charges
\be \label{27082021-man02-13-a1}
\sigma_a = m_a\,, \ \lambda_a\,, \ u_a\,,
\ee
and indices $a,b=1,\ldots,n$ are used to label superfields entering $n$-point vertices.
Throughout this paper the $p_a$ as argument of delta functions \rf{27082021-man02-07} and superfields \rf{27082021-man02-13} stands for the momenta $p_a$ and $\beta_a$.
Sometimes, the quantities $p_{\sigma_1\ldots\sigma_n}^-$, $q_{\sigma_1\ldots\sigma_n}^{-\Rsm,\Lsm}$, and $j_{\sigma_1\ldots\sigma_n}^{-1}$ \rf{27082021-man02-02}-\rf{27082021-man02-05}, will shortly be denoted as $g_\smpn$,
\be \label{27082021-man02-14}
g_\smpn = p_{\sigma_1\ldots\sigma_n}^-,\quad
q_{\sigma_1\ldots\sigma_n}^{-\Rsm},\quad
q_{\sigma_1\ldots\sigma_n}^{-\Lsm},\quad
j_{\sigma_1\ldots\sigma_n}^{-1}\,.
\ee
The quantities $g_\smpn$ \rf{27082021-man02-14} will be referred to as $n$-point densities. %
Often the density $p_\smpn^-$ will be referred to as an $n$-point interaction vertex, while, for $n=3$, the density $p_\smp3^-$ will be referred to as cubic interaction vertex.

In general, the $n$-point densities $g_\smpn$ \rf{27082021-man02-14} depend on the momenta $p_a$, $\beta_a$, Grassmann momenta $p_{\theta_a}$ $p_{\eta_a}$, masses $m_a$, spins $\lambda_a$, and $U(1)$ charges $u_a$, $a=1,2\ldots,n$,
\be \label{27082021-man02-15}
g_\smpn = g_\smpn(p_a,p_{\theta_a},p_{\eta_a},\beta_a)\,.
\ee
Also note that, in \rf{27082021-man02-05}, the differential operators $\Xbf_\smpn^1$, $\Xbf_{\smpn\,\theta}$ act only on the arguments of the superfields. For example, the expression $\langle \Xbf^1 \Theta_\smpn^* | g_\smpn \rangle$ should read as
\be \label{27082021-man02-16}
\langle \Xbf^1 \Theta_\smpn^* | g_\smpn\rangle =  \sum_{\sigma_1,\ldots \sigma_n} (\Xbf^1\Theta_{\sigma_1\ldots\sigma_n}^* ) g_{\sigma_1\ldots\sigma_n}\,.
\ee

\noindent {\bf Kinematical $J^{+-}$-symmetry}. Making use of the commutators  between the dynamical generators $P^-$, $Q^{-\Rsm,\Lsm}$, $J^{-1}$ and the kinematical generator $J^{+-}$, we find the equations for the densities given by:
\beq
\label{27082021-man02-17} && \hspace{-1.5cm} \sum_{a=1}^n  \big( \beta_a\partial_{\beta_a} + \half N_{ p_{\theta a} } + \half N_{ p_{\eta a} } + \half e_{\lambda_a} \big) g_\smpn =   ( n- \half) g_\smpn\,, \ \ \hbox{ for }\ g_\smpn = p_\smpn^-\,, j_\smpn^{-1}\,,\quad
\\
\label{27082021-man02-18} && \hspace{-1.5cm} \sum_{a=1}^n  \big( \beta_a\partial_{\beta_a} + \half N_{ p_{\theta a} } + \half N_{ p_{\eta a} } + \half e_{\lambda_a} \big) g_\smpn =  n  g_\smpn\,, \hspace{1.4cm} \hbox{ for }\ g_\smpn = q_\smpn^{-\Rsm,\Lsm}\,,\quad
\\
\label{27082021-man02-19} &&  N_{p_{\theta a}}\equiv p_{\theta_a}\partial_{p_{\theta_a}}\,, \hspace{1cm}   N_{p_{\eta a} } \equiv  p_{\eta_a} \partial_{p_{\eta_a}}\,.
\eeq

\noindent {\bf Kinematical $J^{+1}$, $Q^{+\Rsm}$-symmetries}. Making use of the (anti)commutators  between the dynamical generators $P^-$, $Q^{-\Rsm,\Lsm}$, $J^{-1}$ and the kinematical generators $J^{+1}$, $Q^{+\Rsm}$, we verify that the dependence of the densities $g_\smpn$ \rf{27082021-man02-14} on the momenta $p_a$ and the Grassmann momenta $p_{\theta_a}$  is realized through new momentum variables $\Po_{ab}$ and $\Po_{\theta\, ab}$  defined by the relations
\be \label{27082021-man02-20}
\Po_{ab} \equiv p_a \beta_b - p_b \beta_a\,, \qquad
\Po_{\theta\, ab} \equiv p_{\theta_a} \beta_b - p_{\theta_b} \beta_a\,.
\ee
Thus the densities $g_\smpn$ \rf{27082021-man02-15} turn out to be functions of the new momenta $\Po_{ab}$ and $\Po_{\theta\, ab}$ in place of the generic momenta $p_a$, $p_{\theta_a}$,
\be \label{27082021-man02-21}
g_\smpn = g_\smpn (\Po_{ab},\, \Po_{\theta\,ab},p_{\eta_a}, \beta_a)\,.
\ee

\noindent {\bf Kinematical $U(1)$-symmetry}. Making use of the commutators  between the dynamical generators $P^-$, $Q^{-\Rsm,\Lsm}$, $J^{-1}$ and the kinematical generator $U$, we find the equations for the densities given by:
\beq
\label{27082021-man02-22} && \sum_{a=1}^n (N_{p_{\theta a}} + N_{p_{\eta a}} +  u_a) p_\smpn^- = (n - 1) g_\smpn^-\,, \hspace{1cm} \hbox{ for } \quad g_\smpn= p_\smpn^-\,,\ j_\smpn^{-1}\,,
\\
\label{27082021-man02-23} && \sum_{a=1}^n (N_{p_{\theta a}} + N_{p_{\eta a}} +  u_a) q_\smpn^{-\Rsm} = (n - 2) q_\smpn^{-\Rsm}\,,\qquad  \sum_{a=1}^n (N_{p_{\theta a}} + N_{p_{\eta a}} +  u_a) q_\smpn^{-\Lsm} = n q_\smpn^{-\Lsm}\,,\qquad \qquad
\eeq
where we use the notation given in \rf{27082021-man02-19}.

Let us now summarize the kinematical symmetry restrictions on the $n$-point densities.

\noindent \ibf) The kinematical $P^1$,  $P^+$, and $Q^{+\Lsm}$ symmetries lead to  delta-functions in expressions for the dynamical generators $P^-$, $Q^{-\Rsm,\Lsm}$, $J^{-1}$ \rf{27082021-man02-02}-\rf{27082021-man02-05}. These symmetries imply the conservation laws for the  momenta $p_a$, $\beta_a$ and the Grassmann momenta $p_{\theta_a}$.

\noindent \iibf) The kinematical $J^{+-}$ symmetry leads to the differential equations given in \rf{27082021-man02-17},\rf{27082021-man02-18}.

\noindent \iiibf)  The kinematical $J^{+1}$ and $Q^{+\Rsm}$ symmetries imply that the $n$-point densities $p_\smpn^-$, $q_\smpn^{-\Rsm,\Lsm}$, $j_\smpn^{-1}$  turn out to be functions that depend on the new momenta $\Po_{ab}$ and $\Po_{\theta ab}$ \rf{27082021-man02-18} in place of the generic momenta $p_a$ and $p_{\theta_a}$ respectively. Conservation laws for the momenta $p_a$, $\beta_a$, $p_{\theta_a}$ imply  that there are $n-2$ independent momenta $\Po_{ab}$ and $n-2$ independent Grassmann momenta $\Po_{\theta ab}$ \rf{27082021-man02-18}. For example, if $n=3$, then there is one independent momentum $\Po$ and one independent Grassmann momentum $\Po_\theta$ (see below).

\newsection{ \large N=2 Poincar\'e superalgebra restrictions for cubic vertex and light-cone gauge dynamical principle } \label{sec-05}

In this Section, firstly, we represent kinematical $J^{+-}$ symmetry equations \rf{27082021-man02-17},\rf{27082021-man02-18} in terms of the momenta $\Po_{ab}$ and $\Po_{ab\, \theta}$. Secondly, we find restrictions imposed by dynamical symmetries.  Finally, we discuss light-cone gauge dynamical principle and write down the complete system equations which allows us to determine the cubic vertices unambiguously.

\noindent {\bf Kinematical symmetries of the cubic densities}. Making use of the momentum conservation laws
\be \label{27082021-man02-30}
p_1 + p_2 + p_3 = 0\,, \quad \beta_1 +\beta_2 +\beta_3 =0 \,,\quad p_{\theta_1} + p_{\theta_2} + p_{\theta_3}=0\,,
\ee
it is easy to check that $\Po_{12}$, $\Po_{23}$, $\Po_{31}$ and Grassmann momenta $\Po_{\theta\, 12}$, $\Po_{\theta\, 23}$, $\Po_{\theta\, 31}$ can be expressed in terms of momenta $\Po$, $\Po_\theta$,
\be \label{27082021-man02-31}
\Po_{12} =\Po_{23} = \Po_{31} = \Po \,,\qquad
\Po_{\theta\, 12} =\Po_{\theta\, 23} = \Po_{\theta\, 31} = \Po_\theta \,,
\ee
where the momenta $\Po$ and $\Po_\theta$ are defined by the relations
\be \label{27082021-man02-32}
\Po \equiv \frac{1}{3}\sum_{a=1,2,3} \betach_a p_a\,, \qquad  \Po_\theta \equiv \frac{1}{3}\sum_{a=1,2,3} \betach_a p_{\theta_a}\,, \qquad \betach_a\equiv \beta_{a+1}-\beta_{a+2}\,, \quad \beta_a\equiv
\beta_{a+3}\,.
\ee
The use of the momenta \rf{27082021-man02-32} is convenient for us because these momenta turn out to manifestly invariant upon cyclic permutations of the external line indices
$1,2,3$. Making use of the simplified notation for the cubic densities,
\be \label{27082021-man02-33}
p_\smp3^- = p_{\sigma_1\sigma_2\sigma_3}^- \,, \qquad  q_\smp3^{-\Rsm,\Lsm} = q_{\sigma_1\sigma_2\sigma_3}^{-\Rsm,\Lsm}\,, \qquad  j_\smp3^{-1}  = j_{\sigma_1\sigma_2\sigma_3}^{-1}\,,
\ee
we note that the cubic densities $p_\smp3^-$, $q_\smp3^{-\Rsm,\Lsm}$, and $j_\smp3^{-1}$ are functions of the momenta $\beta_a$, $\Po$, the Grassmann momenta $\Po_\theta$, $p_{\eta_a}$ and $\sigma_a = m_a, \lambda_a, u_a$, $a=1,2,3$,
\beq
\label{27082021-man02-34} && p_\smp3^- = p_{\sigma_1\sigma_2\sigma_3}^-(\Po,\Po_\theta, p_{\eta_a}, \beta_a)\,, \qquad  q_\smp3^{-\Rsm,\Lsm} = q_{\sigma_1\sigma_2\sigma_3}^{-\Rsm,\Lsm}(\Po,\Po_\theta, p_{\eta_a}, \beta_a)\,, \quad
\nonumber\\
&& j_\smp3^{-1} = j_{\sigma_1\sigma_2\sigma_3}^{-\Rsm,\Lsm}(\Po,\Po_\theta, p_{\eta_a}, \beta_a)\,.
\eeq
Thus we see that the dependence of the cubic densities on the momenta $p_a$ and $p_{\theta_a}$ is realized  through the respective momenta $\Po$ and $\Po_\theta$. In view of this, the study of cubic densities is considerably simplified. Let us now represent  equations given in \rf{27082021-man02-17},\rf{27082021-man02-18} and \rf{27082021-man02-22},\rf{27082021-man02-23} in terms of the cubic densities given in \rf{27082021-man02-34}.

\noindent {\bf $J^{+-}$-symmetry equations}: Making use of the representation for cubic densities in \rf{27082021-man02-34}, we verify that, for $n=3$,  equations \rf{27082021-man02-17},\rf{27082021-man02-18} can be represented as
\be \label{27082021-man02-35}
(\Jbf^{+-} - \frac{5}{2}) p_\smp3^- = 0\,,  \hspace{0.7cm} (\Jbf^{+-} - 3) q_\smp3^{-\Rsm,\Lsm} = 0\,, \hspace{0.7cm} (\Jbf^{+-} -  \frac{5}{2}) j_\smp3^{-1} = 0\,,\qquad
\ee
where $\Jbf^{+-}$ is defined by the relations
\beq
\label{27082021-man02-36} && \Jbf^{+-} \equiv    N_\Po + \frac{3}{2} N_{\Po_\theta} +  \sum_{a=1,2,3}  \big( \beta_a \partial_{\beta_a}  + \half N_{ p_{\eta a} } + \half e_{\lambda_a}\big)\,,
\\
\label{27082021-man02-37} && N_\Po \equiv \Po\partial_\Po\,, \hspace{1cm} \hspace{1cm} N_{\Po_\theta} \equiv \Po_\theta \partial_{\Po_\theta}\,, \qquad
N_{p_{\eta a}} \equiv p_{\eta_a} \partial_{p_{\eta a}}\,.\qquad
\eeq
\noindent {\bf $U(1)$-symmetry equations}: Making use of the representation for the cubic densities in \rf{27082021-man02-34}, we verify that, for $n=3$,  equations \rf{27082021-man02-22},\rf{27082021-man02-23} can be represented as
\beq
\label{27082021-man02-38} && \hspace{-1cm} \big( N_{\Po_\theta} - 2+ \sum_{a=1,2,3} (N_{p_{\eta a}} +  u_a)\big)  g_\smp3^- = 0\,,\hspace{1cm} \hbox{ for } \ \ g_\smp3 = p_\smp3^-\,, \ \ j_\smp3^{-1}\,,
\\
\label{27082021-man02-39} && \hspace{-1cm} \big(N_{\Po_\theta} -1 + \sum_{a=1,2,3} (N_{p_{\eta a}} +  u_a) \big) q_\smp3^{-\Rsm} = 0 \,,\qquad  \big(N_{\Po_\theta} -3 + \sum_{a=1,2,3} ( N_{p_{\eta a}} +  u_a) \big) q_\smp3^{-\Lsm} = 0\,,\qquad \qquad
\eeq
where we use the notation as in \rf{27082021-man02-37}.

We now proceed with studying the restrictions imposed by the dynamical symmetries.

\noindent {\bf Dynamical symmetries restrictions on cubic densities}. Restrictions on the cubic densities obtained from (anti)commutators between the dynamical generators will be referred to as dynamical symmetries restrictions in this paper.
This is to say that now we discuss restrictions obtained from the following
(anti)commutators:
\beq
 \label{27082021-man02-40} && [P^-,J^{-1}]=0\,, \hspace{1.8cm}  [P^-,Q^{-\Rsm,\Lsm}]=0\,,
\\
\label{27082021-man02-41} && [Q^{-\Rsm,\Lsm},J^{-1}]=0\,,     \hspace{1.3cm} \{Q^{-\Rsm},Q^{-\Lsm} \} = - P^-\,.\hspace{1cm}
\eeq
Considering the commutators in \rf{27082021-man02-40}, we find in the cubic approximation the following relations:
\be  \label{27082021-man02-42}
[P_\smpt^- ,J_\smp3^{-1}] + [P_\smp3^-,J_\smpt^{-1}]=0\,, \qquad [P_\smpt^-,Q_\smp3^{-\Rsm,\Lsm}] + [P_\smp3^-, Q_\smpt^{-\Rsm,\Lsm}]=0\,.
\ee
Making use of the commutators \rf{27082021-man02-42}, we find the representation for the densities $q_\smp3^{-\Rsm,\Lsm}$ and $j_\smp3^{-1}$ in terms of the cubic vertex $p_\smp3^-$,
\beq
\label{27082021-man02-43} && \Pbf^- q_\smp3^{-\Rsm} = -  \frac{\varepsilon}{\sqrt{2}} \Big( \Po \partial_{\Po_\theta} + \sum_{a=1,2,3}  \irm m_a \partial_{p_{\eta a}} \Big) p_\smp3^-\,,
\\
\label{27082021-man02-44} && \Pbf^- q_\smp3^{-\Lsm} =   \frac{\varepsilon}{\sqrt{2}} \Big( \frac{\Po\Po_\theta}{\beta}  + \sum_{a=1,2,3}  \frac{\irm m_a}{\beta_a} p_{\eta a} \Big) p_\smp3^-\,,
\\
\label{27082021-man02-45} && \Pbf^- j_\smp3^{-1} = - \Jbf^{-1} p_\smp3^-\,,
\eeq
where in \rf{27082021-man02-43}-\rf{27082021-man02-45} we use the notation
\beq
\label{27082021-man02-45-a1} && \hspace{-1cm} \Pbf^- \equiv \frac{\Po^2}{2\beta} - \sum_{a=1,2,3} \frac{m_a^2}{2\beta_a}\,,
\\
\label{27082021-man02-45-a2} && \hspace{-1cm}  \Jbf^{-1} \equiv  - \frac{\Po}{\beta} \No_\beta^{\eta E}  + \sum_{a=1,2,3} \frac{\betach_a}{6\beta_a} m_a^2 \partial_{\Po} -\MM
+  \sum_{a=1,2,3} \irm m_a\big(\frac{ \betach_a }{6\beta} \Po_\theta \partial_{p_{\eta a}}   -  \frac{\betach_a  }{6\beta_a} p_{\eta a} \partial_{\Po_\theta} \big)\,,\qquad
\\
\label{27082021-man02-46} &&  \hspace{1cm} \No_\beta^{\eta E} \equiv \No_\beta  + \sum_{a=1,2,3} \frac{1}{6}\betach_a (N_{p_{\eta a}} + e_{\lambda_a})
\,, \hspace{1cm} \No_\beta \equiv \frac{1}{3} \sum_{a=1,2,3} \betach_a \beta_a \partial_{\beta_a}\,, \qquad
\\
\label{27082021-man02-47} && \hspace{1cm} \MM \equiv \sum_{a=1,2,3} \frac{\irm m_a \lambda_a}{\beta_a}\,, \hspace{0.9cm}
\\
\label{27082021-man02-48} && \hspace{1cm} \beta \equiv \beta_1 \beta_2 \beta_3\,,\hspace{0.9cm} \betach_a \equiv \beta_{a+1}- \beta_{a+2}\,,
\\
\label{27082021-man02-49} && \hspace{1cm} \varepsilon \equiv (-)^{\Ebf_\lambda}\,, \hspace{1.2cm} \Ebf_\lambda \equiv  e_{\lambda_1} + e_{\lambda_2} + e_{\lambda_3}  \,,  \qquad
\eeq
while the symbol $e_\lambda$ entering \rf{27082021-man02-49} is defined in \rf{26082021-man02-34}.

Making use of the solution for $q_\smp3^{-\Rsm,\Lsm}$ and $j_\smp3^{-1}$ in \rf{27082021-man02-43}-\rf{27082021-man02-45}, one can easily check that the kinematical symmetry equations for $q_\smp3^{-\Rsm,\Lsm}$ and $j_\smp3^{-1}$  in \rf{27082021-man02-35}-\rf{27082021-man02-39} are automatically satisfied provided the cubic vertex $p_\smp3^-$ satisfies the kinematical symmetry equations given in \rf{27082021-man02-35},\rf{27082021-man02-38}. Moreover, by using solutions for $q_\smp3^{-\Rsm,\Lsm}$ and $j_\smp3^{-1}$ in \rf{27082021-man02-43}-\rf{27082021-man02-45}, one can check that, in the cubic approximation, the (anti)commutators in \rf{27082021-man02-41} are also automatically satisfied.
Thus, in the cubic approximation, the kinematical symmetry equations for $p_\smp3^-$ in \rf{27082021-man02-35},\rf{27082021-man02-38} and equations \rf{27082021-man02-43}-\rf{27082021-man02-45} exhaust all restrictions imposed by the (anti)commutators of the N=2 Poincar\'e superalgebra.

\noindent {\bf Light-cone gauge dynamical principle}. Equations in \rf{27082021-man02-35},\rf{27082021-man02-38} and \rf{27082021-man02-43}-\rf{27082021-man02-45} do not fix the cubic densities unambiguously. In order to fix the cubic densities unambiguously we should impose some additional restrictions on the cubic densities. Such additional restrictions will be referred to as light-cone gauge dynamical principle. We formulated the light-cone gauge dynamical principle as follows:

\noindent \ibf) All cubic densities $p_\smp3^-$, $q_\smp3^{-\Rsm,\Lsm}$, $j_\smp3^{-1}$ should  be polynomial in the momentum  $\Po$ and $\beta$-analytic;%
\footnote{ If function $f=f(\beta_1,\beta_2,\beta_3)$ can be presented as $f=g/h$   where $g$, $h$ are polynomials in the momenta $\beta_1 ,\beta_2, \beta_3$, then we say the function $f$ is $\beta$-analytic. Otherwise, we say that the function $f$ is $\beta$-nonanalytic. Let us consider
Taylor series expansion for the vertex $p_\smp3^- = \sum_{n=0}^K f_n\Po^n$. If all $f_n$ are $\beta$-analytic, then we say the vertex $p_\smp3^-$ as $\beta$-analytic. Otherwise, we say that the vertex $p_\smp3^-$ is $\beta$-nonanalytic.
}

\noindent \iibf) The cubic vertex $p_\smp3^-$ should satisfy the following restriction
\be \label{27082021-man02-50}
p_\smp3^-  \ne  \Pbf^- W\,, \quad W \ \hbox{is polynomial in } \Po \hbox{ and $\beta$-analytic}\,,
\ee
where $\Pbf^-$ is defined in \rf{27082021-man02-45-a1}. We note that the restriction in \rf{27082021-man02-50} is
motivated by the freedom of field redefinitions. Namely, upon field redefinitions, the cubic vertex $p_\smp3^-$ is changed by terms proportional to $\Pbf^-$ (see, e.g., Appendix B in Ref.\cite{Metsaev:2005ar}). Therefore ignoring restriction \rf{27082021-man02-50} leads to cubic vertex $p_\smp3^-$  which can be removed by field redefinitions. However we are interested in the cubic vertex $p_\smp3^-$
that cannot be removed by field redefinitions.  For this reason we impose the restriction \rf{27082021-man02-50}. For the reader convenience we note that the restriction in \ibf) is the light-cone counterpart of locality condition used in Lorentz covariant and gauge invariant  formulations.

\noindent {\bf Complete system of equations for cubic vertex}. To summarize the discussion in this section, we note that, for the cubic vertex given by
\be \label{27082021-man02-51}
p_\smp3^- = p_\smp3^-(\Po,\Po_\theta, p_{\eta_a}\,, \beta_a)\,,
\ee
the complete system of equations which remains to be analysed is given by
\beq
&& \hspace{-1.6cm} \hbox{ \sf \small N=2 Poincar\'e superalgebra kinematical and dynamical restrictions:}
\\
\label{27082021-man02-52} && \hbox{Equation for $p_\smp3^-$ in \rf{27082021-man02-35}} \,, \hspace{2.5cm} \hbox{kinematical } \  J^{+-} \hbox{ symmetry};
\\
\label{27082021-man02-53} &&  \hbox{Equation for $p_\smp3^-$ in \rf{27082021-man02-38} }\,, \hspace{2.4cm} \hbox{kinematical } \  U(1) \hbox{ symmetry;}
\\
\label{27082021-man02-54} &&  \hbox{Equation \rf{27082021-man02-45}}\,, \hspace{3.8cm} \hbox{ dynamical } P^-, J^{-1} \hbox{ symmetries;}\qquad
\\
\label{27082021-man02-55} && \hbox{Equations \rf{27082021-man02-43} and \rf{27082021-man02-44}}\,, \hspace{1.8cm} \hbox{ dynamical } P^-, Q^{-\Rsm,\Lsm} \hbox{ symmetries;}\qquad
\\
&& \hspace{-1.6cm} \hbox{ \sf\small Light-cone gauge dynamical principle:}
\nonumber\\
\label{27082021-man02-56} && p_\smp3^-\,, \ q_\smp3^{-\Rsm,\Lsm}\,,  \ j_\smp3^{-1} \hspace{0.5cm} \hbox{ should be polynomial in } \Po \hbox{ and $\beta$-analytic};
\\
\label{27082021-man02-57} && p_\smp3^- \ne \Pbf^- W, \hspace{1cm} W  \hbox{ should be polynomial in } \Po \hbox{ and $\beta$-analytic.} \qquad
\eeq
It is the equations \rf{27082021-man02-52}-\rf{27082021-man02-57} that constitute our basic complete system of equations. If some vertex $p_{\smp3\, \fix}^-$ satisfies our system of equations, then the vertex $p_\smp3^-$ obtained from $p_{\smp3\, \fix}^-$ via the field redefinition $p_{\smp3}^- = p_{\smp3\, \fix}^-  + \Pbf^- f$ also satisfies our system of equations. This implies that in order to fix the vertex uniquely we should chose some representative of the vertex by using field redefinitions.  Only after choosing a representative of the vertex our system of equations allows us to find all possible solutions for the cubic vertex $p_\smp3^-$ and the densities $q_\smp3^{-\Rsm,\Lsm}$, $j_\smp3^{-1}$ uniquely. Below, we use a representative of cubic vertex which we refer to as first-derivative representation of the vertex.

\newsection{ \large Superspace first-derivative representation for cubic vertices  }\label{sec-06}

\noinbf{Superspace first-derivative representation for cubic vertex}. In general, the vertex $p_\smp3^-$ is a degree-$K$ polynomial in the momentum $\Po$, where $K<\infty$. However, as noted in Ref.\cite{Metsaev:2020gmb}, in the framework of light-cone gauge approach in $3d$, by making use of field redefinitions, any such vertex $p_\smp3^-$ can be cast into a degree-1 polynomial in the momentum $\Po$. Throughout this paper, the representation for the vertex $p_\smp3^-$ in terms of degree-1 polynomial in the momentum $\Po$ is referred to as first-derivative representation of vertex. Thus our first-derivative vertices are polynomials of degree-1 in the momentum $\Po$. In this section, we present our superspace first-derivative representation for all cubic vertices $p_\smp3^-$ and the corresponding densities $q_\smp3^{-\Rsm,\Lsm}$, $j_\smp3^{-1}$.
In this section and Secs.\ref{sec-07},\ref{sec-08}, we consider cubic vertices which involve at least one massive superfield. Cubic vertices for the case of three massless superfields are considered separately in Sec.\ref{sec-09}. Derivation of the results below presented may be found in Appendix C.  We now present our results.

We find that equations \rf{27082021-man02-43}-\rf{27082021-man02-45} lead to three types of cubic vertices which we refer as type-$k$ cubic vertices, where  $k=1,2,3$. We note that the type-$k$ cubic vertex is realized as a homogeneous degree-$k$ polynomial in the Grassmann momenta $\Po_\theta$ and $p_{\eta_1}$, $p_{\eta_2}$, $p_{\eta_3}$,
\be  \label{01082021-man02-00}
(N_{\Po_\theta} + \sum_{a=1,2,3} N_{ p_{\eta_a} } \big) p_\smp3^- = k p_\smp3^-\,, \qquad \hbox{ for type-$k$ vertex},
\ee
where we use the notation as in \rf{27082021-man02-37}.%
\footnote{ The fact that the solutions for the vertex $p_\smp3^-$ can be classified as homogeneous polynomials in the $\Po_\theta$ and $p_{\eta_1}$, $p_{\eta_2}$, $p_{\eta_3}$ is obvious in view of the $U(1)$ symmetry restrictions in \rf{27082021-man02-38}, \rf{27082021-man02-39}.
}

To clarify further our classification of vertices we use requirement that the Hamiltonian $P_\smp3^-$ \rf{27082021-man02-02} should be Grassmann even. To this end we note the following relations:
\be \label{01082021-man02-01-a1}
\GP(d\Gamma_\smp3^{p_\eta\,p_\theta}) = 1\,, \qquad \GP(\Theta_\smp3^*) = \Ebf_\lambda\,, \qquad \GP(p_\smp3^-) = \Ebf_\lambda + 1\,.
\ee
The 1st relation in \rf{01082021-man02-01-a1} tells us that the integration measure $d\Gamma_\smp3^{p_\theta\,p_\eta}$ \rf{27082021-man02-06} is Grassmann odd, while the 2nd relation for the Grassmann parity of the product of three superfields $\Theta_\smp3^*$ \rf{27082021-man02-13} is obtained by using \rf{01092021-man02-45}. It is the 3rd relation in \rf{01082021-man02-01-a1} that is obtained by requiring $\GP(P_\smp3^-)=0$. Note however that the Grassmann parity of the type-$k$ cubic vertex is equal to $\GP(p_\smp3^-)=k$. Equating this relation to the 3rd relation \rf{01082021-man02-01-a1},  we find the restriction
\be \label{01082021-man02-01-a3}
\Ebf_\lambda + 1 = k\!\!\! \mod 2\,.
\ee
Using \rf{01082021-man02-01-a3}, we can clarify further our classification of vertices. Namely,
from \rf{01092021-man02-43}-\rf{01092021-man02-45}, we note the relations $\GP(\Phi^*)=0$, $\GP(\Psi^*)=1$. Using these Grassmann parities of the superfields $\Phi^*$, $\Psi^*$, we conclude that, for $k=1,3$, relation \rf{01082021-man02-01-a3} is satisfied by the product $\Phi^*\Phi^*\Phi^*$ and $\Psi^*\Psi^*\Phi^*$ while, for $k=2$, relation \rf{01082021-man02-01-a3} is satisfied by the product $\Phi^*\Phi^*\Psi^*$ and $\Psi^*\Psi^*\Psi^*$. Omitting the asterisk of the superfields, we then conclude that we deal with the following vertices:
\beq
\label{01082021-man02-01-a4} \hbox{ type-1 $\Phi\Phi\Phi$ and $\Psi\Psi\Phi$ vertices;}
\\
\label{01082021-man02-01-a5} \hbox{ type-3 $\Phi\Phi\Phi$ and $\Psi\Psi\Phi$ vertices;}
\\
\label{01082021-man02-01-a6} \hbox{ type-2 $\Phi\Phi\Psi$ and $\Psi\Psi\Psi$ vertices.}
\eeq

As the type-1 and type-3 cubic vertices turn out to be related by hermitian conjugation, we first discuss these vertices and after that we discuss the type-2 cubic vertex.

\noinbf{Type-1 $\Phi\Phi\Phi$ and $\Psi\Psi\Phi$ vertices}. Superspace first-derivative  representation for the  type-1 cubic vertices $\Phi\Phi\Phi$ and $\Psi\Psi\Phi$ takes the form
\beq
\label{01082021-man02-01}  p_\smp3^- & = & \half \big(1 + \frac{\irm \Po}{\kappa}\big) \big (\kappa \Po_\theta - \beta \sum_{a=1,2,3}\frac{m_a}{\beta_a} p_{\eta a} \big) V
\nonumber\\
& - &  \half \big(1 - \frac{\irm \Po}{\kappa}\big) (\kappa \Po_\theta + \beta \sum_{a=1,2,3}\frac{m_a}{\beta_a} p_{\eta a} \big) \Vb\,,
\eeq
where a quantity $\kappa$ is defined as
\be \label{kappa}
\kappa^2 \equiv -\beta \sum_{a=1,2,3}\frac{m_a^2}{\beta_a}\,, \qquad \beta\equiv \beta_1\beta_2\beta_3\,,
\ee
while $V$ and $\Vb$ are vertices that depend only on the $\beta$-momenta $\beta_1$, $\beta_2$, $\beta_3$ and the labels $\sigma_1$, $\sigma_2$, $\sigma_3$ defined in \rf{27082021-man02-13-a1}. The vertices $V$ and $\Vb$  satisfy the following decoupled differential equations:
\beq
\label{01082021-man02-02} && \frac{\kappa}{\beta}  \No_\beta^E   V +  \irm \MM  V  = 0\,, \hspace{1cm} \Jbf_\beta V= 0\,,
\\
\label{01082021-man02-03} && \frac{\kappa}{\beta}  \No_\beta^E   \Vb -  \irm \MM  \Vb= 0\,, \hspace{1cm}   \Jbf_\beta \Vb = 0\,,
\eeq
where we use the notation
\be \label{01082021-man02-04}
\No_\beta^E \equiv \No_\beta  + \frac{1}{6} \sum_{a=1,2,3} \betach_a e_{\lambda_a}\,, \hspace{1cm} \Jbf_\beta \equiv \sum_{a=1,2,3} (\beta_a \partial_{\beta_a} + \half e_{\lambda_a})\,,
\ee
while the quantities $\No_\beta$, $\MM$, $\beta$ are defined in \rf{27082021-man02-46}-\rf{27082021-man02-48}.

Thus, for the type-1 cubic vertex, equations \rf{27082021-man02-43}-\rf{27082021-man02-45} amount to superspace representation in \rf{01082021-man02-01} and first-order differential equations in \rf{01082021-man02-02},\rf{01082021-man02-03}.  As we discuss below, the equations \rf{01082021-man02-02},\rf{01082021-man02-03} allow us to fix vertices $V$, $\Vb$ uniquely up to coupling constants.  Also we note that equations  \rf{01082021-man02-02},\rf{01082021-man02-03} on an equal footing describe the type-1 $\Phi\Phi\Phi$ and the type-1 $\Psi\Psi\Phi$  vertices.
Note however that, for the superfield $\Phi_{m,\lambda,u}^*$, $\lambda=s$, we use $e_\lambda=0$, while, for the superfield $\Psi_{m,\lambda,u}^*$, $\lambda=s+\half$, we use $e_\lambda=1$. For this reason, the particular forms of  equations in \rf{01082021-man02-02},\rf{01082021-man02-03} for $\Phi\Phi\Phi$ and type-1 $\Psi\Psi\Phi$  vertices are different. For example, to get  equations \rf{01082021-man02-02},\rf{01082021-man02-03} for the type-1 $\Phi\Phi\Phi$ vertex, we
use the values $\lambda_1=s_1$, $\lambda_2=s_2$, $\lambda_3=s_3$ to fix the $\MM$ \rf{27082021-man02-47} and the values $e_{\lambda_1}=0$, $e_{\lambda_2}=0$, $e_{\lambda_3}=0$ to fix the operators in \rf{01082021-man02-04}, while, for the type-1 $\Psi\Psi\Phi$ vertex, we
use the values $\lambda_1=s_1+\half$, $\lambda_2=s_2+\half$, $\lambda_3=s_3$ to fix the $\MM$ \rf{27082021-man02-47} and the values $e_{\lambda_1}=1$, $e_{\lambda_2}=1$, $e_{\lambda_3}=0$ to fix the operators in \rf{01082021-man02-04}.

\noinbf{Type-3 $\Phi\Phi\Phi$ and $\Psi\Psi\Phi$ vertices}. Superspace first-derivative representation for the type-3 $\Phi\Phi\Phi$ and $\Psi\Psi\Phi$ cubic vertices takes the form
\beq
\label{01082021-man02-05} p_\smp3^- & = & \frac{1}{2\kappa} \big(1 + \frac{\irm \Po}{\kappa}\big) \big (\kappa \Po_\theta - \beta \sum_{a=1,2,3}\frac{m_a}{\beta_a} p_{\eta a} \big) \sum_{b=1,2,3} m_b p_{\eta b+1}p_{\eta b+2} V'
\nonumber\\
& + &  \frac{1}{2\kappa} \big(1 - \frac{\irm \Po}{\kappa}\big) (\kappa \Po_\theta + \beta \sum_{a=1,2,3}\frac{m_a}{\beta_a} p_{\eta a} \big) \sum_{b=1,2,3} m_b p_{\eta b+1}p_{\eta b+2} \Vb'\,.
\eeq
The vertices $V'$ and $\Vb'$  depend only on the momenta $\beta_1$, $\beta_2$, $\beta_3$ and the labels $\sigma_1$, $\sigma_2$, $\sigma_3$ defined in \rf{27082021-man02-13-a1}. These two vertices satisfy the following decoupled equations:
\beq
\label{01082021-man02-06} && \frac{\kappa}{\beta}  \No_\beta^E   V' +  \irm \MM  V'  = 0\,, \hspace{2cm} \Jbf_\beta V' = 0\,,
\\
\label{01082021-man02-07} && \frac{\kappa}{\beta}  \No_\beta^E   \Vb' -  \irm \MM  \Vb'= 0\,, \hspace{2cm} \Jbf_\beta  \Vb' = 0\,,
\eeq
where we use the notation as in \rf{01082021-man02-04}, while the quantities $\No_\beta$, $\MM$, $\beta$ are defined in \rf{27082021-man02-46}-\rf{27082021-man02-48}. We see that though the expressions for the type-1 and type-3 cubic vertices $p_\smp3^-$ are different, the vertices $V$, $\Vb$ and their primed cousins satisfy the same equations (see \rf{01082021-man02-02}, \rf{01082021-man02-03} and \rf{01082021-man02-06}, \rf{01082021-man02-07}).
Also we see that equations  \rf{01082021-man02-06},\rf{01082021-man02-07} on an equal footing describe the type-3 $\Phi\Phi\Phi$ and the type-3 $\Psi\Psi\Phi$  vertices.

\noinbf{Type-2 $\Phi\Phi\Psi$ and $\Psi\Psi\Psi$ vertices:}. Superspace first-derivative representation for the type-2 cubic vertices $\Phi\Phi\Psi$ and $\Psi\Psi\Psi$ takes the form
\beq
\label{01082021-man02-08} p_\smp3^- & = & \half \big(1 + \frac{\irm \Po}{\kappa}\big) \big (\kappa \Po_\theta - \beta \sum_{a=1,2,3}\frac{m_a}{\beta_a} p_{\eta a} \big) \sum_{b=1,2,3} p_{\eta b} V_b
\nonumber\\
& - &  \half \big(1 - \frac{\irm \Po}{\kappa}\big) (\kappa \Po_\theta + \beta \sum_{a=1,2,3}\frac{m_a}{\beta_a} p_{\eta a} \big) \sum_{b=1,2,3} p_{\eta b} \Vb_b\,,
\eeq
where $V_a$ and $\Vb_a$, $a=1,2,3$, are vertices that depend only on the momenta $\beta_1$, $\beta_2$, $\beta_3$  and the labels $\sigma_1$, $\sigma_2$, $\sigma_3$  \rf{27082021-man02-13-a1}. Equations for the three vertices $V_a$ and their bared cousins are given  by
\beq
\label{01082021-man02-09} && \frac{\kappa}{\beta} \big(\No_\beta^E + \frac{\betach_a}{6}\big) V_a + \irm \MM V_a  +  \frac{m_a}{6\kappa\beta_a} \sum_{b=1,2,3} \betach_b m_b V_b=0\,, \hspace{2cm} a=1,2,3;\qquad \qquad
\\
\label{01082021-man02-10} && \big( \Jbf_\beta + \half \big) V_a = 0\,, \quad a=1,2,3; \hspace{2cm} \sum_{a=1,2,3} m_a V_a = 0 \,,
\\
\label{01082021-man02-11} && \frac{\kappa}{\beta} \big(\No_\beta^E + \frac{\betach_a}{6}\big) \Vb_a - \irm \MM \Vb_a + \frac{m_a}{6\kappa\beta_a} \sum_{b=1,2,3} \betach_b m_b \Vb_b=0\,, \hspace{2cm} a=1,2,3;\qquad
\\
\label{01082021-man02-12} && \big( \Jbf_\beta + \half \big) \Vb_a = 0\,, \quad a=1,2,3; \hspace{2cm}   \sum_{a=1,2,3} m_a \Vb_a = 0 \,, \qquad
\eeq
where we use the notation as in \rf{01082021-man02-04} and \rf{27082021-man02-46}-\rf{27082021-man02-48}.
We see that equations for the vertices $V_a$ in \rf{01082021-man02-09},\rf{01082021-man02-10} are decoupled from the ones for the vertices $\Vb_a$ in \rf{01082021-man02-11},\rf{01082021-man02-12}.  Note that equations  \rf{01082021-man02-09}-\rf{01082021-man02-12} on an equal footing describe both the type-2 $\Phi\Phi\Psi$ and type-2 $\Psi\Psi\Psi$  vertices. Particular form of these equations corresponding to the type-2 $\Phi\Phi\Psi$ and type-2 $\Psi\Psi\Psi$  vertices is obtained by plugging the suitable values of $\lambda_a$ and $e_{\lambda_a}$ into \rf{01082021-man02-09}-\rf{01082021-man02-12}.

\noinbf{Superspace representation for densities $j_\smp3^{-1}$, $q_\smp3^{-\Rsm,\Lsm}$}. We now discuss a superspace representation for the densities $j_\smp3^{-1}$, $q_\smp3^{-\Rsm,\Lsm}$}. For all cubic vertices, expression for the density $j_\smp3^{-1}$ takes the form
\be \label{01082021-man02-13}
j_\smp3^{-1} = 2\irm \No_\beta^{\eta E} V_\Po\,,
\ee
where the operator $ \No_\beta^{\eta E}$ is given in \rf{27082021-man02-46}, while a vertex $V_\Po$  is expressible in terms of the $V$-vertices.
We now present explicit form of the vertex $V_\Po$, and the supercharge
densities $q_\smp3^{-\Rsm,\Lsm}$ for all type-$k$ vertices. Note that the quantity $\beta$, $\varepsilon$ appearing below are defined in \rf{27082021-man02-48}, \rf{27082021-man02-49}.
\beq
&& \hspace{-2cm} \hbox{\bf $V_\Po$ and $q_\smp3^{-\Rsm,\Lsm}$ for type-1 $\Phi\Phi\Phi$ and $\Psi\Psi\Phi$ vertices:}
\\
\label{01082021-man02-14} && V_\Po = \frac{1}{2\kappa}(\kappa \Po_\theta - \beta\sum_{a=1,2,3} \frac{m_a}{\beta_a} p_{\eta_a}\big)V + \,\, \frac{1}{2\kappa}(\kappa \Po_\theta + \beta\sum_{a=1,2,3} \frac{m_a}{\beta_a} p_{\eta_a}\big) \Vb\,,
\\
\label{01082021-man02-15} && q_\smp3^{-\Rsm} = -\frac{\irm \varepsilon\beta}{\sqrt{2}} (V+\Vb)\,,
\\
\label{01082021-man02-16} && q_\smp3^{-\Lsm} = -\frac{\irm \varepsilon\beta }{\sqrt{2}\,\kappa} \Po_\theta \sum_{a=1,2,3}\frac{m_a}{\beta_a} p_{\eta_a} (V-\Vb)\,.
\\[5pt]
&& \hspace{-2cm} \hbox{\bf $V_\Po$ and $q_\smp3^{-\Rsm,\Lsm}$ for  type-3 $\Phi\Phi\Phi$ and $\Psi\Psi\Phi$ vertices:}
\\
\label{01082021-man02-17} && V_\Po = \frac{1}{2\kappa^2}(\kappa \Po_\theta - \beta\sum_{a=1,2,3} \frac{m_a}{\beta_a} p_{\eta_a}\big) \sum_{b=1,2,3} m_b p_{\eta b+1}p_{\eta b+2}  V'
\nonumber\\
&& \hspace{0.7cm} - \,\, \frac{1}{2\kappa^2}(\kappa \Po_\theta + \beta\sum_{b=1,2,3} \frac{m_a}{\beta_a} p_{\eta_a}\big)  \sum_{b=1,2,3} m_b p_{\eta b+1}p_{\eta b+2} \Vb'\,,
\\
\label{01082021-man02-18} && q_\smp3^{-\Rsm} = -\frac{\irm \varepsilon\beta}{\sqrt{2}\,\kappa} \sum_{a=1,2,3} m_a p_{\eta_{a+1}} p_{\eta_{a+2}} (V'-\Vb')\,,
\\
\label{01082021-man02-19}  && q_\smp3^{-\Lsm} = -\frac{\irm \varepsilon\beta }{\sqrt{2}\,\kappa^2} \Po_\theta \sum_{a=1,2,3}\frac{m_a}{\beta_a}p_{\eta_a} \sum_{b=1,2,3} m_b p_{\eta_{b+1}} p_{\eta_{b+2}} (V'+\Vb')\,.
\\[5pt]
&& \hspace{-2cm} \hbox{\bf $V_\Po$ and $q_\smp3^{-\Rsm,\Lsm}$ for type type-2 $\Phi\Phi\Psi$ and $\Psi\Psi\Psi$ vertices:}
\\
\label{01082021-man02-20} && V_\Po = \frac{1}{2\kappa}(\kappa \Po_\theta - \beta\sum_{a=1,2,3} \frac{m_a}{\beta_a} p_{\eta_a}\big) \sum_{b=1,2,3} p_{\eta_b} V_b
\nonumber\\
&& \hspace{0.7cm} + \,\, \frac{1}{2\kappa}(\kappa \Po_\theta + \beta\sum_{a=1,2,3} \frac{m_a}{\beta_a} p_{\eta_a}\big) \sum_{b=1,2,3} p_{\eta_b} \Vb_b\,,
\\
\label{01082021-man02-21} && q_\smp3^{-\Rsm} = -\frac{\irm \varepsilon\beta}{\sqrt{2}} \sum_{b=1,2,3} p_{\eta_b}  (V_b+\Vb_b)\,,
\\
\label{01082021-man02-22} && q_\smp3^{-\Lsm} = -\frac{\irm \varepsilon\beta }{\sqrt{2}\,\kappa} \Po_\theta \sum_{a=1,2,3} \frac{m_a}{\beta_a}p_{\eta_a} \sum_{b=1,2,3}  p_{\eta_b} (V_b-\Vb_b)\,.
\eeq

For critical and non-critical masses, a structure of solutions of the equations for the $V$-vertices is different. Therefore, before discussing solutions of the equations for the $V$-vertices, we provide the definition of critical and non-critical masses we use in this paper.

\noinbf{Critical and non-critical mass values}. Consider a cubic vertex for three superfields which have masses $m_1$, $m_2$, $m_3$. Let us introduce quantities $D$, $\Pbf_{\epsilon m}$ defined as
\beq
\label{01082021-man02-23} && D \equiv m_1^4 + m_2^4 + m_3^4 - 2m_1^2m_2^2 -2 m_2^2 m_3^2 - 2 m_3^2 m_1^2\,,
\\
\label{01082021-man02-24} && \Pbf_{\epsilon m}\equiv \sum_{a=1,2,3} \epsilon_a m_a  \,,
\\
\label{01082021-man02-25} && \epsilon_1^2 =1\,, \quad \epsilon_2^2 =1\,, \quad \epsilon_3^2 =1\,,
\\
\label{01082021-man02-26} && \hspace{1cm}D = (m_1 + m_2 + m_3)(m_1 + m_2 - m_3)(m_1 - m_2 + m_3)(m_1 - m_2 - m_3)\,, \qquad
\eeq
where, relation \rf{01082021-man02-26} provides us an alternative representation for the  $D$ defined in \rf{01082021-man02-23} . If, for masses $m_1$, $m_2$, $m_3$, we meet the inequality $D\ne 0$, then we refer to such masses as non-critical masses, while, if, for masses $m_1$, $m_2$, $m_3$, we meet the equality $D=0$,  then such masses are referred to as critical masses. Note that the relation $\Pbf_{\epsilon m}=0$ implies the relation $D=0$, while the relation $D=0$ implies that there exist $\epsilon_1$, $\epsilon_2$, $\epsilon_3$ \rf{01082021-man02-25} such that the relation $\Pbf_{\epsilon m}=0$ holds true. Thus we have the relations
\beq
\label{01082021-man02-27} && D \ne 0\,,\hspace{4.3cm} \hbox{ for non-critical masses;}
\\
\label{01082021-man02-28} && D=0\,, \qquad \Pbf_{\epsilon m} =0\,,\hspace{1.7cm}  \hbox{ for critical masses.}
\eeq

\noinbf{General structure of solutions to $V$-vertices}. We find that, for the non-critical masses \rf{01082021-man02-27}, general structure of $\beta$-analytical solutions to vertices $V$, $V'$, $V_a$, and their bared cousins takes the form
\beq
&& \hspace{-1cm} \hbox{\small \sf For non-critical masses:}
\nonumber\\
\label{01082021-man02-29} && V = C V_\kappa\,, \hspace{1cm} \Vb = C V_{-\kappa}\,, \hspace{1.4cm}  \hbox{for type-1 $\Phi\Phi\Phi$ and $\Psi\Psi\Phi$ vertices;}\qquad
\\
\label{01082021-man02-30} && V' = C' V_\kappa\,, \hspace{1cm} \Vb' = C' V_{-\kappa}\,,  \hspace{1cm}  \hbox{for type-3 $\Phi\Phi\Phi$ and $\Psi\Psi\Phi$ vertices;}\qquad
\\
\label{01082021-man02-31}  && V_a  = C^\smone  V_{a,\kappa}^\smone + C^\smtwo  V_{a,\kappa}^\smtwo\,, \hspace{1cm} \Vb_a =  C^\smone  V_{a,-\kappa}^\smone + C^\smtwo  V_{a,-\kappa}^\smtwo\,,\qquad a=1,2,3\,, \hspace{1.3cm}
\nonumber\\
&& \hspace{6.5cm}  \hbox{for type-2 $\Phi\Phi\Psi$ and $\Psi\Psi\Psi$ vertices;}\qquad
\eeq
where $C$, $C'$, $C^\smone$, $C^\smtwo$ are coupling constants, while $V_\kappa$, $V_{a,\kappa}^{\smone,\smtwo}$ are some fixed functions of $\beta_1$, $\beta_2$, $\beta_3$. Explicit expressions for these functions are given in  Secs.\ref{subsec-71}-\ref{subsec-73}.
For the hermitian Hamiltonian $P_\smp3^-$, the coupling constants entering vertices in \rf{01082021-man02-29}-\rf{01082021-man02-31} should satisfy the following rules:
\beq
\label{03082021-man02-03-a1} && C^* = I_uC'\,,
\\
\label{03082021-man02-03-a2} && C^{\smone *} = I_u C^\smtwo\,,
\eeq
where $I_u$ stands for sign-inversion operator of the $U(1)$ charges (see \rf{07082021-man02-01}), while the asterisk stands for the complex conjugation.%
\footnote{ The coupling constants depend, among other things, on the $U(1)$ charges. Showing explicit dependence of the coupling constants on the $U(1)$ charges,  $C=C_{u_1,u_2,u_3}$,  $C'=C_{u_1,u_2,u_3}'$ we can represent, for example the relation in \rf{03082021-man02-03-a1}, as $C_{u_1,u_2,u_3}^*=C_{-u_1,-u_2,-u_3}'$.} Let us make comments on the solution in \rf{01082021-man02-29}-\rf{01082021-man02-31}.

\noinbf{i}) From \rf{01082021-man02-02}, we see that the vertex $V$ satisfies the two first-order differential equations with respect to $\beta$-momenta $\beta_1$, $\beta_2$, $\beta_3$. Note however that, in view of the conservation law $\beta_1+\beta_2+\beta_3=0$, the vertex $V$ depends only on two $\beta$-momenta. This implies that, up to one coupling constant $C$ the vertex $V$ is determined uniquely. The same holds true for the vertex $\Vb$.

\noinbf{ii}) Comparing \rf{01082021-man02-06},\rf{01082021-man02-07} and \rf{01082021-man02-02},\rf{01082021-man02-03}, we see that the vertices $V'$ and $\Vb'$  satisfy the same equations as the respective vertices $V$ and $\Vb$. For this reason, by module of the coupling constants, the solution for the vertices $V'$ and $\Vb'$ \rf{01082021-man02-30} takes the same form as the one for the respective vertices $V$ and $\Vb$ \rf{01082021-man02-29}.

\noinbf{iii}) From \rf{01082021-man02-09},\rf{01082021-man02-10}, we see that the three vertices $V_a$ satisfy the six first-order differential equations with respect to $\beta$-momenta $\beta_1$, $\beta_2$, $\beta_3$ and one algebraic constraint. Note however that, in view of the conservation law $\beta_1+\beta_2+\beta_3=0$, the vertices $V_a$ depend only on two $\beta$-momenta. This implies that, the three vertices $V_a$ are determined uniquely up to two coupling constants denoted as $C^\smone$ and $C^\smtwo$ in \rf{01082021-man02-31} . The same holds true for the vertices $\Vb_a$.

\noinbf{iv}) Equations for $\Vb$ \rf{01082021-man02-03} are obtained from the ones for $V$ \rf{01082021-man02-02} by the replacement $\kappa\rightarrow -\kappa$. Therefore, if the solution for the vertex $V$ is described by the function $V_\kappa$, then the solution for the vertex $\Vb$ is given by the $V_{-\kappa}$. The same holds true also for the vertices $V'$, $\Vb'$ and $V_a$, $\Vb_a$.

\noinbf{v}) From \rf{01082021-man02-29}, we see that the vertices $V$ and $\Vb$ enter in the game with the one and the same coupling constant $C$. It is the requirement of the $\beta$ analycity that leads to the one and same coupling constant for the vertices $V$ and $\Vb$. Namely, ignoring the $\beta$-analycity, we find that the general solution for the vertex $\Vb$ takes the form $\Vb= \Cb V_{-\kappa}$. If $\Cb\ne C$, then the vertex $p_\smp3^-$ is $\beta$-non-analytic, while, if $\Cb = C$, then the vertex $p_\smp3^-$ becomes even function of $\kappa$, $p_\smp3^-(\kappa)=p_\smp3^-(-\kappa)$. Explicit expressions for the $V_\kappa$ given in the next sections tell us that, in general, the $V_\kappa$ is a rational function of $\kappa$. This implies that, for $\Cb=C$, the vertex $p_\smp3^-$ depends on $\kappa^2$ and hence is $\beta$-analytic. The same holds true also for the vertices $V'$, $\Vb'$ and $V_a$, $\Vb_a$ in \rf{01082021-man02-30},\rf{01082021-man02-31}.

We now discuss vertices for the critical masses.
We find that, for the critical masses \rf{01082021-man02-28}, general structure of $\beta$-analytical solution to the vertices $V$, $V'$, $V_a$, and their bared cousins takes the form
\vspace{-0.2cm}
\beq
&& \hspace{-1.3cm} \hbox{\small \sf For critical masses:}
\nonumber\\[-3pt]
\label{01082021-man02-32} && V = C V_\epsilon\,, \hspace{1cm} \Vb = \Cb V_{-\epsilon}\,, \hspace{1.4cm}  \hbox{for type-1 $\Phi\Phi\Phi$ and $\Psi\Psi\Phi$ vertices;}\qquad
\\
\label{01082021-man02-33} && V' = C' V_\epsilon\,, \hspace{1cm} \Vb' = \Cb' V_{-\epsilon}\,,\hspace{1cm}  \hbox{for type-3 $\Phi\Phi\Phi$ and $\Psi\Psi\Phi$ vertices;}\qquad %
\\
\label{01082021-man02-34} && V_a  = C^\smone  V_{a,\epsilon}^\smone + C^\smtwo  V_{a,\epsilon}^\smtwo\,, \hspace{1cm} \Vb_a =  \Cb^\smone  V_{a,-\epsilon}^\smone + \Cb^\smtwo  V_{a,-\epsilon}^\smtwo\,,
\nonumber\\
&& \hspace{6.5cm}  \hbox{for type-2 $\Phi\Phi\Psi$ and $\Psi\Psi\Psi$ vertices;}\qquad
\eeq
where quantities $C$, $C'$, $C^\smone$, $C^\smtwo$ and their bared cousins $\Cb$, $\Cb'$, $\Cb^\smone$, $\Cb^\smtwo$ are coupling constants, while $V_\epsilon$, $V_{a,\epsilon}^{\smone,\smtwo}$ are some fixed functions of $\beta_1$, $\beta_2$, $\beta_3$.
Explicit expressions for these functions are given in Secs.\ref{subsec-81} and \ref{subsec-82}. For the hermitian Hamiltonian $P_\smp3^-$, the coupling constants entering vertices in \rf{01082021-man02-32}-\rf{01082021-man02-34}, unless otherwise specified, should satisfy the following rules:
\beq
\label{01082021-man02-34-b1} && C^* = I_uC'\,, \hspace{1.4cm} \Cb^* = I_u\Cb'\,,
\\
\label{01082021-man02-34-b2} && C^{\smone *} = I_uC^\smtwo\,, \qquad
\Cb^{\smone *} = I_u\Cb^\smtwo\,.
\eeq

Let us make comments on the expressions in \rf{01082021-man02-32}-\rf{01082021-man02-34}.

\noinbf{i}) For the critical masses, in view of the relation
\be \label{01082021-man02-34-a1}
-\beta \sum_{a=1,2,3}\frac{m_a^2}{\beta_a} = \Po_{\epsilon m}^2\,,  \hspace{1cm}  \Po_{\epsilon m} \equiv \frac{1}{3} \sum_{a=1,2,3} \betach_a \epsilon_a m_a\,, \qquad \betach_a \equiv \beta_{a+1}-\beta_{a+2}\,,
\ee
the $\kappa$ defined in \rf{kappa} can be chosen as
\be \label{01082021-man02-34-a2}
\kappa= \Po_{\epsilon m} \hspace{1cm} \hbox{for critical masses}.
\ee
Taking into account \rf{01082021-man02-34-a1},\rf{01082021-man02-34-a2}, we see that equations for $\Vb$ \rf{01082021-man02-03} are obtained from the ones for $V$ \rf{01082021-man02-02} by using the replacement $\epsilon_1,\epsilon_2,\epsilon_3 \rightarrow -\epsilon_1,-\epsilon_2,-\epsilon_3$. Therefore, if a solution for the vertex $V$ is described by the function $V_\epsilon$, then a solution for the vertex $\Vb$ is given by the $V_{-\epsilon}$. The same holds true also for the vertices $V'$, $\Vb'$ and $V_a$, $\Vb_a$.

\noinbf{ii}) The $\kappa$ given in \rf{01082021-man02-34-a2} is $\beta$-analytic. Solutions for the $V$-vertices and their bared cousins turn also to be
$\beta$-analytic. Therefore, in contrast to the relations in \rf{01082021-man02-29}-\rf{01082021-man02-31}, there is no need to tune the coupling constants $C$, $C'$, $C^\smone$, $C^\smtwo$ and their bared cousins in \rf{01082021-man02-32}-\rf{01082021-man02-34}.

As we said, for hermitian Hamiltonian $P_\smp3^-$, the type-1 and type-3 vertices are related by hermitian conjugation rules. These rules are given explicitly by the relations \rf{28082021-man02-13},\rf{28082021-man02-14} in the Appendix B. For the type-2 vertices, hermicity of the Hamiltonian leads to the rules given in \rf{28082021-man02-15}-\rf{28082021-man02-17}. Hermitian conjugation rules in \rf{28082021-man02-13}-\rf{28082021-man02-17} are given in terms of the vertices. In terms of the coupling constants these rules are given in \rf{03082021-man02-03-a1},\rf{03082021-man02-03-a2} and \rf{01082021-man02-34-b1},\rf{01082021-man02-34-b2}.

\noinbf{U(1) symmetry restrictions for type-$k$ vertices}. For the type-$k$ vertices, by using the $U(1)$ symmetry restriction \rf{27082021-man02-38}, we can obtain constraints on the labels $u_1$, $u_2$, $u_3$  of the cubic vertex $p_\smp3^- = p_{\sigma_1\sigma_2\sigma_3}^-$, where $\sigma_a$ is defined in \rf{27082021-man02-13-a1}. Namely, using \rf{27082021-man02-38}, we find that the $u_1$, $u_2$, $u_3$  of the cubic vertex $p_{\sigma_1\sigma_2\sigma_3}^-$ should satisfy the restrictions:
\beq
\label{01082021-man02-01-a7} && \Ubf = 1 \hspace{0.5cm} \hbox{ for type-1 vertices;} \hspace{1.5cm} \Ubf = -1  \hspace{0.5cm} \hbox{ for type-3 vertices;}
\\
\label{01082021-man02-01-a9} && \Ubf = 0  \hspace{0.6cm}  \hbox{ for type-2 vertices;}
\\
\label{01082021-man02-01-a10} && \Ubf \equiv u_1 + u_2 + u_3\,.
\eeq
In other words, the type-$k$ cubic vertices are non-trivial only for the values of $\Ubf$ given in \rf{01082021-man02-01-a7},\rf{01082021-man02-01-a9}. The vertex $p_\smp3^- = p_{\sigma_1\sigma_2\sigma_3}^-$ and the product of three superfields $\Theta_{\sigma_1}^*\Theta_{\sigma_2}^*\Theta_{\sigma_3}^*$ entering the Hamiltonian \rf{27082021-man02-02} have the same labels $\sigma_a$, \rf{27082021-man02-10}, \rf{27082021-man02-13-a1}. Therefore restrictions \rf{01082021-man02-01-a7} are valid not only for the $u$-labels of cubic vertex but also for the $u$-labels of the product of three superfields $\Theta_{\sigma_1}^*\Theta_{\sigma_2}^*\Theta_{\sigma_3}^*$.

\newsection{ \large $V_\kappa$-vertices for superfields with non-critical masses}\label{sec-07}

In this Section,we deal with the non-critical masses defined by relations \rf{01082021-man02-27}, where $D$ is given in \rf{01082021-man02-23}. Taking into account that some masses may be equal to zero, we will consider the following cases in turn:
\beq
\label{03082021-man02-01}  && m_1 \ne 0\,, \qquad m_2\ne 0\,, \qquad m_3 \ne 0\,, \hspace{1cm} D \ne 0;
\\
\label{03082021-man02-02} && m_1 \ne 0\,, \qquad m_2\ne 0\,, \qquad m_3 = 0\,, \hspace{1cm} |m_1| \ne |m_2|\,,
\\
\label{03082021-man02-03} && m_1 = 0\,, \qquad m_2 = 0\,, \qquad m_3 \ne 0\,.
\eeq
Note that, for $m_3=0$, the inequality $D\ne 0$ amounts to the inequality $|m_1|\ne |m_2|$ in \rf{03082021-man02-02}, while, for $m_1=0$, $m_2=0$, the inequality $D\ne 0$ amounts to the inequality $m_3\ne 0$ in \rf{03082021-man02-03}.

General expressions for the vertices we deal in this section are given in \rf{01082021-man02-29}-\rf{01082021-man02-31}. From these expressions, we see that all that is required is to provide explicit expressions for the functions $V_\kappa$ and $V_{a,\kappa}^{\smone,\smtwo}$.
Such expressions for the  $V_\kappa$ and $V_{a,\kappa}^{\smone,\smtwo}$ are given below in Secs.\ref{subsec-71}-\ref{subsec-73}.

\subsection{ \large $V_\kappa$-vertices for three massive superfields }\label{subsec-71}

\noinbf{$\Phi\Phi\Phi$ vertex. Three arbitrary spin massive superfields $\Phi$}. We consider a cubic vertex for the following three massive superfields:
\beq
\label{02032020-man-10} && \Phi_{m_1,s_1,u_1}-\Phi_{m_2,s_2,u_2}-\Phi_{m_3,s_3,u_3}\,, \hspace{1cm} s_1,s_2,s_3 \in \No_0\,,
\nonumber\\
\label{02032020-man-11} &&   m_1 \ne 0\,,\quad m_2 \ne 0\,,\quad m_3 \ne 0\,, \hspace{1cm} D\ne 0\,.
\eeq
Solution for the $V_\kappa$ entering the vertices $V$ and $V'$ \rf{01082021-man02-29},\rf{01082021-man02-30} takes the form
\beq
\label{02032020-man-12} && V_\kappa =   L_{\kappa,1}^{s_1} L_{\kappa,2}^{s_2} L_{\kappa,3}^{s_3}\,,
\\
\label{02032020-man-13} && \hspace{1cm} L_{\kappa,a} \equiv  \frac{\kappa}{\beta_a} + \frac{\betach_a}{2\beta_a} m_a + \frac{m_{a+1}^2 -m_{a+2}^2}{2m_a}  \,.
\eeq
We note that the helpful relations for $L_{\kappa,a}$ \rf{02032020-man-13} may be found in Appendix A in Ref.\cite{Metsaev:2020gmb}.

\noinbf{$\Psi\Psi\Phi$ vertex. Two arbitrary spin massive superfields $\Psi$ and one arbitrary spin massive superfield $\Phi$}.
We consider a cubic vertex for the following three massive superfields:
\beq
\label{02032020-man-14} && \Psi_{m_1,s_1+\half,u_1}-\Psi_{m_2,s_2+\half,u_2}-\Phi_{m_3,s_3,u_3}\,, \hspace{1cm} s_1,s_2,s_3 \in \No_0\,,
\nonumber\\
&&   m_1 \ne 0\,,\quad m_2 \ne 0\,,\quad m_3 \ne 0\,, \hspace{1cm} D\ne 0\,.
\eeq
Solution for the $V_\kappa$ entering the vertices $V$ and $V'$ \rf{01082021-man02-29},\rf{01082021-man02-30} takes the form
\be \label{02032020-man-15}
V_\kappa =  K_\kappa L_{\kappa,1}^{s_1} L_{\kappa,2}^{s_2} L_{\kappa,3}^{s_3}\,,
\hspace{1cm} K_\kappa \equiv \frac{1}{\beta_1\beta_2}(\kappa+ m_1\beta_2 -m_2\beta_1)\,,
\ee
where $L_{\kappa,a}$ are given in \rf{02032020-man-13}.

\noinbf{$\Phi\Phi\Psi$ vertex. Two arbitrary spin massive superfields $\Phi$  and one arbitrary spin massive superfield $\Psi$}. We consider a cubic vertex for the following three massive superfields:
\beq
\label{02032020-man-16} && \Phi_{m_1,s_1,u_1}-\Phi_{m_2,s_2,u_2}-\Psi_{m_3,s_3+\half,u_3}\,, \hspace{1cm} s_1,s_2,s_3 \in \No_0\,,
\nonumber\\
&&   m_1 \ne 0\,,\quad m_2 \ne 0\,,\quad m_3 \ne 0\,, \hspace{1cm} D\ne 0\,.
\eeq
Solution for the functions $V_{\kappa,a}^{\smone,\smtwo}$ entering the vertices $V_a$  \rf{01082021-man02-31} takes the form
\be \label{02032020-man-17}
V_{a,\kappa}^{\smone,\smtwo} = \VV_{a,\kappa}^{\smone,\smtwo} L_{\kappa, 1}^{s_1}  L_{\kappa, 2}^{s_2}  L_{\kappa, 3}^{s_3}\,, \hspace{1cm} a=1,2,3\,,
\ee
where $L_{\kappa,a}$ are given in \rf{02032020-man-13} and we use the notation
\beq
\label{02032020-man-18} && \VV_{1,\kappa}^\smone  = \frac{1}{\kappa}\Big(  L_{\kappa,3} - \frac{m_1^2+m_2^2-m_3^2}{2m_3} \big)\,,
\\
\label{02032020-man-19} && \VV_{2,\kappa}^\smone =    \frac{m_1 m_2}{\kappa m_3}\,,
\\
\label{02032020-man-20} && \VV_{3,\kappa}^\smone = \frac{m_1}{\kappa m_3}  \big( - L_{\kappa,3} + \frac{m_1^2-m_2^2-m_3^2}{2m_3}\big)\,,
\\
\label{02032020-man-21} && \VV_{1,\kappa}^\smtwo = -\frac{m_1m_2}{\kappa m_3}\,,
\\
\label{02032020-man-22} && \VV_{2,\kappa}^\smtwo =  \frac{1}{\kappa}\Big(  L_{\kappa,3} + \frac{m_1^2+m_2^2-m_3^2}{2m_3}\big)\,,
\\
\label{02032020-man-23} && \VV_{3,\kappa}^\smtwo = \frac{m_2}{\kappa m_3}  \big( - L_{\kappa,3} + \frac{m_1^2-m_2^2+m_3^2}{2m_3}\big)\,.
\eeq

\noinbf{$\Psi\Psi\Psi$ vertex. Three arbitrary spin massive superfields $\Psi$}.  We consider a cubic vertex for the following three massive superfields:
\beq
\label{02032020-man-24} && \Psi_{m_1,s_1+\half,u_1}-\Psi_{m_2,s_2+\half,u_2}-\Psi_{m_3,s_3+\half,u_3}\,, \hspace{1cm} s_1,s_2,s_3 \in \No_0\,,
\nonumber\\
&&   m_1 \ne 0\,,\quad m_2 \ne 0\,,\quad m_3 \ne 0\,, \hspace{1cm} D\ne 0\,.
\eeq
Solution for the $V_{\kappa,a}^{\smone,\smtwo}$ entering the vertices $V_a$ \rf{01082021-man02-31} takes the form
\be \label{02032020-man-25}
V_{a,\kappa}^{\smone,\smtwo} = K_\kappa \VV_{a,\kappa}^{\smone,\smtwo} L_{\kappa, 1}^{s_1}  L_{\kappa, 2}^{s_2}  L_{\kappa, 3}^{s_3}\,, \hspace{1cm} K_\kappa \equiv \frac{1}{\beta_1\beta_2} (\kappa + m_1 \beta_2 - m_2\beta_1)\,,\qquad
\ee
where $L_{\kappa,a}$ are given in \rf{02032020-man-13}, while $\VV_{a,\kappa}^{\smone,\smtwo}$ take the same form as in \rf{02032020-man-18}-\rf{02032020-man-23}

\subsection{ \large $V_\kappa$-vertices for two massive superfields and one massless superfield} \label{subsec-72}

\noinbf{$\Phi\Phi\Phi$ vertex. Two arbitrary spin massive superfields $\Phi$ with masses $|m_1|\ne |m_2|$ and one massless superfield $\Phi$}. We consider a cubic vertex for the following three superfields:
\beq
\label{02032020-man-26} && \Phi_{m_1,s_1,u_1}-\Phi_{m_2,s_2,u_2}-\Phi_{0,0,u_3}\,, \hspace{1cm} s_1,s_2 \in \No_0\,,
\nonumber\\
&&   m_1 \ne 0\,,\quad m_2 \ne 0\,,\quad m_3=0\,,\quad |m_1| \ne |m_2|.
\eeq
Solution for the $V_\kappa$ entering the vertices $V$ and $V'$ \rf{01082021-man02-29},\rf{01082021-man02-30} takes the form
\beq
\label{02032020-man-27} && V_\kappa =   L_{\kappa,1}^{s_1} L_{\kappa,2}^{s_2} \,, \hspace{1cm}  \kappa^2 = - \beta_2 \beta_3 m_1^2 - \beta_1\beta_3 m_2^2\,,
\\
\label{02032020-man-28} && L_{\kappa,1} \equiv  \frac{ \kappa}{\beta_1} + \frac{\betach_1}{2\beta_1} m_1 + \frac{m_2^2}{2m_1} \,, \hspace{1cm} L_{\kappa,2} \equiv   \frac{ \kappa}{\beta_2} +\frac{\betach_2}{2\beta_2} m_2 - \frac{m_1^2}{2m_2} \,.
\eeq

\noinbf{$\Psi\Psi\Phi$ vertex. Two arbitrary spin massive superfields $\Psi$ with masses $|m_1|\ne |m_2|$ and one massless superfield  $\Phi$}. We consider a cubic vertex for the following three superfields:
\beq
\label{03032020-man-25} && \Psi_{m_1,s_1+\half,u_3}-\Psi_{m_2,s_2+\half,u_2}-\Phi_{0,0,u_3}\,, \hspace{1cm} s_1,s_2 \in \No_0\,,
\nonumber\\
\label{03032020-man-26} && m_1 \ne 0,\quad m_2\ne 0,\quad m_3=0\,,\quad |m_1|\ne |m_2|\,.
\eeq
Solution for the $V_\kappa$ entering the vertices $V$ and $V'$ \rf{01082021-man02-29},\rf{01082021-man02-30} takes the form
\be \label{03032020-man-28}
V_\kappa =  K_\kappa L_{\kappa,1}^{s_1} L_{\kappa,2}^{s_2} \,, \hspace{1cm} K_\kappa \equiv \frac{1}{\beta_1\beta_2}(\kappa+ m_1\beta_2 -m_2\beta_1)\,,
\ee
where $L_{\kappa, 1}$, $L_{\kappa, 2}$, and $\kappa$ take the same form as in \rf{02032020-man-27}, \rf{02032020-man-28}.

\noinbf {\bf $\Psi\Psi\Phi$-vertex. One massless superfield $\Psi$, one arbitrary spin massive superfield $\Psi$ and one arbitrary spin massive superfield $\Phi$ with masses $|m_2|\ne |m_3|$}. We consider a cubic vertex for the following three superfields:
\beq
\label{02032020-man-29} && \Psi_{0,\half,u_1}-\Psi_{m_2,s_2+\half,u_2}-\Phi_{m_3,s_3,u_3}\,, \hspace{1cm} s_2,s_3 \in \No_0\,,
\nonumber\\
&& m_1=0\,,\quad m_2\ne 0,\quad \quad m_3 \ne 0 \,, \quad |m_2|\ne |m_3|\,.
\eeq
Solution for the $V_\kappa$ entering the vertices $V$ and $V'$ \rf{01082021-man02-29},\rf{01082021-man02-30} takes the form
\beq
\label{02032020-man-30} && V_\kappa =  K_\kappa L_{\kappa,2}^{s_2} L_{\kappa,3}^{s_3} \,, \hspace{1cm}  K_\kappa \equiv \frac{1}{\beta_1\beta_2}(\kappa - m_2\beta_1)\,,
\\
\label{02032020-man-31} && L_{\kappa,2} \equiv   \frac{ \kappa}{\beta_2} + \frac{\betach_2}{2\beta_2} m_2 + \frac{m_3^2}{2m_2}  \,, \hspace{1cm} L_{\kappa,3} \equiv  \frac{ \kappa}{\beta_3} + \frac{\betach_3}{2\beta_3} m_3 - \frac{m_2^2}{2m_3}  \,,
\\
\label{02032020-man-32} &&  \kappa^2 = - \beta_1 \beta_3 m_2^2 - \beta_1\beta_2 m_3^2\,.
\eeq

\noinbf{$\Phi\Phi\Psi$ vertex. Two arbitrary spin massive superfields $\Phi$ with masses $|m_1|\ne |m_2|$ and one massless superfield $\Psi$ }. We consider a cubic vertex for the following three superfields:
\beq
\label{02032020-man-33} && \Phi_{m_1,s_1,u_1}-\Phi_{m_2,s_2,u_2}-\Psi_{0,\half,u_3}\,, \hspace{1cm} s_1,s_2 \in \No_0\,,
\nonumber\\
&&   m_1 \ne 0\,,\quad m_2 \ne 0\,,\quad m_3=0\,,\quad |m_1| \ne |m_2|\,.
\eeq
Solution for the $V_{a,\kappa}^{\smone,\smtwo}$ entering the vertices $V_a$ in \rf{01082021-man02-31} takes the form
\beq
\label{02032020-man-34} && V_{1,\kappa}^\smone  = \frac{m_2}{\kappa} B_\kappa\,, \qquad V_{2,\kappa}^\smone = -\frac{m_1}{\kappa} B_\kappa\,,\qquad  V_{3,\kappa}^\smone = 0 \,,\qquad
\nonumber\\
\label{02032020-man-35} && V_{1,\kappa}^\smtwo =0\,, \hspace{1.8cm}  V_{2,\kappa}^\smtwo = 0 \,, \hspace{2cm} V_{3,\kappa}^\smtwo = \frac{1}{\beta_3} B_\kappa\,, \qquad B_\kappa \equiv  L_{\kappa, 1}^{s_1} L_{\kappa, 2}^{s_2} \,;\hspace{2cm}
\eeq
where $L_{\kappa, 1}$, $L_{\kappa, 2}$, and $\kappa$ take the same form as in \rf{02032020-man-27}, \rf{02032020-man-28}.

\noinbf{$\Phi\Phi\Psi$ vertex. One massless superfield $\Phi$, one arbitrary spin massive superfield $\Phi$ and one arbitrary spin massive superfield  $\Psi$ with masses $|m_2|\ne |m_3|$}. We consider a cubic vertex for the following three superfields:
\beq
\label{02032020-man-36} && \Phi_{0,0,u_1}-\Phi_{m_2,s_2,u_2}-\Psi_{m_3,s_3+\half,u_3}\,, \hspace{1cm} s_2,s_3 \in \No_0\,,
\nonumber\\
&&  m_1=0\,,\quad  m_2 \ne 0\,,\quad m_3 \ne 0\,,\quad |m_2| \ne |m_3|\,.
\eeq
Solution for the $V_{a,\kappa}^{\smone,\smtwo}$ entering the vertices $V_a$ in \rf{01082021-man02-31} takes the form
\beq
\label{02032020-man-37} && V_{1,\kappa}^\smone = \frac{1}{\beta_1\beta_3} F_\kappa\,,\qquad V_{2,\kappa}^\smone = 0 \,, \hspace{2.3cm} V_{3,\kappa}^\smone = 0;
\nonumber\\
\label{02032020-man-38} && V_{1,\kappa}^\smtwo=0\,, \hspace{2cm} V_{2,\kappa}^\smtwo = \frac{m_3}{\beta_3 \kappa} F_\kappa\,, \hspace{1.4cm} V_{3,\kappa}^\smtwo =- \frac{m_2}{\beta_3 \kappa} F_\kappa\,,\qquad
\nonumber\\
&& \hspace{3.5cm} F_\kappa \equiv (\kappa + m_3 \beta_1) L_{\kappa,2}^{s_2} L_{\kappa,3}^{s_3} \,,
\eeq
where $L_{\kappa,2}$, $L_{\kappa,3}$, and $\kappa$ take the same form as in \rf{02032020-man-31}, \rf{02032020-man-32}.

\noinbf{$\Psi\Psi\Psi$ vertex. Two arbitrary spin massive superfields $\Psi$ with masses $|m_1|\ne |m_2|$ and one massless  superfield  $\Psi$}. We consider a cubic vertex for the following three superfields:
\beq
\label{02032020-man-39} && \Psi_{m_1,s_1+\half,u_1}-\Psi_{m_2,s_2+\half,u_2}-\Psi_{0,\half,u_3}\,, \hspace{1cm} s_1,s_2 \in \No_0\,,
\nonumber\\
&&   m_1 \ne 0\,,\quad m_2 \ne 0\,,\quad m_3=0\,, \quad |m_1| \ne |m_2|\,.
\eeq
Solution for the $V_{a,\kappa}^{\smone,\smtwo}$ entering vertices $V_a$ in \rf{01082021-man02-31} takes the form
\beq
\label{02032020-man-40} && V_{1,\kappa}^\smone = \frac{m_2}{\kappa \beta_1\beta_2} F_\kappa\,, \qquad V_{2,\kappa}^\smone = - \frac{m_1}{\kappa \beta_1\beta_2} F_\kappa\,,\qquad V_{3,\kappa}^\smone = 0\,,
\nonumber\\
\label{02032020-man-41} && V_{1,\kappa}^\smtwo = 0 \,, \hspace{2.3cm}  V_{2,\kappa}^\smtwo = 0 \,,\hspace{2.6cm}  V_{3,\kappa}^\smtwo = \frac{1}{\beta_1\beta_2\beta_3} F_\kappa\,,
\nonumber\\
&& \hspace{1.8cm} F_\kappa \equiv  (\kappa + m_1 \beta_2 - m_2\beta_1) L_{\kappa,1}^{s_1} L_{\kappa,2}^{s_2} \,,
\eeq
where $L_{\kappa,1}$, $L_{\kappa,2}$, and $\kappa$ take the same form as in \rf{02032020-man-27}, \rf{02032020-man-28}.

\subsection{ \large $V_\kappa$-vertices for two massless superfields and one massive superfield} \label{subsec-73}

\noinbf{$\Phi\Phi\Phi$ vertex. Two massless superfields $\Phi$ and one arbitrary spin massive superfield $\Phi$}. We consider a cubic vertex for the following three superfields:
\beq
\label{02032020-man-42} && \Phi_{0,0,u_1}-\Phi_{0,0,u_2}-\Phi_{m_3,s_3,u_3}\,,  \hspace{1cm} s_3 \in \No_0\,,
\nonumber\\
&& m_1=0\,, \quad m_2 = 0 \,,  \quad m_3 \ne 0\,,
\eeq
Solution for the $V_\kappa$ entering the vertices $V$ and $V'$ \rf{01082021-man02-29},\rf{01082021-man02-30} takes the form
\be \label{02032020-man-44}
V_\kappa = L_{\kappa,3}^{s_3}\,, \hspace{1cm}  L_{\kappa,3} \equiv   \frac{ \kappa}{\beta_3} + \frac{\betach_3}{2\beta_3} m_3  \,, \hspace{1cm} \kappa^2 = -\beta_1\beta_2 m_3^2\,.
\ee

\noinbf{$\Psi\Psi\Phi$ vertex. Two massless superfields $\Psi$ and one arbitrary spin massive superfield $\Phi$}. We consider a cubic vertex for the following three superfields:
\beq
\label{02032020-man-44-a1} && \Psi_{0,\half,u_1}-\Psi_{0,\half,u_2}-\Phi_{m_3,s_3,u_3}\,,  \hspace{1cm} s_3 \in \No_0\,.
\nonumber\\
&& m_1=0\,, \quad m_2 = 0 \,,  \quad m_3 \ne 0\,.
\eeq
Solution for the $V_\kappa$ entering the vertices $V$ and $V'$  \rf{01082021-man02-29},\rf{01082021-man02-30} takes the form
\be \label{02032020-man-44-a2}
V_\kappa =  \frac{\kappa}{\beta_1\beta_2} L_{\kappa,3}^{s_3}\,, \qquad L_{\kappa,3} = \frac{ \kappa}{\beta_3}  + \frac{\betach_3}{2\beta_3} m_3\,, \qquad \kappa^2 = - \beta_1 \beta_2 m_3^2 \,.
\ee

\noinbf{$\Psi\Psi\Phi$ vertex. One arbitrary spin massive superfield $\Psi$, one massless superfield $\Psi$ and one massless superfield $\Phi$}. We consider a cubic vertex for the following three superfields:
\beq
\label{02032020-man-44-a3} && \Psi_{m_1,s_1+\half,u_1}-\Psi_{0,\half,u_2}-\Phi_{0,0,u_3}\,,  \hspace{1cm} s_1 \in \No_0\,,
\nonumber\\
&& m_1 \ne 0\,,\quad m_2=0\,,\quad m_3 = 0 \,.
\eeq
Solution for the $V_\kappa$ entering the vertices $V$ and $V'$  \rf{01082021-man02-29},\rf{01082021-man02-30} takes the form
\be
V_\kappa = K_\kappa L_{\kappa,1}^{s_1}\,, \quad L_{\kappa,1} = \frac{ \kappa}{\beta_1}  + \frac{\betach_1}{2\beta_1} m_1\,, \quad K_\kappa = \frac{1}{\beta_1\beta_2}(\kappa + m_1\beta_2)\,, \quad \kappa^2 = - \beta_2 \beta_3 m_1^2\,.
\ee

\noinbf{$\Phi\Phi\Psi$ vertex. Two massless superfields $\Phi$ and one arbitrary spin massive superfield $\Psi$}. We consider a cubic vertex for the following three superfields:
\beq
&& \Phi_{0,0,u_1}-\Phi_{0,0,u_2}-\Psi_{m_3,s_3+\half,u_3}\,, \hspace{1cm} s_3 \in \No_0\,,
\nonumber\\
&& m_1=0\,,\quad m_2=0\,,\quad m_3 \ne 0 \,.
\eeq
Solution for the $V_{a,\kappa}^{\smone,\smtwo}$ entering the vertices $V_a$ \rf{01082021-man02-31} takes the form
\beq
&& V_{1,\kappa}^\smone = \frac{1}{\beta_1\beta_3} (\kappa + m_3 \beta_1)  L_{\kappa,3}^{s_3}\,, \hspace{1cm} V_{2,\kappa}^\smone = 0 \,, \qquad V_{3,\kappa}^\smone= 0\,;
\nonumber\\
&& V_{1,\kappa}^\smtwo=0\,, \qquad V_{2,\kappa}^\smtwo  = \frac{m_3}{\beta_3\kappa} (\kappa + m_3 \beta_1) L_{\kappa, 3}^{s_3} \,,\hspace{1cm} V_{3,\kappa}^\smtwo =0\,;
\eeq
where $L_{\kappa,3}$ and $\kappa$ take the same form as in \rf{02032020-man-44}.

\noinbf{$\Phi\Phi\Psi$ vertex. One massless superfield $\Phi$, one arbitrary spin massive superfield $\Phi$ and one massless superfield $\Psi$}. We consider a cubic vertex for the following three superfields:
\beq
&& \Phi_{0,0,u_1}-\Phi_{m_2,s_2,u_2}-\Psi_{0,\half,u_3}\,,  \hspace{1cm} s_2 \in \No_0\,,
\nonumber\\
&& m_1=0\,,\quad m_2\ne 0\,,\quad m_3 =0 \,.
\eeq
Solution for the $V_{a,\kappa}^{\smone,\smtwo}$ entering the vertices $V_a$ \rf{01082021-man02-31} takes the form
\beq
&& V_{1,\kappa}^\smone = \frac{m_2}{\kappa} L_{\kappa, 2}^{s_2} \,, \qquad V_{2,\kappa}^\smone = 0\,,\qquad  V_{3,\kappa}^\smone = 0 \,;
\nonumber\\
&& V_{1,\kappa}^\smtwo =0\,, \hspace{1.5cm} V_{2,\kappa}^\smtwo = 0 \,, \hspace{0.9cm}  V_{3,\kappa}^\smtwo = \frac{1}{\beta_3} L_{\kappa, 2}^{s_2}\,;
\nonumber\\
&& \hspace{2cm} L_{\kappa,2} \equiv   \frac{ \kappa}{\beta_2} + \frac{\betach_2}{2\beta_2} m_2  \,, \hspace{1cm}\kappa^2 = -\beta_3\beta_1 m_2^2\,. \qquad \qquad
\eeq

\noinbf{$\Psi\Psi\Psi$ vertex. Two massless superfields $\Psi$ and one arbitrary spin massive superfield $\Psi$}. We consider a cubic vertex for the following three superfields:
\beq
&& \Psi_{0,\half,u_1}-\Psi_{0,\half,u_2}-\Psi_{m_3,s_3+\half,u_3}\,, \hspace{1cm} s_3 \in \No_0\,,
\nonumber\\
&& m_1=0\,,\quad m_2=0\,,\quad m_3 \ne 0 \,.
\eeq
Solution for the $V_{a,\kappa}^{\smone,\smtwo}$ entering the vertices $V_a$ \rf{01082021-man02-31} takes the form
\beq
&& V_{1,\kappa}^\smone =  \frac{m_3}{\kappa\,\beta_1\beta_3} (\kappa + m_3 \beta_1 ) L_{\kappa, 3}^{s_3} \,, \hspace{1cm} V_{2,\kappa}^\smone = 0 \,, \hspace{2.4cm} V_{3,\kappa}^\smone  = 0 \,,
\nonumber\\
&& V_{1,\kappa}^\smtwo = 0 \,, \hspace{1cm} V_{2,\kappa}^\smtwo =-  \frac{1}{\beta_1\beta_2\beta_3}(\kappa + m_3 \beta_1) L_{\kappa,3}^{s_3}\,, \hspace{2cm} V_{3,\kappa}^\smtwo = 0 \,,\qquad \qquad
\eeq
where $L_{\kappa, 3}$ and $\kappa$ take the same form as in \rf{02032020-man-44}.

\newsection{ \large $V_\kappa$-vertices for superfields with critical masses }\label{sec-08}

In this Section, we deal with the critical masses defined by relations \rf{01082021-man02-28}. Taking into account that some masses may be equal to zero, we will consider the following cases in turn:
\beq
\label{05082021-man02-01}  && m_1 \ne 0\,, \qquad m_2\ne 0\,, \qquad m_3 \ne 0\,, \hspace{1cm} \Pbf_{\epsilon m} = 0;
\\
\label{05082021-man02-02} && m_1 \ne 0\,, \qquad m_2\ne 0\,, \qquad m_3 = 0\,, \hspace{1cm} |m_1| =|m_2|\,,
\eeq
where $\Pbf_{\epsilon m}$ is defined in \rf{01082021-man02-24},\rf{01082021-man02-25}.
Note that, for $m_3=0$, the relation $\Pbf_{\epsilon m} = 0$ amounts to the relation $|m_1| = |m_2|$ in \rf{05082021-man02-02}.

General expressions for the vertices we deal in this section are given in \rf{01082021-man02-32}-\rf{01082021-man02-34}. From these expressions, we see that all that is required is to provide explicit expressions for the functions $V_\epsilon$ and $V_{a,\epsilon}^{\smone,\smtwo}$.
Such expressions for the  $V_\epsilon$ and $V_{a,\epsilon}^{\smone,\smtwo}$ are given below in Secs.\ref{subsec-81},\ref{subsec-82}. Throughout this section we use $\Pbf_{\epsilon m}$ and $\epsilon_1$, $\epsilon_2$, $\epsilon_3$ defined in \rf{01082021-man02-24},\rf{01082021-man02-25}.

\subsection{ \large $V_\kappa$-vertices for three massive superfields } \label{subsec-81}

\noinbf{$\Phi\Phi\Phi$ vertex. Three arbitrary spin massive superfields $\Phi$}. We consider a cubic vertex for the following three massive superfields:
\beq
\label{05082021-man02-03} && \Phi_{m_1,s_1,u_1}-\Phi_{m_2,s_2,u_2}-\Phi_{m_3,s_3,u_3}\,, \hspace{1cm} s_1,s_2,s_3 \in \No_0\,,
\nonumber\\
&&   m_1 \ne 0\,,\quad m_2 \ne 0\,,\quad m_3 \ne 0; \hspace{1.4cm} \Pbf_{\epsilon m} = 0\,.
\eeq
The $V_\epsilon$ entering the vertices $V$ and $V'$ \rf{01082021-man02-32},\rf{01082021-man02-33} is given by
\beq
\label{05082021-man02-04} && V_\epsilon = \Po_{\epsilon m}^{\, \Sbf_\epsilon } \prod_{a=1,2,3} \beta_a^{-\epsilon_a s_ a}\,,
\\
\label{05082021-man02-05} && \Po_{\epsilon m} \equiv \frac{1}{3}\sum_{a=1,2,3} \betach_a \epsilon_a m_a\,,  \hspace{1cm}  \Sbf_\epsilon \equiv \sum_{a=1,2,3} \epsilon_a s_ a\,,
\eeq
where $\betach_a$ is defined in \rf{27082021-man02-32}. We recall that the $V_{-\epsilon}$ entering  the vertices $\Vb$, $\Vb'$ \rf{01082021-man02-32},\rf{01082021-man02-33} is obtained from the $V_\epsilon$ \rf{05082021-man02-04} by using the replacement $\epsilon_1$,$\epsilon_2$,$\epsilon_3$ $\rightarrow$ $-\epsilon_1$,$-\epsilon_2$,$-\epsilon_3$.

\noinbf{$\Psi\Psi\Phi$ vertex. Two arbitrary spin massive superfields $\Psi$ and one arbitrary spin massive superfield $\Phi$}. We consider a cubic vertex for the following three massive superfields:
\beq
\label{05082021-man02-06} && \Psi_{m_1,s_1+\half,u_1}-\Psi_{m_2,s_2+\half,u_2}-\Phi_{m_3,s_3,u_3}\,, \hspace{1cm} s_1,s_2,s_3 \in \No_0\,,
\nonumber\\
&&   m_1 \ne 0\,,\quad m_2 \ne 0\,,\quad m_3 \ne 0; \hspace{1cm} \Pbf_{\epsilon m} = 0\,.
\eeq
Solution for the $V_\epsilon$ entering the vertices $V$ and $V'$ \rf{01082021-man02-32},\rf{01082021-man02-33} takes the form
\be \label{05082021-man02-07}
V_\epsilon =  K_\epsilon \Po_{\epsilon m}^{\, \Sbf_\epsilon } \prod_{a=1,2,3} \beta_a^{- \epsilon_a s_ a}\,, \hspace{1cm} K_\epsilon \equiv  \Po_{\epsilon m}^{\half (\epsilon_1 + \epsilon_2) } \beta_1^{- \half(1+\epsilon_1)} \beta_2^{-\half(1+\epsilon_2)} \,,
\ee
where we use the notation as in \rf{05082021-man02-05}.

\noinbf{$\Phi\Phi\Psi$ vertex. Two arbitrary spin massive superfields $\Phi$ and one arbitrary spin massive superfield $\Psi$}. We consider a cubic vertex for the following three massive superfields:
\beq
\label{05082021-man02-08} && \Phi_{m_1,s_1,u_1}-\Phi_{m_2,s_2,u_2}-\Psi_{m_3,s_3+\half,u_3}\,, \hspace{1cm} s_1,s_2,s_3 \in \No_0\,,
\nonumber\\
&&   m_1 \ne 0\,,\quad m_2 \ne 0\,,\quad m_3 \ne 0; \hspace{1cm} \Pbf_{\epsilon m} = 0\,.
\eeq
Solution for the $V_{a,\epsilon}^{\smone,\smtwo}$ entering the vertices $V_a$ \rf{01082021-man02-34} takes the form
\be \label{05082021-man02-09}
V_{a,\epsilon}^{\smone,\smtwo} = \VV_{a,\epsilon}^{\smone,\smtwo} \Po_{\epsilon m}^{\Sbf_\epsilon} \prod_{a=1,2,3} \beta_a^{-\epsilon_a s_a}\,,
\ee
where we use the notation as in \rf{05082021-man02-05}, while $\VV_{a,\epsilon}^{\smone,\smtwo}$ are defined as

\beq
\label{05082021-man02-10} && \VV_{1,\epsilon}^{\smone} = \frac{1}{\Po_{\epsilon m} } \big( \epsilon_1 \frac{\Po_{\epsilon m}^{\epsilon_3}}{\beta_3^{\epsilon_3}} - \frac{\epsilon_2 m_3 }{m_1m_2} \pi_{3-}\big)\,,
\\
\label{05082021-man02-11} && \VV_{2,\epsilon}^{\smone} =  \frac{1}{\Po_{\epsilon m} } \big(  \epsilon_2 \frac{\Po_{\epsilon m}^{\epsilon_3}}{\beta_3^{\epsilon_3}} + \frac{\epsilon_1 m_3 }{m_1m_2} \pi_{3-}\big) \,,
\\
\label{05082021-man02-12} && \VV_{3,\epsilon}^{\smone} =  \frac{1}{\Po_{\epsilon m} } \big(\epsilon_3 \frac{\Po_{\epsilon m}^{\epsilon_3}}{\beta_3^{\epsilon_3}} +  \frac{\epsilon_2 m_1 - \epsilon_1 m_2 }{m_1m_2}\pi_{3-}   \big)\,,
\\
\label{05082021-man02-13} && \VV_{1,\epsilon}^{\smtwo} = \frac{1}{\Po_{\epsilon m} } \big( \epsilon_1 + \frac{\epsilon_2m_3}{m_1m_2} \frac{\Po_{\epsilon m} }{\beta_3} \pi_{3+}\big) \,,
\\
\label{05082021-man02-14} && \VV_{2,\epsilon}^{\smtwo} = \frac{1}{\Po_{\epsilon m} } \big( \epsilon_2 - \frac{\epsilon_1m_3}{m_1m_2} \frac{\Po_{\epsilon m} }{\beta_3} \pi_{3+}   \big) \,,
\\
\label{05082021-man02-14-01} && \VV_{3,\epsilon}^{\smtwo} = \frac{1}{\Po_{\epsilon m} } \big( \epsilon_3 - \frac{\epsilon_2 m_1-\epsilon_1 m_2}{m_1m_2}\frac{\Po_{\epsilon m} }{\beta_3}  \pi_{3+} \big) \,,
\\
\label{05082021-man02-15}  && \pi_{3\pm} \equiv \frac{1\pm \epsilon_3}{2}\,.
\eeq
Requiring hermicity of the Hamiltonian, we get the restrictions on the coupling constants,
\beq
\label{05082021-man02-15-a1} && C^{\smone *} = \epsilon_1\epsilon_2 I_u C^{\smone}\,, \quad C^{\smtwo *} = - \epsilon_1\epsilon_2 I_u C^{\smtwo}\,,
\nonumber\\
&& \Cb^{\smone *} = \epsilon_1\epsilon_2 I_u \Cb^{\smone}\,, \quad \Cb^{\smtwo *} = - \epsilon_1\epsilon_2 I_u \Cb^{\smtwo}\,.
\eeq

\noinbf{$\Psi\Psi\Psi$ vertex. Three arbitrary spin massive superfields $\Psi$}. We consider a cubic vertex for the following three massive superfields:
\beq
&& \Psi_{m_1,s_1+\half,u_1}-\Psi_{m_2,s_2+\half,u_2}-\Psi_{m_3,s_3+\half,u_3}\,, \hspace{1cm} s_1,s_2,s_3 \in \No_0\,,
\nonumber\\
&&   m_1 \ne 0\,,\quad m_2 \ne 0\,,\quad m_3 \ne 0; \hspace{1cm} \Pbf_{\epsilon m} = 0\,.
\eeq
Solution for the $V_{a,\epsilon}^{\smone,\smtwo}$ entering the vertices  $V_a$ \rf{01082021-man02-34} takes the form
\be
V_{a,\epsilon}^{\smone,\smtwo} = K_\epsilon \VV_{a,\epsilon}^{\smone,\smtwo} \Po_{\epsilon m}^{\Sbf_\epsilon} \prod_{a=1,2,3} \beta_a^{-\epsilon_a s_a}\,, \hspace{1cm} K_\epsilon = \beta_1^{-\half (1 +\epsilon_1)} \beta_2^{-\half (1 +\epsilon_2)} \Po_{\epsilon m}^{\half (\epsilon_1+\epsilon_2)}\,,
\ee
where $\Po_{\epsilon m}$, $\Sbf_\epsilon$ are defined as in \rf{05082021-man02-05}, while $\VV_{a,\epsilon}^{\smone,\smtwo}$ are defined as in \rf{05082021-man02-10}-\rf{05082021-man02-15}. To ensure hermicity of the Hamiltonian the coupling constants should satisfy the relations \rf{05082021-man02-15-a1}.

\subsection{ \large $V_\kappa$-vertices for two massive superfields and one massless superfield} \label{subsec-82}

\noinbf{$\Phi\Phi\Phi$ vertex. Two arbitrary spin massive superfields $\Phi$ and one massless superfield $\Phi$}. We consider a cubic vertex for the following three superfields:
\beq
\label{09082021-man02-01} && \Phi_{m_1,s_1,u_1}-\Phi_{m_2,s_2,u_2}-\Phi_{0,0,u_3}\,, \hspace{1cm} s_1,s_2 \in \No_0\,,
\nonumber\\
&&   m_1 \ne 0\,,\quad m_2 \ne 0\,,\quad m_3=0\,,\hspace{1cm} \epsilon_1 m_1 + \epsilon_2 m_2 =0\,,\qquad \epsilon_1^2=1\,,\qquad \epsilon_2^2=1\,.\qquad \qquad
\eeq
Solution for the $V_\epsilon$ entering the vertices $V$ and $V'$ \rf{01082021-man02-32},\rf{01082021-man02-33} takes the form
\be \label{09082021-man02-02}
V_\epsilon  = \Po_{\epsilon m}^{\epsilon_1 s_1 + \epsilon_2 s_2 } \beta_1^{-\epsilon_1 s_1}\beta_2^{- \epsilon_2 s_2} \,, \qquad \Po_{\epsilon m}= \epsilon_2 m_2 \beta_3 \,.\qquad\qquad
\ee

\noinbf{$\Psi\Psi\Phi$ vertex. Two arbitrary spin massive superfields $\Psi$ and one massless superfield $\Phi$}. We consider a cubic vertex for the following three superfields:
\beq
\label{09082021-man02-04} && \Psi_{m_1,s_1+\half,u_1}-\Psi_{m_2,s_2+\half,u_2}-\Phi_{0,0,u_3}\,, \hspace{1cm} s_1,s_2 \in \No_0\,,
\nonumber\\
&&   m_1 \ne 0\,,\quad m_2 \ne 0\,,\quad m_3=0\,, \hspace{1cm} \epsilon_1 m_1 + \epsilon_2 m_2 =0\,, \hspace{0.5cm} \epsilon_1^2=1\,, \qquad \epsilon_2^2=1\,.\qquad\qquad
\eeq
Solution for the $V_\epsilon$ entering the vertices $V$ and $V'$ \rf{01082021-man02-32},\rf{01082021-man02-33} takes the form
\beq
\label{09082021-man02-05} && V_\epsilon  = K_\epsilon \Po_{\epsilon m}^{\epsilon_1 s_1 + \epsilon_2 s_2 } \beta_1^{-\epsilon_1 s_1}\beta_2^{- \epsilon_2 s_2} \,,
\nonumber\\
&&\hspace{1cm}  K_\epsilon = \beta_1^{-\half(1+\epsilon_1)} \beta_2^{-\half(1+\epsilon_2)}\beta_3^{\half(\epsilon_1 + \epsilon_2)} \,, \hspace{0.5cm} \Po_{\epsilon m} = \epsilon_2 m_2 \beta_3 \,.
\eeq

\noinbf{$\Psi\Psi\Phi$ vertex. One massless superfield $\Psi$, one arbitrary spin massive superfield $\Psi$ and one arbitrary spin massive superfield $\Phi$}. We consider a cubic vertex for the following superfields:
\beq
\label{09082021-man02-06} && \Psi_{0,\half,u_1}-\Psi_{m_2,s_2+\half,u_2}-\Phi_{m_3,s_3,u_3}\,, \hspace{1cm} s_1,s_2 \in \No_0\,,
\nonumber\\
&&   m_1=0\,,\quad m_2 \ne 0\,,\quad m_3 \ne 0\,, \hspace{1cm} \epsilon_2 m_2 + \epsilon_3 m_3 =0\,, \hspace{0.5cm} \epsilon_2^2=1\,, \qquad \epsilon_3^2=1\,.\qquad \qquad
\eeq
Solution for the $V_\epsilon$ entering the vertices $V$ and $V'$ \rf{01082021-man02-32},\rf{01082021-man02-33} takes the form
\beq
\label{09082021-man02-07} && V_\epsilon  = K_\epsilon \Po_{\epsilon m}^{\epsilon_2 s_2 + \epsilon_3 s_3 } \beta_2^{-\epsilon_2 s_2}\beta_3^{- \epsilon_3 s_3} \,, \qquad
\nonumber\\
&& K_\epsilon = \beta_1^{-\half(1-\epsilon_2)} \beta_2^{-\half(1+\epsilon_2)}\,, \hspace{1cm} \Po_{\epsilon m} = \epsilon_3 m_3 \beta_1 \,.
\eeq

\noinbf{$\Phi\Phi\Psi$ vertex. Two arbitrary spin massive superfields $\Phi$ and one massless superfield $\Psi$}. We consider a cubic vertex for the following three superfields:
\beq
\label{09082021-man02-08} && \Phi_{m_1,s_1,u_1}-\Phi_{m_2,s_2,u_2}-\Psi_{0,\half,u_3}\,, \hspace{1cm} s_1,s_2 \in \No_0\,,
\nonumber\\
&&   m_1 \ne 0\,,\quad m_2 \ne 0\,,\quad m_3=0\,, \hspace{1cm} \epsilon_1 m_1 + \epsilon_2 m_2 =0\,, \hspace{0.5cm} \epsilon_1^2=1\,, \qquad \epsilon_2^2=1\,.\qquad\qquad
\eeq
Solution for the $V_{a,\epsilon}^{\smone,\smtwo}$ entering the vertices $V_a$  \rf{01082021-man02-34} takes the form
\beq
\label{09082021-man02-09} && V_{1,\epsilon}^\smone = \epsilon_1 F_\epsilon\,, \qquad V_{2,\epsilon}^\smone = \epsilon_2  F_\epsilon\,,\qquad  V_{3,\epsilon}^\smone = 0 \,;
\nonumber\\
&& V_{1,\epsilon}^\smtwo =0\,, \hspace{1.4cm} V_{2,\epsilon}^\smtwo = 0 \,, \hspace{1.3cm} V_{3,\epsilon}^\smtwo = \epsilon_1\epsilon_2 F_\epsilon \,;
\nonumber\\
&& F_\epsilon \equiv \frac{1}{\beta_3} \Po_{\epsilon m}^{\epsilon_1 s_1 + \epsilon_2 s_2 } \beta_1^{-\epsilon_1 s_1}\beta_2^{- \epsilon_2 s_2} \,, \hspace{1cm}  \Po_{\epsilon m}= \epsilon_2 m_2 \beta_3 \,.
\eeq

\noinbf{$\Phi\Phi\Psi$ vertex. One massless superfield $\Phi$, one arbitrary spin massive superfield $\Phi$ and one arbitrary spin  massive superfield $\Psi$}. We consider a cubic vertex for the following three superfields:
\beq
\label{09082021-man02-10} && \Phi_{0,0,u_1}-\Phi_{m_2,s_2,u_2}-\Psi_{m_3,s_3+\half,u_3}\,, \hspace{1cm} s_2,s_3 \in \No_0\,,
\nonumber\\
&& m_1=0\,,\quad  m_2 \ne 0\,,\quad m_3 \ne 0\,, \hspace{1cm}
\epsilon_2 m_2 + \epsilon_3 m_3 = 0\,, \hspace{0.5cm} \epsilon_2^2=1\,, \qquad \epsilon_3^2=1\,.\qquad
\eeq
Solution for the $V_{a,\epsilon}^{\smone,\smtwo}$ entering the vertices $V_a$ \rf{01082021-man02-34} takes the form
\beq
\label{09082021-man02-11} &&  V_{1,\epsilon}^\smone =  \epsilon_2\epsilon_3 F_\epsilon\hspace{1cm}  V_{2,\epsilon}^\smone =0\,, \hspace{1.3cm} V_{3,\epsilon}^\smone = 0 \,; \hspace{1cm}
\nonumber\\
&& V_{1,\epsilon}^\smtwo=0\,, \qquad V_{2,\epsilon}^\smtwo = \epsilon_2 F_\epsilon\,, \qquad V_{3,\epsilon}^\smtwo = \epsilon_3  F_\epsilon\,;
\nonumber\\
&& F_\epsilon \equiv \beta_1^{-\half(1-\epsilon_3)} \beta_3^{-\half(1+\epsilon_3)}  \Po_{\epsilon m}^{\epsilon_2 s_2 + \epsilon_3 s_3 } \beta_2^{-\epsilon_2 s_2}\beta_3^{- \epsilon_3 s_3} \,,
\hspace{1cm} \Po_{\epsilon m}= \epsilon_3 m_3 \beta_1\,.
\eeq

\noinbf{$\Psi\Psi\Psi$ vertex. Two arbitrary spin massive superfields $\Psi$ and one massless superfield $\Psi$}. We consider a cubic vertex for the following three superfields:
\beq
\label{09082021-man02-12} && \Psi_{m_1,s_1+\half,u_1}-\Psi_{m_2,s_2+\half,u_2}-\Psi_{0,\half,u_3}\,, \hspace{1cm} s_1,s_2 \in \No_0\,,
\nonumber\\
&& m_1 \ne 0\,,\quad m_2 \ne 0\,, \quad m_3=0\,, \hspace{1cm}  \epsilon_1 m_1 + \epsilon_2 m_2  = 0\,, \hspace{0.5cm} \epsilon_1^2=1\,, \qquad \epsilon_2^2=1\,.\qquad \qquad
\eeq
Solution for the $V_{a,\epsilon}^{\smone,\smtwo}$ entering the vertices $V_a$  \rf{01082021-man02-34} takes the form
\beq
\label{09082021-man02-13} && V_{1,\epsilon}^\smone = \epsilon_1 F_\epsilon\,, \qquad V_{2,\epsilon}^\smone = \epsilon_2 F_\epsilon\,,\qquad  V_{3,\kappa}^\smone = 0 \,;
\nonumber\\
&&  V_{1,\kappa}^\smtwo =0\,, \hspace{1.4cm} V_{2,\kappa}^\smtwo = 0 \,,\hspace{1.4cm}   V_{3,\epsilon}^\smtwo = \epsilon_1 \epsilon_2 F_\epsilon\,;
\nonumber\\
&& F_\epsilon\equiv \beta_1^{-\half(1+\epsilon_1)} \beta_2^{-\half(1+\epsilon_2)} \beta_3^{\half(\epsilon_1+\epsilon_2)-1}  \Po_{\epsilon m}^{\epsilon_1 s_1 + \epsilon_2 s_2 } \beta_1^{-\epsilon_1 s_1}\beta_2^{- \epsilon_2 s_2} \,,\hspace{1cm} \Po_{\epsilon m}= \epsilon_2 m_2 \beta_3 \,.\qquad
\eeq

\newsection{ \large Superspace first-derivative cubic vertices for massless fields }\label{sec-09}

We now present cubic vertices of three massless superfields, $m_1 = 0$, $m_2 = 0$, $m_3 = 0$. We find that, for the three massless superfields, equations \rf{27082021-man02-43}-\rf{27082021-man02-45} lead to four types of cubic vertices which we refer as type-$k$ cubic vertices, where  $k=1,2,3,4$. By definition, the type-$k$ cubic vertex is realized as homogeneous degree-$k$ polynomial in the Grassmann momenta $\Po_\theta$ and $p_{\eta_a}$.%

Let us first present relations for the $p_\smp3^-$ and $j_\smp3^{-1}$ which are valid for all types of cubic vertices. Namely, all cubic vertices $p_\smp3^-$ and the corresponding densities $j_\smp3^{-1}$ can be presented as
\beq
\label{06082021-man02-01} && p_\smp3^- = \irm \Po V_\Po\,,
\\
\label{06082021-man02-02} && j_\smp3^{-1} = 2\irm \No_\beta^{\eta E} V_\Po\,,
\eeq
where $V_\Po$ is a vertex that depends only on the momenta $\beta_1$, $\beta_2$, $\beta_3$ and the Grassmann momenta $\Po_\theta$, $p_{\eta_1}$, $p_{\eta_2}$, $p_{\eta_3}$, while the operator is given in $\No_\beta^{\eta E}$ is defined in \rf{27082021-man02-46}. From \rf{06082021-man02-01},\rf{06082021-man02-02}, we see that in order to get the cubic densities we should fix expressions for the $V_\Po$ and the supercharge densities $q_\smp3^{-\Rsm,\Lsm}$. The expressions for $V_\Po$ and $q_\smp3^{-\Rsm,\Lsm}$ we find are presented below in \rf{06082021-man02-03}-\rf{06082021-man02-11}.%
\footnote{As compared to massive fields in the Sec. \ref{sec-06}, we note the appearance of the additional type-0 and type-4 vertices for massless fields in \rf{06082021-man02-09}-\rf{06082021-man02-11}.
}
Outline of the derivation of relations in \rf{06082021-man02-03}-\rf{06082021-man02-11} may be found at the end of Appendix C.
\beq
&& \hspace{-1cm} \hbox{\small Type-\,1 $\Phi\Phi\Phi$ and $\Psi\Psi\Phi$ vertices:}
\nonumber\\[-3pt]
\label{06082021-man02-03} &&   V_\Po = \Po_\theta X + \sum_{a=1,2,3} \frac{p_{\eta_a}}{\beta_a} X_a\,,
\\
\label{06082021-man02-04} && q_\smp3^{-\Rsm} = -  \irm \varepsilon\sqrt{2}\, \beta  X\,, \hspace{1cm} q_\smp3^{-\Lsm} =  \irm \varepsilon \sqrt{2}\,  \Po_\theta \sum_{a=1,2,3}\frac{p_{\eta_a}}{\beta_a} X_a\,;
\\
&& \hspace{-1cm} \hbox{\small Type-\,3 $\Phi\Phi\Phi$ and $\Psi\Psi\Phi$ vertices:}
\nonumber\\[-3pt]
\label{06082021-man02-05} && V_\Po = \frac{1}{\beta} \Po_\theta \sum_{a=1,2,3}  p_{\eta_{a+1}}  p_{\eta_{a+2}} \Xb_a + p_{\eta_1}p_{\eta_2}p_{\eta_3} \Xb\,,
\\
\label{06082021-man02-06} && q_\smp3^{-\Rsm} = -  \irm \varepsilon\sqrt{2}\, \sum_{a=1,2,3} p_{\eta_{a+1}} p_{\eta_{a+2}} X_a\,, \hspace{1cm} q_\smp3^{-\Lsm} =  \irm \varepsilon \sqrt{2}\,  \Po_\theta \sum_{a=1,2,3} p_{\eta_1} p_{\eta_2} p_{\eta_3}  \Xb\,;\hspace{1.5cm}
\\
&& \hspace{-1cm} \hbox{\small Type-\,2 $\Phi\Phi\Psi$ and $\Psi\Psi\Psi$ vertices:}
\nonumber\\[-3pt]
\label{06082021-man02-07} && V_\Po = \Po_\theta \sum_{a=1,2,3}  \frac{p_{\eta_a}}{\beta_a} Y_a + \sum_{a=1,2,3} p_{\eta_{a+1}}p_{\eta_{a+2}} \Yb_a\,,
\\
\label{06082021-man02-08} && q_\smp3^{-\Rsm} = -  \irm \varepsilon\sqrt{2}\, \beta  \sum_{a=1,2,3}\frac{p_{\eta_a}}{\beta_a} Y_a\,, \hspace{1cm} q_\smp3^{-\Lsm} =  \irm \varepsilon \sqrt{2}\,  \Po_\theta \sum_{a=1,2,3} p_{\eta_{a+1}} p_{\eta_{a+2}}  \Yb_a\,;\hspace{1cm}
\\
&& \hspace{-1cm} \hbox{\small Type-\,0 $\Phi\Phi\Psi$ and $\Psi\Psi\Psi$ vertices:}
\nonumber\\[-3pt]
\label{06082021-man02-09} &&  V_\Po = Y \,, \hspace{1cm} q_\smp3^{-\Rsm} = 0\,, \hspace{1cm} q_\smp3^{-\Lsm} =  \irm \varepsilon \sqrt{2}\,  \Po_\theta Y\,,
\\
&& \hspace{-1cm} \hbox{\small Type-\,4 $\Phi\Phi\Psi$ and $\Psi\Psi\Psi$ vertices:}
\nonumber\\[-3pt]
\label{06082021-man02-10} && V_\Po = \frac{1}{\beta}\Po_\theta p_{\eta_1}p_{\eta_2}p_{\eta_3} \Yb\,,
\\
\label{06082021-man02-11} && q_\smp3^{-\Rsm} = -  \irm \varepsilon\sqrt{2}\, p_{\eta_1} p_{\eta_2} p_{\eta_3} \Yb\,, \hspace{1cm}  q_\smp3^{-\Lsm} =  0.
\eeq
We recall that $\beta$ and $\varepsilon$ are defined in \rf{27082021-man02-48},\rf{27082021-man02-49}.
In \rf{06082021-man02-03}-\rf{06082021-man02-11}, the quantities $X$, $Y$, $X_a$, $Y_a$ and their bared cousins are functions of the momenta $\beta_1$, $\beta_2$, $\beta_3$.
We find the following $\beta$-homogeneity equations
\beq
\label{06082021-man02-12} && \Jbf_\beta X =0\,, \hspace{1cm} (-2 + \Jbf_\beta) X_a=0\,,
\\
\label{06082021-man02-13} && (-\frac{3}{2} + \Jbf_\beta) Y=0\,, \hspace{2cm}  (-\half + \Jbf_\beta) Y_a=0\,,
\eeq
and identical equations for $\Xb$,\,$\Yb$,\,$\Xb_a$,\,$\Yb_a$. Solution to $\beta$-homogeneity equations \rf{06082021-man02-12},\rf{06082021-man02-13} can easily be presented, for example, in the following way:
\beq
\label{06082021-man02-14} && X = v_{\phi\phi\phi} (\frac{\beta_1}{\beta_2})\,, \hspace{1.5cm} X_a = \frac{1}{\beta_1\beta_2}v_{a,\phi\phi\phi}(\frac{\beta_1}{\beta_2})\,, \hspace{1cm} \hbox{\small for type-1 $\Phi\Phi\Phi$ vertex;}\qquad
\\
\label{06082021-man02-15} && X = \frac{1}{\beta_3} v_{\psi\psi\phi}(\frac{\beta_1}{\beta_2})\,, \hspace{1cm} X_a = \beta_3 v_{a,\psi\psi\phi}(\frac{\beta_1}{\beta_2})\,, \hspace{1.4cm} \hbox{\small for type-1 $\Psi\Psi\Phi$ vertex};
\\
\label{06082021-man02-16} && Y = \beta_3 v_{\phi\phi\psi}(\frac{\beta_1}{\beta_2})\,, \hspace{6.1cm} \hbox{\small for type-0 $\Phi\Phi\Psi$ vertex};
\\
\label{06082021-man02-17} && Y = v_{\psi\psi\psi}(\frac{\beta_1}{\beta_2})\,, \hspace{6.5cm} \hbox{\small for type-0 $\Psi\Psi\Psi$ vertex;}
\\
\label{06082021-man02-18} && Y_a =  v_{a,\phi\phi\psi}(\frac{\beta_1}{\beta_2})\,, \hspace{6.3cm} \hbox{\small for type-2 $\Phi\Phi\Psi$ vertex;}
\\
\label{06082021-man02-19} && Y_a = \frac{1}{\beta_3} v_{a,\psi\psi\psi}(\frac{\beta_1}{\beta_2})\,, \hspace{5.8cm} \hbox{\small for type-2 $\Psi\Psi\Psi$ vertex;}
\eeq
where quantities $v_{\phi\phi\phi}$, ..., $v_{a,\psi\psi\psi}$ appearing in \rf{06082021-man02-14}-\rf{06082021-man02-19} are functions of the ratio $\beta_1/\beta_2$. These functions, which we refer to as $v$-functions, are taken to be $\beta$-analytic. Solution for $\Xb$, $\Yb$, $\Xb_a$, $\Yb_a$ takes the same form as \rf{06082021-man02-14}-\rf{06082021-man02-19}, where we should replace the $v$-functions by their bared cousins $\vb_{\phi\phi\phi}$, ..., $\vb_{a,\psi\psi\psi}$, which we refer to as $\vb$-functions. If we ignore the hermicity of the Hamiltonian $P_\smp3^-$, then the $v$ and $\vb$-functions are independent.
Requiring that the Hamiltonian $P_\smp3^-$ be hermitian, we find that the $v$ and $\vb$-functions are related by the hermitian conjugation rules. In terms of the quantities  $X$,\,$Y$,\,$X_a$,\,$Y_a$, and their bared cousins, those hermitian conjugation rules take the form
\beq
\label{06082021-man02-20} && X^* = - I_\beta I_u \Xb\,, \hspace{1cm}  X_a^* = - I_\beta I_u \Xb_a\,, \hspace{1cm} \hbox{ for $\Phi\Phi\Phi$ vertex;}
\\
\label{06082021-man02-21} && X^* =  I_\beta I_u \Xb\,, \hspace{1.4cm}  X_a^* = I_\beta I_u \Xb_a\,, \hspace{1.3cm} \hbox{ for $\Psi\Psi\Phi$ vertex;}
\\
\label{06082021-man02-22} && Y^* =  - I_\beta I_u \Yb\,, \hspace{1.2cm}  Y_a^* = - I_\beta I_u \Yb_a\,, \hspace{1.1cm} \hbox{ for $\Phi\Phi\Psi$ vertex;}
\\
\label{06082021-man02-23} && Y^* =  I_\beta I_u \Yb\,, \hspace{1.6cm}  Y_a^* = I_\beta I_u \Yb_a\,, \hspace{1.4cm} \hbox{ for $\Psi\Psi\Psi$ vertex;}
\eeq
where, in \rf{06082021-man02-20}-\rf{06082021-man02-23}, the asterisk stands for the complex conjugation, while the definition of the operator $I_x$ is given in \rf{07082021-man02-01} in Appendix A.

\noinbf{U(1) symmetry restrictions for type-$k$ vertices}. Using $U(1)$ symmetry  equations in \rf{27082021-man02-38}, \rf{27082021-man02-53}, we can straightforwardly obtain constraints on the labels $u_1$, $u_2$, $u_3$  of the cubic vertex $p_\smp3^-$. Namely, using \rf{27082021-man02-38}, we find that the $u_1$, $u_2$, $u_3$  of the cubic vertex $p_\smp3^-$ should satisfy the restrictions:
\beq
&& \Ubf = 1 \hspace{0.5cm} \hbox{ for type-1 vertices;}\hspace{1.5cm} \Ubf = -1  \hspace{0.5cm} \hbox{ for type-3 vertices;}
\\
&& \Ubf = 2 \hspace{0.5cm} \hbox{ for type-0 vertices;}\hspace{1.5cm} \Ubf = -2  \hspace{0.5cm} \hbox{ for type-4 vertices;}
\\
&& \Ubf = 0  \hspace{0.6cm}  \hbox{ for type-2 vertices;}
\eeq
where $\Ubf$ is defined as in \rf{01082021-man02-01-a10}.

In conclusion of this section, we recall that, for the massive and massless fields having $\kappa\equiv\hspace{-0.4cm}/\hspace{0.2cm} 0$, we found a finite number of the cubic vertices in Secs.\ref{sec-06},\ref{sec-07},\ref{sec-08}. In contrast to this, for the three massless fields, we obtained infinite number of light-cone gauge cubic vertices \rf{06082021-man02-03}-\rf{06082021-man02-11} which are parametrized by the $v$- and $\vb$-functions. To our knowledge, an infinite number of Lorentz covariant vertices associated with our infinite number light-cone gauge vertices are not available in the literature at the present time. This is to say that we encounter a mismatch between classifications of Lorentz covariant  and light-cone gauge cubic vertices. This mismatch for the massless fields in $3d$ was noticed in Ref.\cite{Metsaev:2020gmb}.
We recall that, for the case of massless fields in $4d$, the mismatch between classifications of Lorentz covariant cubic vertices and light-cone gauge cubic vertices is well known (see, e.g., Ref.\cite{Conde:2016izb}).%
\footnote{ Recent study of new Lorentz covariant vertices in the framework of $4d$ chiral higher-spin gravity may be found in Ref.\cite{Krasnov:2021nsq}-\cite{Tran:2021ukl}. We expect that adaptation of the twistor methods in these references to $3d$ may be helpful for the study of yet unknown new Lorentz covariant vertices in $3d$. Recent discussion of twistors in $3d$ may be found in Refs.\cite{Kuzenko:2021vmh}. For interesting application of twistors to the study of a superparticle in $AdS_5\times S^5$, see Refs.\cite{Uvarov:2020wqd} .
}
%

\newsection{ \large Conclusions}\label{sec-10}

In this paper, we applied a superfield light-cone gauge approach for the study of the interacting $N=2$ massive arbitrary  (integer and half-integer) spin supermultiplets and massless (spin-0 and spin-$\half$)  supermultiplets in $3d$ flat space. Using the light-cone momentum superspace, we developed the light-cone gauge formulation in terms of the unconstrained superfields. Using our superfield formulation, we built all cubic vertices that describe interactions of the massive and massless supermultiplets under consideration. We believe that development made in this paper might be helpful for the  following applications and generalizations.

\noinbf{i}) In this paper, we studied cubic vertices for the N=2 massive and massless supermultiplets  in {\it the flat $3d$ space}. Light-cone gauge action for arbitrary spin free massive fields in $AdS_3$ space was obtained in Ref.\cite{Metsaev:2000qb}, while the method for studying cubic vertices of light-cone gauge higher-spin massless AdS fields was developed in Ref.\cite{Metsaev:2018xip}.%
\footnote{ We note the Ref.\cite{Skvortsov:2018uru} where it was shown that the light-cone gauge action turns out to be helpful for studying boundary 3-point correlation functions.
}
Generalization of our studies in this paper and in Refs.\cite{Metsaev:2000qb,Metsaev:2018xip} to the case of light-cone gauge interacting N=2 massive and massless supermultiplets in {\it $AdS_3$ space} could be of great interest. For the reader convenience, we note the recent Ref.\cite{Zinoviev:2021cmi} devoted to studying cubic vertices for fields in $AdS_3$ space by using covariant frame-like formulation.
For discussion of methods for the investigation of supersymmetric theories in $3d$, see  Refs.\cite{Kuzenko:2016qwo}-\cite{Kuzenko:2018lru} (and references therein). Various studies of supersymmetric theories of massless fields in $4d$ which could be helpful for the investigation of interacting massive supermultiplets in $3d$ may be found, e.g., in Refs.\cite{Buchbinder:2017nuc}-\cite{Bonora:2020aqp}.%
\footnote{ For $d>3$, we mention various methods of building cubic vertices in Refs.\cite{Metsaev:2005ar}-\cite{Boulanger:2012dx} and BRST method in Refs.\cite{Fotopoulos:2010ay}-\cite{Buchbinder:2021xbk}. For the studying supersymmetric interactions in $AdS_5$, see Refs.\cite{Alkalaev:2002rq}. The use of the BRST method in $3d$ may be found, e.g., in Ref.\cite{Rahman:2019mra}.}

\noinbf{ii}) Conformal higher-spin theories in $3d$ have actively been studied in the recent time (see, e.g., Refs.\cite{Nilsson:2015pua}-\cite{Ponomarev:2021xdq}  and references therein). We note that the ordinary-derivative formulation of conformal fields in Refs.\cite{Metsaev:2007rw,Metsaev:2016oic} and Stueckelberg formulation of massive fields share some common features. The light-cone gauge approach considerably simplifies the whole analysis of interacting massive fields. We expect therefore that the light-cone gauge formulation of conformal fields in  Ref.\cite{Metsaev:2016rpa} will also be helpful for the study of interacting light-cone gauge conformal fields in $3d$.

\noinbf{iii}) $4d$ chiral higher-spin gravity (CHSG) proposed in Ref.\cite{Ponomarev:2016lrm}, upon dimensional reduction  to $3d$, leads to $3d$ massive CHSG discussed in Refs.\cite{Metsaev:2020gmb,Skvortsov:2020pnk}. We note that
$4d$, N=1 supersymmetric CHSG discussed in Ref.\cite{Metsaev:2019dqt}, upon dimensional reduction to $3d$, leads to N=2 supersymmetric massive CHSG in $3d$ with central charges. In this paper, we studied N=2 supersymmetric theories in $3d$ without central charges. It would be then interesting to understand whether exist or not N=2 massive supersymmetric CHSG in $3d$ without central charges. In this respect, the light-cone gauge method discussed in Refs.\cite{Metsaev:1991mt,Metsaev:1991nb} could be helpful.

\medskip

{\bf Acknowledgments}. This work was supported by the RFBR Grant No.20-02-00193.

\setcounter{section}{0}\setcounter{subsection}{0}
\appendix{ \large Notation and conventions  } \label{app-01}

For any quantity $A$ that depends on three variables $x_1$, $x_2$, $x_3$, we introduce an operation $I_x$ which is defined by the relations
\be \label{07082021-man02-01}
A \equiv A(x_1,x_2,x_3)\,, \hspace{1cm} I_x A \equiv A(-x_1,-x_2,-x_3)\,.
\ee

We use Grassmann momenta denoted as $p_\theta$, $p_\eta$, while left derivatives of the $p_\theta$  and $p_\eta$ are denoted as $\partial_{p_\theta}$ and $\partial_{p_\eta}$ respectively. Integrals over $p_\theta$ and $p_\eta$ are defined to be $\int dp_\theta p_\theta =1$ and $\int dp_\eta p_\eta =1$. The basic integral is normalized to be
\be \label{07082021-man02-02}
\int dp_\eta dp_\theta p_\theta p_\eta =1\,.
\ee
We note the following helpful Grassmann integrals
\beq
\label{07082021-man02-03} && \int dp_\eta^\dagger dp_\theta^\dagger\, e^{\frac{1}{\beta}(p_\theta^\dagger p_\theta + p_\eta^\dagger p_\eta)} = - \frac{1}{\beta^2} p_\theta p_\eta\,,
\\
\label{07082021-man02-04} && \int dp_\eta^\dagger dp_\theta^\dagger\, e^{\frac{1}{\beta}(p_\theta^\dagger p_\theta + p_\eta^\dagger p_\eta)} p_\theta^\dagger= \frac{1}{\beta} p_\eta\,,
\\
\label{07082021-man02-05} && \int dp_\eta^\dagger dp_\theta^\dagger\, e^{\frac{1}{\beta}(p_\theta^\dagger p_\theta + p_\eta^\dagger p_\eta)} p_\eta^\dagger=  -\frac{1}{\beta} p_\theta \,,
\\
\label{07082021-man02-06} && \int dp_\eta^\dagger dp_\theta^\dagger\, e^{\frac{1}{\beta}(p_\theta^\dagger p_\theta + p_\eta^\dagger p_\eta)} p_\theta^\dagger p_\eta^\dagger = 1\,.
\eeq

Ghost parities of $p_\theta$, $p_\eta$ and  $dp_\theta$, $dp_\eta$ are given by
\be \label{07082021-man02-07}
\GP(p_\theta) = 1\,, \qquad \GP(p_\eta) = 1\,,\quad  \GP(dp_\theta) = 1\,,\quad  \GP(dp_\eta) = 1\,,
\ee
and we assume the conventions $AB= (-)^{\varepsilon_A\varepsilon_B}BA$, $\varepsilon_{_X}\equiv \GP(X)$. Hermitian conjugation is defined as $(AB)^\dagger = B^\dagger A^\dagger$, where two quantities $A$, $B$ have arbitrary ghost parity.

\noinbf{Berezin integrals for 3-point vertices}. For any quantity $A$ that depends on the Grassmann momenta $p_{\theta_a}$, $p_{\eta_a}$, $A = A(p_{\theta_a}, p_{\eta_a})$, we introduce a quantity $\langle A(p_\theta,p_\eta)\rangle$ defined by the relation
\be \label{07082021-man02-08}
\langle A(p_\theta,p_\eta)\rangle \equiv \int \prod_{a=1,2,3} dp_{\eta a}^\dagger dp_{\theta a}^\dagger \delta(\Pbf_\theta^\dagger)\,\exp(\sum_{b=1,2,3}  \frac{1}{\beta_b} (p_{\theta b}^\dagger p_{\theta b}^{\vphantom{5pt}}
+ p_{\eta b}^\dagger p_{\eta b}^{\vphantom{5pt}}) \big)  A(p_\theta^\dagger,p_\eta^\dagger)\,,
\ee
where $\delta(\Pbf_\theta^\dagger)=p_{\theta_1}^\dagger+p_{\theta_2}^\dagger+p_{\theta_3}^\dagger$. Using \rf{07082021-man02-03}-\rf{07082021-man02-06}, we get then straightforwardly the following basic relations
\beq
\label{07082021-man02-09} && \langle 1 \rangle = -\frac{1}{\beta^2} \delta(\Pbf_\theta) \Po_\theta p_{\eta 1} p_{\eta 2} p_{\eta 3} \,,
\\
\label{07082021-man02-10} && \langle \Po_\theta \rangle = - \frac{1}{\beta} \delta(\Pbf_\theta)p_{\eta 1} p_{\eta 2} p_{\eta 3}\,,
\\
\label{07082021-man02-11} && \langle  p_{\eta a} \rangle = - \frac{\beta_a}{\beta^2} \delta(\Pbf_\theta) \Po_\theta  p_{\eta a+1} p_{\eta a+2} \,,
\\
\label{07082021-man02-12} && \langle \Po_\theta p_{\eta a} \rangle = - \frac{\beta_a}{\beta} \delta(\Pbf_\theta) p_{\eta a+1} p_{\eta a+2} \,,
\\
\label{07082021-man02-13} && \langle p_{\eta_{a+1}} p_{\eta_{a+2}} \rangle = \frac{1}{\beta} \delta(\Pbf_\theta) \Po_\theta \frac{p_{\eta a}}{\beta_a} \,,
\\
\label{07082021-man02-14} && \langle \Po_\theta p_{\eta a+1} p_{\eta a+2} \rangle  = \delta(\Pbf_\theta) \frac{p_{\eta a}}{\beta_a} \,,
\\
\label{07082021-man02-15} && \langle  p_{\eta 1} p_{\eta 2} p_{\eta 3}\rangle =\frac{1}{\beta} \delta(\Pbf_\theta) \Po_\theta\,,
\\
\label{07082021-man02-16} && \langle \Po_\theta p_{\eta 1} p_{\eta 2} p_{\eta 3}\rangle = \delta(\Pbf_\theta) \,,
\eeq
where $\beta\equiv\beta_1\beta_2\beta_3$, while $\Po_\theta$ is defined in \rf{27082021-man02-32}.
In turn, by using relations \rf{07082021-man02-09}-\rf{07082021-man02-16}, we derive the relations which are helpful for the studying hermicity properties of the cubic vertices
\beq
\label{07082021-man02-17} && \langle \big( \kappa \Po_\theta - \beta \sum_{a=1,2,3}  \frac{m_a}{\beta_a} p_{\eta a} \big) \rangle
\nonumber\\
&& \hspace{2cm} =\,\,   \frac{1}{\kappa \beta} \delta(\Pbf_\theta) \big( \kappa \Po_\theta + \beta \sum_{a=1,2,3} \frac{m_a}{\beta_a} p_{\eta a} \big) \sum_{b=1,2,3} m_b p_{\eta b+1} p_{\eta b+2} \,,
\\
\label{07082021-man02-18} && \langle \big( \kappa \Po_\theta -\beta  \sum_{a=1,2,3} \frac{m_a}{\beta_a} p_{\eta a}\big) \sum_{b=1,2,3} p_{\eta b} V_b\rangle
\nonumber\\
&& \hspace{1cm} =\,\,  -\frac{\kappa}{\beta^2} \delta(\Pbf_\theta) \big( \kappa \Po_\theta + \beta  \sum_{a=1,2,3} \frac{m_a}{\beta_a} p_{\eta a}\big) \sum_{b=1,2,3} p_{\eta b} f_b\,,
\eeq
where quantities $f_a$ appearing in \rf{07082021-man02-18} are defined in terms of the vertices $V_a$ as
\be \label{07082021-man02-19}
f_a \equiv \frac{\beta}{\kappa^2} \big( m_{a+1} \beta_{a+2} V_{a+2} - m_{a+2} \beta_{a+1} V_{a+1}\big)  \,, \qquad a=1,2,3\,.
\ee
Inverse relations to the ones in \rf{07082021-man02-19} are given by
\be \label{07082021-man02-20}
V_a = \frac{1}{\beta} \big(m_{a+1} \beta_{a+2} f_{a+2} - m_{a+2} \beta_{a+1} f_{a+1}\big)\,,\qquad a=1,2,3\,.
\ee

\appendix{ Properties of superfields  $\Theta_{m,\lambda,u}^*$} \label{app-02}

We find it convenient to use superfields $\Theta_{m,\lambda,u}^*$ for building interaction vertices. These superfields are defined in \rf{01092021-man02-39}-\rf{01092021-man02-44}.

\noindent {\bf Differential operators realization of N=2 Poincar\'e superalgebra on superfield $\Theta_{m,\lambda,u}^*$ }: The realization is given by
\beq
\label{28082021-man02-01} && P^1 = - p\,, \qquad P^+ = - \beta\,, \qquad P^- = - p^-\,, \qquad p^- = -\frac{p^2 + m^2}{2\beta}\,,
\\
\label{28082021-man02-02} && J^{+1}  = \irm x^+ P^1 + \partial_p \beta\,,\hspace{1cm} J^{+-}  = \irm x^+ P^- + \partial_\beta \beta + M^{+-}\,,
\\
\label{28082021-man02-03} && J^{-1} = - \partial_\beta p + \partial_p p^- - \frac{\irm m}{\beta}\big( \lambda  - \half   p_\theta \partial_{p_\eta} - \half p_\eta \partial_{p_\theta}\big) - \frac{p}{\beta} M^{+-}\,,
\\
\label{28082021-man02-04} && U =  1 + u - p_\theta \partial_{p_\theta} -  p_\eta \partial_{p_\eta} \,,
\\
\label{28082021-man02-05} && \hspace{1cm} M^{+-} \equiv \half p_\theta \partial_{p_\theta}  + \half p_\eta \partial_{p_\eta} -\half e_\lambda\,,
\\
\label{28082021-man02-06} && Q^{+\Rsm} = (-)^{e_\lambda}\beta \partial_{p_\theta}\,, \hspace{3cm} Q^{+\Lsm} = (-)^{e_{\lambda+\half} } p_\theta\,,
\\
\label{28082021-man02-08} && Q^{-\Rsm} = \frac{(-)^{e_\lambda}}{\sqrt{2}} \big(p  \partial_{p_\theta} + \irm m \partial_{p_\eta}\big)\,, \hspace{1cm}  Q^{-\Lsm} = \frac{(-)^{e_{\lambda+\half}}}{\sqrt{2}\,\beta} \big( p p_\theta - \irm m p_\eta\big)\,,
\eeq
where the $e_\lambda$ is defined in \rf{26082021-man02-34}.
Note that superfields $\Theta_{m,\lambda,u}^*$ and $\Theta_{m,\lambda,u}$ are related as
\be \label{28082021-man02-10}
\beta \Theta_{m,\lambda,-u}^*(-p,-p_\theta,p_\eta) = \Theta_{m,\lambda,u}(p,p_\theta,p_\eta)\,.
\ee

Making use of the relations \rf{28082021-man02-01}-\rf{28082021-man02-10} and \rf{01092021-man02-46},\rf{01092021-man02-47}, it is easy to check the standard equal-time (anti)commutator between the generators and the superfields $\Theta_{m,\lambda,u}^*$,
\be \label{28082021-man02-11}
[\Theta_{m,\lambda,u}^*,G_\smpt]_{\pm} =  G_\diff \Theta_{m,\lambda,u}^* \,,
\ee
where $G_\diff$ may be found in \rf{28082021-man02-01}-\rf{28082021-man02-08}.
If $\Theta_{m,\lambda,u}^{*\dagger}$ is a hermitian conjugate of the superfield $\Theta_{m,\lambda,u}^*$, then one has the relation
\beq
\label{28082021-man02-12} && \Theta_{m,\lambda,u}^*(p,p_\theta,p_\eta)^\dagger  = \beta \int dp_\eta dp_\theta\, e^{\frac{1}{\beta}(p_\theta p_\theta^\dagger + p_\eta p_\eta^\dagger)} \Theta_{m,\lambda,-u}^*(-p,p_\theta, -p_\eta)\,.
\eeq

\noindent {\bf Hermitian conjugation rules for vertices}. Using relation \rf{28082021-man02-12} and requiring the Hamiltonian $P_\smp3^-$ to be hermitian, we find that the type-1 $\Phi\Phi\Phi$ and $\Psi\Psi\Phi$ vertices $V$, $\Vb$ in \rf{01082021-man02-01} are related to the type-3 $\Phi\Phi\Phi$ and $\Psi\Psi\Phi$ vertices $V'$, $\Vb'$  \rf{01082021-man02-05} in the following way:
\beq
\label{28082021-man02-13} && V^* = \varepsilon_{_{\Theta_1\Theta_2\Theta_3}} I_uI_\beta \Vb'\,, \hspace{0.8cm} \Vb^*  = \varepsilon_{_{\Theta_1\Theta_2\Theta_3}} I_uI_\beta V'\,, \hspace{1cm} \hbox{for non-critical masses;}\qquad
\\
\label{28082021-man02-14} && V^* = \varepsilon_{_{\Theta_1\Theta_2\Theta_3}} I_uI_\beta  V'\,, \qquad \Vb^* =
\varepsilon_{_{\Theta_1\Theta_2\Theta_3}} I_uI_\beta \Vb'\,, \hspace{1cm} \hbox{for critical masses;}
\eeq
while, for the type-2 $\Psi\Psi\Phi$ and $\Psi\Psi\Psi$ vertices $V_a$, $\Vb_a$ in \rf{01082021-man02-01}, we find the relations
\beq
\label{28082021-man02-15} &&  \frac{1}{\kappa} \big(  m_{a+2} \beta_{a+1} V_{a+1}^* - m_{a+1} \beta_{a+2} V_{a+2}^*\big)  = \varepsilon_{_{\Theta_1\Theta_2\Theta_3}} I_\beta I_u\Vb_a\,, \hspace{0.5cm} \hbox{for non-critical masses;} \qquad
\\
\label{28082021-man02-16} && \frac{1}{\kappa} \big(  m_{a+2} \beta_{a+1} V_{a+1}^* - m_{a+1} \beta_{a+2} V_{a+2}^*\big)  = \varepsilon_{_{\Theta_1\Theta_2\Theta_3}} I_\beta I_u V_a\,, \hspace{0.5cm} \hbox{for critical masses;} \qquad
\\
\label{28082021-man02-17} &&  \frac{1}{\kappa} \big(  m_{a+2} \beta_{a+1} \Vb_{a+1}^* - m_{a+1} \beta_{a+2} \Vb_{a+2}^*\big)  = \varepsilon_{_{\Theta_1\Theta_2\Theta_3}} I_\beta I_u\Vb_a\,, \hspace{0.5cm} \hbox{for critical masses;} \qquad
\\
&& \hspace{3cm} \varepsilon_{_{\Theta_1\Theta_2\Theta_3}}   \equiv (-)^{e_{\lambda_1} e_{\lambda_2} + e_{\lambda_2} e_{\lambda_3} + e_{\lambda_3} e_{\lambda_1}}\,,
\eeq
where, in \rf{28082021-man02-15}-\rf{28082021-man02-17}, $a=1,2,3$. In \rf{28082021-man02-13}-\rf{28082021-man02-17}, the asterisk stands for the complex conjugation, while the operators $I_u$, $I_\beta$ are defined as in \rf{07082021-man02-01}.
We recall that, for the non-critical masses, the $\kappa$ is defined in \rf{kappa}, while, for the critical masses, we use $\kappa$ given in \rf{01082021-man02-34-a2}. We recall also that, for the superfield $\Phi$, $e_\lambda=0$, while, for the superfield $\Psi$,   $e_\lambda=1$. This gives the relations
\be
\varepsilon_{_{\Phi_1\Phi_2\Phi_3}}=1\,, \quad \varepsilon_{_{\Phi_1\Phi_2\Psi_3}}=1\,, \quad
\varepsilon_{_{\Psi_1\Psi_2\Phi_3}}=-1\,, \qquad \varepsilon_{_{\Psi_1\Psi_2\Psi_3}}=-1\,.
\ee

\appendix{ Derivation of superspace representation for cubic vertex $p_\smp3^-$ and densities $j_\smp3^{-1}$, $q_\smp3^{-\Rsm,\Lsm}$} \label{app-03}

In this Appendix, firstly, we outline a procedure of the derivation of the superspace representation for the vertices $p_\smp3^-$ given in \rf{01082021-man02-01}, \rf{01082021-man02-05}, \rf{01082021-man02-08} and equations for the $V$-vertices given in \rf{01082021-man02-02},\rf{01082021-man02-03}, \rf{01082021-man02-06},\rf{01082021-man02-07}, and \rf{01082021-man02-09}-\rf{01082021-man02-12}. Secondly, we outline  the derivation of the densities $j_\smp3^{-1}$, $q_\smp3^{-\Rsm,\Lsm}$ given in \rf{01082021-man02-13}-\rf{01082021-man02-22}. Thirdly, we outline a procedure of the derivation of the cubic densities for the three massless fields in Sec.\ref{sec-09}. Our procedure of the derivation of the cubic densities in Sec.\ref{sec-06} is realized in the following five steps.

\noinbf{Step 1}. By definition, the vertex $p_\smp3^-$ is degree-1 polynomial in $\Po$. Expanding then the $p_\smp3^-$ into the $\Po$ and the Grassmann momentum $\Po_\theta$, we get
\beq
\label{29082021-man02-01} && p_\smp3^- = \irm \Po V_\Po + V_0\,,
\\
\label{29082021-man02-02} && V_\Po = \Po_\theta V_{\Po\Po_\theta} + V_{\Po 0}\,, \hspace{1cm} V_0 = \Po_\theta V_{0\Po_\theta} + V_{00}\,,
\eeq
where vertices $V_{\Po\Po_\theta}$, $V_{\Po 0}$, $V_{0\Po_\theta}$, $V_{00}$ depend on the Grassmann momenta $p_{\eta_1}$, $p_{\eta_2}$, $p_{\eta_3}$ and the momenta $\beta_1$, $\beta_2$, $\beta_3$.

\noinbf{Step 2}. At this step we consider restrictions imposed by the equations \rf{27082021-man02-43},\rf{27082021-man02-44}. Plugging \rf{29082021-man02-01}
into \rf{27082021-man02-43},\rf{27082021-man02-44} and comparing terms of various powers of the momenta $\Po$ and $\Po_\theta$, we get expressions for the supercharge densities given by
\be \label{29082021-man02-03}
q_\smp3^{-\Rsm} = -  \irm \varepsilon\sqrt{2}\, \beta  V_{\Po\Po_\theta}\,, \hspace{1cm} q_\smp3^{-\Lsm} =  \irm \varepsilon \sqrt{2}\,  \Po_\theta V_{\Po 0}\,.
\ee
and the following restrictions on the vertices appearing in \rf{29082021-man02-02},
\beq
\label{29082021-man02-04}  &&  V_{0\Po_\theta} - \sum_{a=1,2,3}  m_a \partial_{p_{\eta a}} V_{\Po 0} = 0 \,, \hspace{2.9cm} \sum_{a=1,2,3}  m_a \partial_{p_{\eta a}} V_{\Po\Po_\theta}   = 0 \,,
\\
\label{29082021-man02-06} && \sum_{a=1,2,3}  m_a \partial_{p_{\eta a}} V_{00} + \beta \sum_{a=1,2,3} \frac{m_a^2}{\beta_a} V_{\Po\Po_\theta} = 0 \,,  \hspace{1cm}  \sum_{a=1,2,3}  m_a \partial_{p_{\eta a}} V_{0\Po_\theta}   = 0 \,,
\\
\label{29082021-man02-08} &&  \frac{1}{\beta} V_{00} + \sum_{a=1,2,3}  \frac{m_a}{\beta_a} p_{\eta a} V_{\Po\Po_\theta} = 0 \,,  \hspace{2.8cm}  \sum_{a=1,2,3}  \frac{m_a}{\beta_a} p_{\eta a} V_{\Po 0}   = 0 \,,
\\
\label{29082021-man02-10} && \sum_{a=1,2,3}  \frac{ m_a}{\beta_a} p_{\eta a} V_{0\Po_\theta} - \sum_{a=1,2,3} \frac{m_a^2}{\beta_a} V_{\Po 0} = 0 \,,  \hspace{1.5cm}  \sum_{a=1,2,3}  \frac{m_a}{\beta_a} p_{\eta a} V_{00}   = 0 \,.
\eeq
We recall that $\beta$ and $\varepsilon$ are defined in \rf{27082021-man02-48},\rf{27082021-man02-49}.
Expanding the vertices $V_{\Po\Po_\theta}$, $V_{\Po 0}$, $V_{0\Po_\theta}$, and $V_{00}$
into the Grassmann momenta $p_{\eta_a}$, we find that the general solution to equations \rf{29082021-man02-04}-\rf{29082021-man02-10} can be presented as
\beq
\label{29082021-man02-12}  && V_{\Po\Po_\theta} =  X + \sum_{a=1,2,3} X_a p_{\eta a} + Y' \sum_{a=1,2,3} m_a p_{\eta a+1}p_{\eta a+2}\,,
\\
\label{29082021-man02-13} && V_{00} = - \beta  X \sum_{b=1,2,3} \frac{m_b}{\beta_b}p_{\eta b} - \beta  \sum_{b=1,2,3} \frac{m_b}{\beta_b}p_{\eta b}\sum_{a=1,2,3} X_a p_{\eta a} + \kappa^2 Y' p_{\eta1} p_{\eta2}p_{\eta3}\,,
\\
\label{29082021-man02-14} && V_{0\Po_\theta} = \kappa^2 Y + \kappa^2 \sum_{a=1,2,3} Y_a p_{\eta a} + X' \sum_{a=1,2,3} m_a p_{\eta a+1}p_{\eta a+2}\,,
\\
\label{29082021-man02-15}  && V_{\Po 0} = - \beta Y \sum_{b=1,2,3} \frac{m_b}{\beta_b}p_{\eta b} - \beta \sum_{b=1,2,3} \frac{m_b}{\beta_b} p_{\eta b}\sum_{a=1,2,3} Y_a p_{\eta a} + X' p_{\eta1} p_{\eta2}p_{\eta3}\,,
\eeq
where vertices $X_a,Y_a$ satisfy the algebraic constraints
\be \label{29082021-man02-16}
\sum_{ a=1,2,3} m_a X_a = 0 \,, \hspace{1cm} \sum_{ a=1,2,3} m_a Y_a = 0 \,,
\ee
and vertices $X,Y$, $X',Y'$, $X_a,Y_a$, $a=1,2,3$, depend only on the momenta $\beta_1$, $\beta_2$, $\beta_3$. We recall that the $\kappa$ appearing in \rf{29082021-man02-13},\rf{29082021-man02-14} is defined in \ref{kappa}.

\noinbf{Step 2}. At this step we consider restrictions imposed by the equations \rf{27082021-man02-45}. Plugging \rf{29082021-man02-01},\rf{29082021-man02-02} into \rf{27082021-man02-45}, we find relation \rf{01082021-man02-13} and the following equations
\beq
\label{29082021-man02-17} &&   \sum_{a=1,2,3}\big( \frac{m_a^2}{\beta_a} \No_\beta^{\eta E} - \frac{\betach_a m_a^2}{6\beta_a}\big) V_{\Po\Po_\theta} - \irm \MM  V_{0\Po_\theta} -  \sum_{a=1,2,3} \frac{ \betach_a }{6\beta} m_a  \partial_{p_{\eta a}} V_{00}   = 0\,,
\\
\label{29082021-man02-18} &&   \sum_{a=1,2,3}\big( \frac{m_a^2}{\beta_a} \No_\beta^{\eta E} - \frac{\betach_a m_a^2}{6\beta_a}\big) V_{\Po 0} - \irm \MM  V_{00} +  \sum_{a=1,2,3}  \frac{ \betach_a }{6\beta_a}  m_a p_{\eta a} V_{0\Po_\theta}   = 0\,,
\\
\label{29082021-man02-19} && \frac{1}{\beta} \No_\beta^{\eta E} V_{0\Po_\theta} + \irm \MM  V_{\Po\Po_\theta} + \sum_{a=1,2,3} \frac{ \betach_a }{6\beta} m_a \partial_{p_{\eta a}}  V_{\Po 0} = 0 \,,
\\
\label{29082021-man02-20} && \frac{1}{\beta} \No_\beta^{\eta E} V_{00} + \irm \MM  V_{\Po 0} - \sum_{a=1,2,3}  \frac{ \betach_a }{6\beta_a} m_a p_{\eta a}  V_{\Po\Po_\theta} = 0 \,.
\eeq
Plugging \rf{29082021-man02-12}-\rf{29082021-man02-15} into \rf{29082021-man02-17}-\rf{29082021-man02-20} and introducing $V$-vertices by the relations
\beq
\label{29082021-man02-21} && X = \half (V+\Vb)\,, \hspace{1.3cm} Y = \frac{1}{2\kappa} (V-\Vb)\,,
\\
\label{29082021-man02-22}  && X' = \half (V'+\Vb')\,, \hspace{1cm}  Y' = \frac{1}{2\kappa} (V'-\Vb')\,,
\\
\label{29082021-man02-23}  && X_a = \half (V_a+\Vb_a)\,, \hspace{1cm}  Y_a = \frac{1}{2\kappa} (V_a-\Vb_a)\,,
\eeq
we find that equations \rf{29082021-man02-17}-\rf{29082021-man02-20} amount to the 1st equations for the $V$-vertices given in \rf{01082021-man02-02},\rf{01082021-man02-03}, \rf{01082021-man02-06},\rf{01082021-man02-07} and equations given in \rf{01082021-man02-09},\rf{01082021-man02-11}.

\noinbf{Step 3}. Using $J^{+-}$ symmetry equation for $p_\smp3^-$ in \rf{27082021-man02-35} and relations \rf{29082021-man02-01},\rf{29082021-man02-02}, \rf{29082021-man02-12}-\rf{29082021-man02-15}, \rf{29082021-man02-21}-\rf{29082021-man02-23}, we obtain the 2nd equations for $V$ vertices in \rf{01082021-man02-02},\rf{01082021-man02-03},  \rf{01082021-man02-06},\rf{01082021-man02-07}, and equations in \rf{01082021-man02-10},\rf{01082021-man02-12}.

\noinbf{Step 4}. Plugging expressions $V_{\Po\Po_\theta}$ \rf{29082021-man02-12}, $V_{\Po 0}$ \rf{29082021-man02-15} into $V_\Po$ \rf{29082021-man02-02} and using \rf{29082021-man02-21}-\rf{29082021-man02-23}, we find expressions for $V_\Po$ given in \rf{01082021-man02-14}, \rf{01082021-man02-17}, \rf{01082021-man02-20}.

\noinbf{Step 5}. Plugging expressions $V_{\Po\Po_\theta}$ \rf{29082021-man02-12}, $V_{\Po 0}$ \rf{29082021-man02-15} into \rf{29082021-man02-03} and using \rf{29082021-man02-21}-\rf{29082021-man02-23}, we find expressions for the supercharge densities $q_\smp3^{-\Rsm,\Lsm}$ given in \rf{01082021-man02-15},\rf{01082021-man02-16}, \rf{01082021-man02-18},\rf{01082021-man02-19}, \rf{01082021-man02-21}, and \rf{01082021-man02-22}.

\noinbf{Cubic densities for massless fields}. For this case, as above, we start with the general expressions given in \rf{29082021-man02-01},\rf{29082021-man02-02}. Setting $m_1=0$, $m_2=0$, $m_3=0$ in \rf{29082021-man02-04},\rf{29082021-man02-08}, we then find
\be \label{29082021-man02-24}
V_{0\Po_\theta}=0\,, \hspace{2cm} V_{00}=0.
\ee
Using \rf{29082021-man02-24} in \rf{29082021-man02-01}, we get $p_\smp3^-$ given in \rf{06082021-man02-01}.
Using $p_\smp3^-$ \rf{06082021-man02-01} in \rf{27082021-man02-45}, we get
$j_\smp3^{-1}$ given in \rf{06082021-man02-02}. After that we find that, for $m_1=0$, $m_2=0$, $m_3=0$, the relations in \rf{29082021-man02-04}-\rf{29082021-man02-10} do not impose constraints for the dependence of the $V_{\Po\Po_\theta}$, $V_{\Po 0}$ on the Grassmann momenta $p_{\eta_a}$. This implies that the Grassmann momenta expansion for the $V_{\Po\Po_\theta}$, $V_{\Po 0}$ takes the most general form given by
\beq
\label{29082021-man02-25} && V_{\Po\Po_\theta} =  X + \sum_{a=1,2,3}  \frac{p_{\eta a}}{\beta_a} Y_a+ \frac{1}{\beta} \sum_{a=1,2,3} p_{\eta a+1}p_{\eta a+2} \Xb_a  +  \frac{1}{\beta} p_{\eta_1}p_{\eta_2}p_{\eta_3} \Yb\,,
\\
\label{29082021-man02-26} && V_{\Po 0} =  Y + \sum_{a=1,2,3} \frac{p_{\eta a}}{\beta_a} X_a  + \sum_{a=1,2,3}  p_{\eta a+1}p_{\eta a+2} \Yb_a + p_{\eta_1}p_{\eta_2}p_{\eta_3} \Xb\,.
\eeq

Using \rf{29082021-man02-25}, \rf{29082021-man02-26} in \rf{29082021-man02-03}, we get expressions for the supercharge densities given in \rf{06082021-man02-04}, \rf{06082021-man02-06}, \rf{06082021-man02-08}, \rf{06082021-man02-09}, and \rf{06082021-man02-11}. Comparing \rf{29082021-man02-25} and \rf{29082021-man02-12}, we see appearance of the additional term of degree-3 in the $p_{\eta_a}$ in \rf{29082021-man02-25}, while, comparing \rf{29082021-man02-26} and \rf{29082021-man02-15}, we see appearance of the additional term of degree-0 in the $p_{\eta_a}$ in \rf{29082021-man02-26}. It is those additional terms that lead to the type-0 vertices \rf{06082021-man02-09} and the type-4 vertices \rf{06082021-man02-10} for massless fields in Sec.\ref{sec-09}.


\end{document}